\documentclass[11pt]{article}
\usepackage{fancyhdr}
\usepackage{isomath}
\usepackage{amsmath}
\usepackage{amsbsy}
\usepackage{amssymb}
\usepackage{amscd}
\usepackage{amsfonts}
\usepackage{graphicx,color}
\usepackage{verbatim}
\usepackage{euscript}
\usepackage{alltt}
\usepackage{stmaryrd}
\usepackage{subfigure}
\usepackage{relsize}

\DeclareGraphicsExtensions{.eps,.pdf}
\usepackage{amsmath}
\usepackage{amsbsy}
\usepackage{amssymb}
\usepackage{amscd}
\usepackage{amsfonts}

\newcommand{\bfe}{{\mathbold e}}

\newcommand{\beq}{\begin{equation}}
\newcommand{\eeq}{\end{equation}}
\newcommand{\beqs}{\begin{eqnarray}}
\newcommand{\eeqs}{\end{eqnarray}}
\newcommand{\beql}{\begin{equation} \label}
\newcommand{\half}{\frac{1}{2}}

\usepackage[margin=1in]{geometry}

\newenvironment{rcases}
  {\left.\begin{aligned}}
  {\end{aligned}\right\rbrace}
\newcommand{\bsl}{\backslash}
\newcommand{\veps}{\varepsilon}

\date{}
\begin{document}
\title{Computing with non-orientable defects: nematics, smectics and natural patterns}
\author{Chiqun Zhang\thanks{Microsoft, CA, email:chiqun0524@gmail.com.} \and Amit Acharya\thanks{Department of Civil \& Environmental Engineering, and Center for Nonlinear Analysis, Carnegie Mellon University, Pittsburgh, PA 15213, email: acharyaamit@cmu.edu.} \and Alan C. Newell \thanks{Department of Mathematics, University of Arizona, Tucson, AZ 85721, email: anewell@math.arizona.edu.} \and Shankar C. Venkataramani\thanks{Department of Mathematics, University of Arizona, Tucson, AZ 85721, email: shankar@math.arizona.edu.}
}
\maketitle

\begin{abstract}

\noindent Defects are a ubiquitous feature of ordered media. They have certain universal features, independent of the underlying physical system, reflecting their topological origins. While the topological properties of defects are robust, they appear as `unphysical' singularities, with non-integrable energy densities in coarse-grained macroscopic models. We  develop a principled approach for enriching coarse-grained theories with enough of the `micro-physics' to obtain thermodynamically consistent, well-set models, that allow for the investigations of dynamics and interactions of defects in extended systems.  We also develop associated numerical methods that are applicable to computing energy driven behaviors of defects across the amorphous-soft-crystalline materials spectrum. Our methods can handle order parameters that have a head-tail symmetry, i.e. director fields, in systems with a continuous translation symmetry, as in nematic liquid crystals, and in systems where the translation symmetry is broken, as in smectics and convection patterns. We illustrate our methods with explicit computations.
\end{abstract}

\section{Introduction}

Macroscopic physical systems consist of large numbers of interacting (microscopic) parts and are described by thermodynamic principles. A central tenet of thermodynamics is that the equilibrium state, and the relaxation to equilibrium, are described by an appropriate free energy \cite{greiner1995thermodynamics}.  While the details differ, free energies describing systems that spontaneously generate ordered/patterned states have certain universal features {\em independent of the underlying physics}. These features are present in free energies that  describe 
many systems including liquid crystals \cite{liquidcrystals-variational}, micro-magnetic devices \cite{desimone2000magnetic} and solid-solid phase transitions \cite{kohn1994surface}. 
They are
\begin{enumerate}
\item nonconvexity of the free energy and the existence of 
multiple ``near" ground states for the system. 
\item regularization by a singular perturbation (an ``ultraviolet cutoff") to preclude the formation of structures on arbitrarily fine scales.
\end{enumerate}
The implications of these features are two fold. The first aspect, non-convexity, naturally leads to ``massive" non-uniqueness/degeneracy of ground states. As we argue below, this behavior can be described in terms of small-scale defects in the system which appear as singularities in macroscopic coarse-grained theories. The second aspect, a reflection of ``microscopic physics" in the system, is necessary for explaining the deterministic physical evolution of systems with defects. Indeed, as we discuss below, defects typically have non-integrable energy densities within the macroscopic theory. Their physical nature is thus only revealed through ``renormalizing" away these infinities \cite{Costello2011Renormalization}. We develop a framework for this procedure by enriching the macroscopic theory to include additional physical fields, that reflect some of the microscopic physics and describe the interaction/dynamics of defects \cite{kleinert1989gauge}. This enrichment is crucial for developing appropriate models and robust numerical methods for the dynamics of defects, which are either degenerate (zero energy) or locked (infinite-energy) from the viewpoint of a ``naive" macroscopic theory.

Why should defects occur in extended systems with non-convex energies? For example, in the case of a large Prandtl number convection, fluid is heated uniformly from below, the emerging flow pattern self-organizes into rolls/stripes with a preferred wavelength. This reflects the breaking of the continuous translation symmetry of the ambient space/forcing 
to a discrete symmetry (translation by one wavelength perpendicular to the stripes). The rotational symmetry of the system is however unbroken so that there is no preferred orientation of the stripes. In large aspect ratio systems, where the box size is very large compared to the preferred wavelength, the local orientation is chosen by local biases (such as boundary effects). Thus, the emerging pattern is a mosaic of patches of stripes with preferred wavelength with different orientations, patches which meet and meld along (in 2D) grain boundaries which themselves meet at points. These lines and points in 2D (planes, loops and points in 3D) constitute the defects in the convection pattern.

This argument is not peculiar to convection patterns. Defects arise as well in crystalline elastic solids, and complex fluids in the nematic and smectic phases, including  
dislocations, disclinations, grain boundaries, and twin (phase) boundaries.
Fundamentally, defects arise in these phases due to the presence of microscopic structural symmetries. In crystalline solids the possibility of non-trivial deformations that preserve lattice periodicity locally gives rise to defects, and in nematic liquid crystal phases it is due to the head-tail symmetry of the director field. 

The macroscopic state of an extended system is therefore best understood as a patchwork of domains that meet at various types of defects \cite{mermin1979topological}: disclinations, dislocations, monopoles, walls, etc. It is thus of 
 interest to 
 develop tools that allow us to understand, predict, control and manipulate  energy driven pattern formation. 
 For energy driven systems, these defects are singular solutions of the {\em order parameter equations} that arise from averaging the free energy over all the microstructures consistent with the macroscopic order. Since these 
 equations depend (largely) only on the relevant broken symmetries,  they are universal, i.e., the same equations arise in a variety of physical contexts \cite{passot1994towards}. The topologies of allowed defects are also universal \cite{kleman1977classification,kleman1995topological}, and they are captured by discontinuous and/or singular `solutions' of the averaged pdes that are initially derived to describe the domains \cite{ercolani2000geometry}. This universality is one motivation for the work presented in this paper, namely, the idea that there  is a \textit{common modeling and computational methodology} which can be applied to \textit{study defects} in systems with \textit{vastly different physics at widely separated scales}.
 
Useful examples for thinking about defects in macroscopic equations are point charges in electrostatics or vortex filaments in the Euler equation. These objects are ``singular" parts of the Laplacian of the electrostatic potential $\rho = -\Delta \phi$ and the vorticity (curl of the velocity field) $\omega = \nabla \times \mathbf{v}$ respectively. A point charge has an infinite electrostatic self-energy and likewise, the velocity of a fluid diverges as one approaches an ideal vortex line. \textit{This signals the existence of new physics that the macroscopic model misses}. The divergence of the electrostatic energy near point charges can be ``fixed" by smearing them out (somewhat arbitrarily). For fluids, the Euler equation is missing one aspect of the physics related to dissipation on microscopic scales, namely the bulk viscosity, which smooths the vorticity distribution. One might even need additional physics beyond bulk viscosity. It is not clear that the presence of bulk viscosity alone can allow for a good model of the evolution of a viscous fluid in high Reynolds number regimes where vortex sheet and filament structures begin to appear, as in Euler flows. We draw the following lessons from these examples: 
\begin{enumerate}
    \item Defects can be represented by the singular parts of derivatives of appropriate order acting on the macroscopic fields, like the electrostatic potential or the fluid velocity.
    \item There is a need for a principled introduction of additional degrees of freedom to the universal macroscopic models in order to reflect those aspects of the microscopic physics that are relevant to the energetics and the thermodynamics of defects in extended system.  These additional fields take the form of smooth representations of `singular' parts of higher gradients of the macroscopic fields, since they naturally encode defect behavior. 
\end{enumerate} 
Using these two principles, and building on earlier work \cite{zhang2015single,zhang2016non}, we develop a general thermodynamic framework to obtain numerically tractable models for defects in extended systems. This is our primary accomplishment in this work. The singularities that arise in this work are `layer fields' and their `terminating discontinuities' (see Fig.~\ref{fig:domain}) which are the analogs  of charge sheets (discontinuities of the electric field $E = - \nabla \phi$) or vortex sheets (discontinuities of the tangential velocity $\mathbf{v}_\parallel$), that are bounded by line charges, respectively, vortex lines.

We conclude this introduction with a guide for the reader. In earlier work by the first two authors, we have discussed the relationship between defects in solids and those in liquid crystals \cite{zhang2015single,zhang2016non}, and used these ideas to develop numerical approaches to computing the behavior of defects in these systems. In this work, we build on these ideas to create a framework that can be applied to many other systems. Defects naturally arise in the solutions of macroscopic coarse-grained theories as 
non-integrable singularities in the energy density. A  primary feature of our approach is the replacement of the singular parts of solutions to `classical' defect models by new independent fields (bearing much similarity to gauge fields) that are nevertheless smooth (but localized). This additional `micro-physics' allows for the utilization of (renormalized) integrable energy densities. This approach is outlined in \S\ref{sec:model}, to model defects in scalar fields by ``borrowing" the perspective of defects in elastic solids leading to plasticity and phase transitions. In \S\ref{sec:numerics}  we demonstrate the scope of our model through numerical computations of equilibrium features of various defect configurations in nematics and in smectics/patterns. We illustrate this circle of ideas in \S\ref{sec:scalar} where we provide a fresh perspective on the topological theory of defects in stripe patterns, and on the importance of correctly identifying {\em all} the relevant micro-scale physics in order to build appropriate long-wave models for the interactions and dynamics of defects. This novel approach complements the ``traditional" use of the Swift-Hohenberg and related equations for natural patterns \cite{swift1977hydrodynamic}. In \S\ref{sec:discuss} we presents a concluding overview of our work along with a discussion of its implications and possible future extensions.

\section{A commmon language for defects in nematics, smectics, natural patterns, and elastic solids }\label{sec:model} \label{sec:wein}

In this section, we develop the theory that leads to our framework for modeling defects and their energetics. We begin, in \S\ref{sec:orderparameters} by discussing various aspects of defects as natural topological objects in macroscopic `order-parameter' theories for elasticity, liquid crystals and pattern formation. We also highlight the similarities and the (significant) differences between the defects in these various systems. In \S\ref{sec:layers} we  demonstrate that the defects can be encoded as the ``singular"-parts of a (higher-order) derivative operator applied to a continuum field in the macroscopic theory. We also outline a mathematical description of defect `strengths' or `charges' in terms of these singular fields.  We then enrich the underlying macroscopic theory by adding new physics, involving smooth `rehabilitated' representations of these singular fields and their thermodynamically conjugate variables \textit{such that the relevant energy density and stresses are locally integrable everywhere}. We carry out this procedure for scalar fields, as applicable to liquid crystals and natural patterns in \S\ref{sec:NPSN}, and for vector and tensor fields, as applicable to elasto-plasticity of solids, in \S\ref{sec:solids}. The models in our framework are constrained by thermodynamics, the balance laws of mechanics, and statements reflecting conservation of topological charge of defects, and have a `good' numerical formulation in terms of smoothed defect fields. We discuss the adequacy of this simplified framework for studying the energetics of patterns, but also of nematic and smectic-A liquid crystals in \S\ref{sec:NPSN}.

\subsection{Order parameters and defects} \label{sec:orderparameters}

\begin{figure}[htb]
\centering
\includegraphics[width=6.5in,height=3.5in]{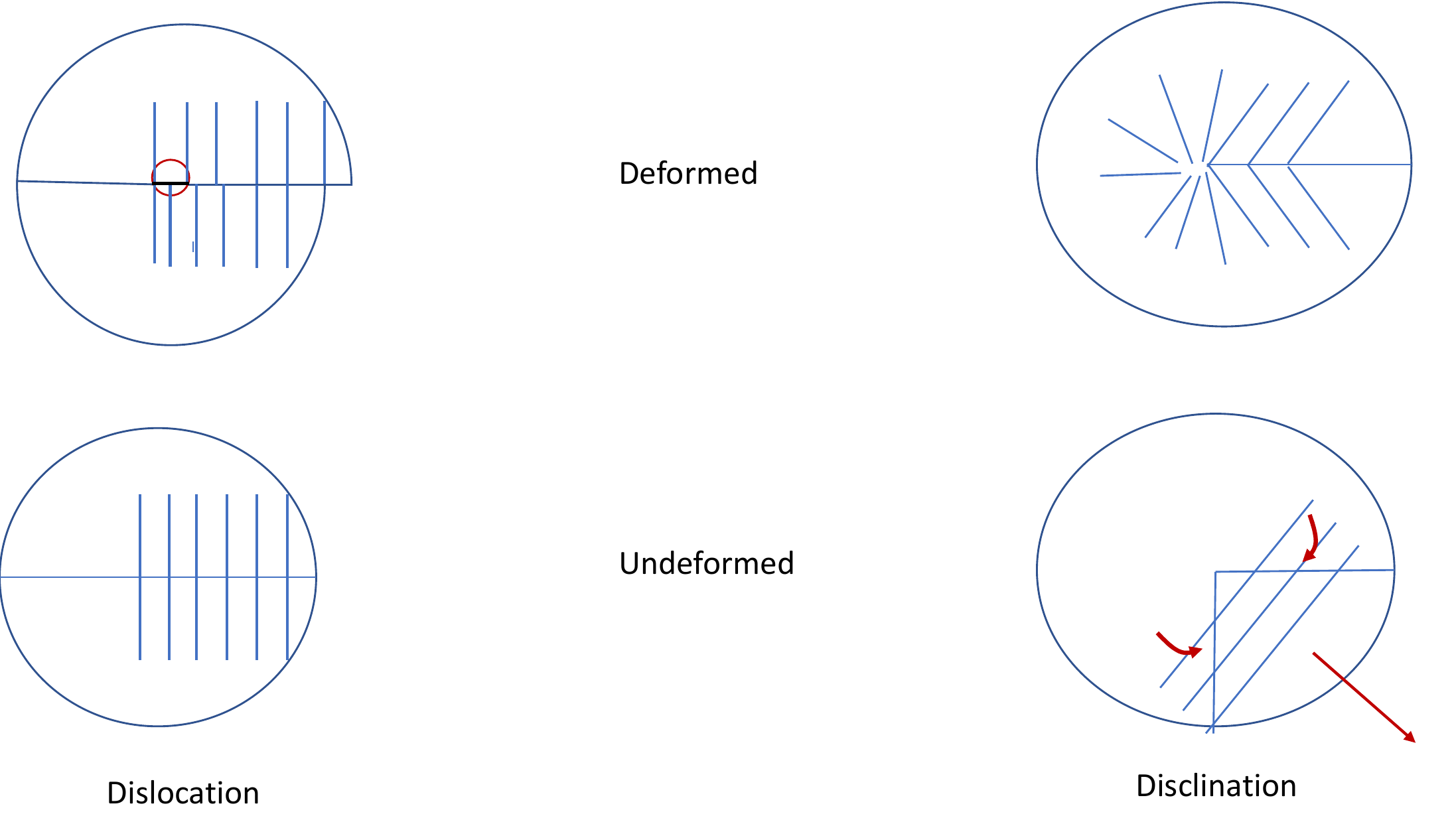}
\caption{Schematic of a dislocation and a disclination in a crystalline solid.}
\label{fig:disloc_disclin}
\end{figure}

In elastic solids, there is a historically systematic way of interpreting the geometric and some elastic (i.e. energetic) aspects of dislocation and disclination defects beginning from Volterra \cite{volterra}, that has been recently generalized and incorporated into modern continuum thermomechanics, generating models for practical application. In this viewpoint, a \textit{disclination} is interpreted as a terminating curve of a surface on which the elastic rotation is discontinuous - this idea has been generalized to consider terminations of surfaces of elastic distortion discontinuity (including strain), with the resulting defect called a \textit{generalized disclination} \cite{zhang2018relevance, zhang2018finite}; a \textit{dislocation} is understood as the terminating curve of a surface on which the (inverse) deformation is discontinuous. This is illustrated in Fig. \ref{fig:disloc_disclin}. The straight lines in both the undeformed configurations are to be interpreted as traces of lattice planes, with the shown spacings that cause no stress in the body and deviations from these spacings resulting in stress.

The dislocation is visualized by making a cut from the right along the horizontal line up to the center of the disk and simply pulling the top of the cut body to the right so that the lattice planes shift by one lattice spacing above the cut, welding the body at the cut and then letting go of the forces that produced the shift, and have the body equilibrate its internal forces by generating further displacements from the welded configuration, under the constraint that at the right end of the cut where it meets the boundary of the body, the shift of the top w.r.t the bottom is 1 lattice spacing.

For the disclination, the wedge shown, obtained by making two cuts, is removed from the body, the two free surfaces brought together and welded and then all applied forces are released and the body allowed to mechanically equilibrate. This results in lattice planes that have the approximate configuration shown in the top figure for the disclination. Clearly the spacings between the lattice planes produced for a body with a disclination are much larger - especially further away from the core of the defect (the center of the body) than in the case of the dislocation (in the sketch of the dislocation the deformations are not obvious but from the procedure for its formation it is clear that there are strains more localized near the core and decaying away from it, in contrast to the situation with the single disclination). By these physical definitions it becomes clear that a dislocation induces a lower energy state in a body than a single disclination. It is no surprise then that dislocations (in the  bulk) and disclination dipoles (at interfaces) are the defects that are most often observed in solids but, as as seen in experiment, individual disclinations in isolation are found in nature at the junction of twin and grain boundaries. 

The core idea above, that line defects of different kinds represent the termination of various types of discontinuities of fields across a two-dimensional surface, lends itself to useful generalization, as has been done to define higher-order \textit{branch point} defects \cite{acharya2020continuum}. It can also be shown, both mathematically and physically \cite{zhang2018relevance}, that a disclination dipole pair in a solid may be interpreted, for small inter-disclination spacing, as a dislocation. 

The relevant field variable for elasticity, in the broader setting including defects, is the {\em inverse deformation} $w_t:\Omega_t \to \mathbb{R}^3$, a vector field on the deforming body $\Omega_t$ at time $t$, and the natural {\em order parameter} is the inverse deformation gradient $Dw_t$ (a $3 \times 3$ matrix). As we discussed above, defects in elastic materials are encoded by discontinuities of $w_t$ and $Dw_t$. The theory of elasticity and of plasticity and other defect mediated behaviors in solids results from defining an appropriate dynamics for the elastic and plastic parts of $Dw_t$, which in essence correspond to the absolutely continuous part and physically mollified, singular part of $Dw_t$, respectively, along with the motion of the body in ambient Euclidean space\footnote{In defect-free elasticity, $w_t$ above is a mapping from the elastically deformed state to a stress-free, time-independent, known \emph{reference} configuration, say $\Omega_R$, and instead of working with $w_t$ and $Dw_t$, one works with the the deformation field $w_t^{-1} = y_t: \Omega_R \to \Omega_t \subset \mathbb{R}^3$ for each $t$ and its deformation gradient $Dy_t$. In the presence of defects and plasticity, there is no distinguished coherent reference available for the body in analogy to a viscous fluid, and nematic and smectic liquid crystals.}. We will use this framework 
as a template to develop models that adequately describe the gross continuum mechanical features of  nematic and smectic liquid crystals. These models can describe the slow evolution of defects\footnote{In elastic-plastic solids, fast motions with significant interactions between defects and material inertia are seamlessly incorporated in such models \cite{zhang2015single, arora2020finite}.} and are thus also useful in studying  the long term coarsening dynamics of natural stripe patterns.

A nematic liquid crystal consists of rod-like molecules and the relevant microscopic field is the (distribution of) orientation of these molecules. The natural order parameter is a director field $n(x)$ that encodes the long-range ordering of the nematic. A smectic is defined by the nematic director $n(x)$ and an additional (nearly periodic) field $\rho(x)$ that describes the maxima and minima of the density of the liquid crystal molecules. While $n$ and $\rho$ are independent in principle, for smectic-A liquid crystals, $n$ is oriented along $\nabla \rho$ motivating de Gennes' definition of a complex order parameter $\psi = A e^{i \theta}$ with $\rho \sim \Re(\psi)$. For stripe patterns, the microscopic field is a locally periodic function $u = u(\theta(x,y)) = A e^{i \theta} + A^* e^{- i \theta}$ . Near defects, $A$ and $\theta$ are independent so the relevant order parameter is $A e^{i \theta}$. Away from defects, however, $A$ is slaved to $\theta$, $A = A(|D \theta|)$ so the director field $k = \pm D\theta$ is the appropriate order parameter $k$.

In all these cases, the order parameter is a director field $n$ or $k$ 
that should be compared with the elastic (or absolutely continuous) part, $W$, of the inverse deformation gradient $Dw_t$ for solids. We can make this analogy if we consider a scalar field $\theta: \Omega_t \to \mathbb{R}$ (instead of a vector field), so that $D \theta$, the analog of $Dw_t$, will give a gradient/vector field whose regular/absolutely continuous part is a representative of the director field $n$ or $k$. At the same time, the dominant elasticity in solids is related to $Dw_t$ whereas that of nematics is related to $Dk$. It is this thread of conceptual unification, while being fully cognizant of the essential physical differences between different systems, that we pursue in this paper, with the formulation of mathematical models that exploit this analogy and the determination of approximate solutions of this model.

What is the motivation for developing a new modeling framework for nematics, smectics and natural patterns? There is, of course, the state-of-the-art model for the mechanics of liquid crystals - the Landau-DeGennes (LDG) model - that can predict a variety of defect related phenomena in liquid crystalline phases \cite{degennes1995book,sonnet2012dissipative}. However, it is fair to say that the LDG model sheds no light on any possible connection that might exist between defects in elasticity of solids and those in liquid crystals, even though such connections have been known to exist from almost the inception of the modern theory of liquid crystals starting with Frank's seminal paper \cite{frank1958liquid} and the fundamental one-to-one correspondence between the elastic fields of the screw dislocation in an elastic solid and the wedge disclination in a nematic liquid crystal. Our paper aims to establish this connection and, in this sense, may be thought to be in the spirit of the work of Kleman \cite{kleman1973defect, kleman2008disclinations}. Abstracting some of the mathematical ideas of defects (and incompatibility) in elastic solids \cite{zhang2015single}, a first successful attempt at representing the energetics and dynamics of a planar director field in nematics along these lines has been demonstrated in \cite{zhang2016non}, which is capable of representing higher-strength defects than $\pm \frac{1}{2}$ as fundamental entities, including their annihilation and dissociation. However, this model relies strongly on the angle parametrization of a planar director field and does not generalize cleanly to 3D, i.e. without any reliance on a specific parametrization of the director field, as shown in Appendix \ref{app:angle_3D}. The full 3-d model we present in this paper overcomes this shortcoming, while being physically more in line with elastic defect theory.

Stripe patterns form in extended systems when a homogeneous state loses stability to periodic state(s) with a preferred wavelength, but no preferred orientation \cite{swift1977hydrodynamic}. As we argued above, in large systems whose size is much bigger than the preferred length scale, defects are both inevitable and ubiquitous. The Cross-Newell equation \cite{cross1984convection}, and its weak/singular solutions \cite{newell1996defects}, describe the long term evolution of stripe patterns and the defects in them. The parallels between defects in stripe patterns and in elastic sheets have been explored in recent work \cite{newell2017elastic}. In this paper we establish a further parallel with the elastic theory of defects in solids, and treat smectic liquid crystals and stripe patterns within a unified framework. Although the defects we consider have universal features, it is important to recognize that defects in different physical systems have important differences as well. There isn't a  single one-size-fits-all model that works for all physical systems with ``defects". Therefore, our goal is to {\em develop a framework that can be adapted to the particular physics of various systems}, and thus has wide applicability.

\subsection{Discontinuities, layer fields and defect charges} \label{sec:layers}

\begin{figure}
\centering
\includegraphics[width=3.5in, height=3.5in]{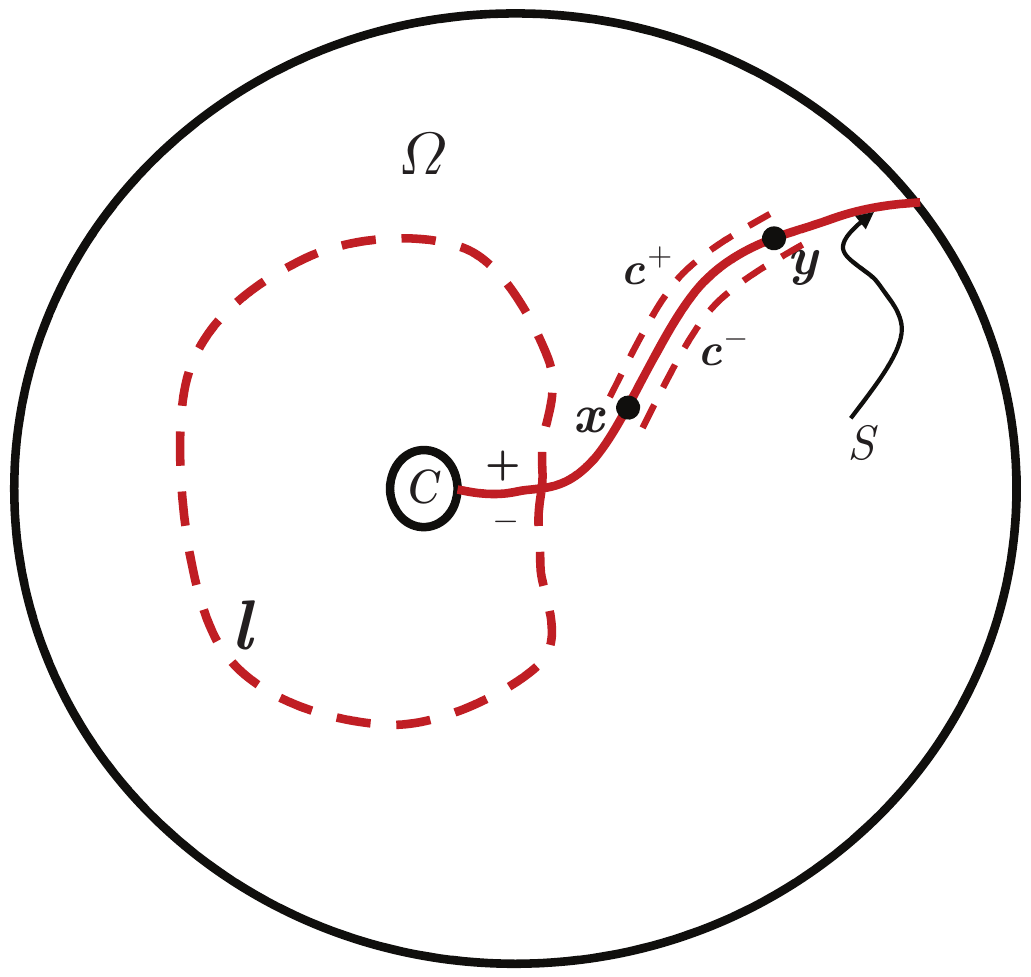}
\caption{Schematic for evaluating jump of phase field}
\label{fig:domain}
\end{figure}

As an illustrative example, we consider a situation with a macroscopic phase (scalar field) $\tilde{\theta}$ whose defects are encoded through its (distributional) second gradient $\tilde{A}_{ij} = \partial_i \partial_j \tilde{\theta}$. With reference to Fig. \ref{fig:domain}, consider the non-simply connected domain $\Omega$ with a through hole. For simplicity, $\Omega$ is depicted as a $2-d$ domain but the argument applies, without loss of generality to a $3-d$ domain containing a through-hole or a toroidal cavity (cf. \cite{zhang2018relevance, acharya2020continuum}). Let $S$ be a surface in $\Omega$ such that $\Omega \bsl S$ is simply connected. Given a smooth, symmetric second-order tensor field $\tilde{A}$ with vanishing $curl$ on $\Omega$, i.e., satisfying $\tilde{A}_{ij} = \tilde{A}_{ji}, \tilde{A}_{ij,k} = \tilde{A}_{ik,j}$, the question is to characterize the jump of the field $\tilde{\theta}$ across $S$ if the second derivative of $\tilde{\theta}$  equals $\tilde{A}$ on $\Omega \bsl S$. The argument here proceeds by considering the field $\tilde{A}$ as given; for the sake of motivation (and only for such purpose), it can be thought of as the absolutely continuous part of the derivative of the absolutely continuous part of the derivative of the scalar field $\tilde{\theta}$ that one wants to construct (when viewed as a field on $\Omega$). Thus, we seek solutions to the system
\begin{equation}\label{eqn:gov_wein}
\begin{rcases}
D \tilde{\theta} &= \tilde{k} \\
D \tilde{k} &= \tilde{A}
\ \end{rcases} \qquad x\in \Omega \bsl S,
\end{equation}
and we are interested in evaluating
\begin{equation*}
\left\llbracket \tilde{\theta} \right\rrbracket (x) := \lim_{x^\pm \to x} \tilde{\theta} \left(x^+\right) - \tilde{\theta}\left(x^-\right), \qquad x \in S,
\end{equation*}
where $x^\pm$ are any sequences of points that approach $x \in S$ from the $\pm$ sides of $S$ at $x$, respectively.

For \emph{any} closed loop $l$ in $\Omega$ that cannot be continuously shrunk to a point while staying within $\Omega$,
\begin{equation}\label{eqn:jump}
\int_l \tilde{A} \, dx = \sf{t}
\end{equation}
is a constant since $\tilde{A}$ is continuous and $curl$-free in $\Omega$. We define the constant ${\sf t}$ to be the \emph{disclination-strength} of the field $\tilde{A}$ in the domain $\Omega$.

Since $\tilde{A}$ is $curl$-free in $\Omega \bsl S$, the latter being simply connected, it is possible to construct a field $\tilde{k}$ on the same domain satisfying \eqref{eqn:gov_wein}$_2$. By \eqref{eqn:jump}, $\tilde{k}$,  in general, has a non-vanishing jump across $S$. For a fixed $S$ and its corresponding $\tilde{k}$, it is further possible to construct a scalar field $\tilde{\theta}$ satisfying \eqref{eqn:gov_wein}$_1$ on $\Omega \bsl S$ due to the symmetry $\tilde{A}_{ij} = \tilde{A}_{ji}$.

Let $c$ be a curve belonging to $S$ joining points $x, y \in S$, and $c^\pm$ two curves near $c$ on the $\pm$ sides of $S$. Noting that 
\[
{\sf t} = \left\llbracket D \tilde{\theta} \right\rrbracket = \left\llbracket \tilde{k} \right\rrbracket
\]
is a constant on $S$, computing $\mathlarger{\int_{c^\pm}} D \tilde{\theta} \, dx$ along the curves $c^\pm$ and considering the limit of the difference of the result as $c^\pm \to c$, we obtain
\begin{equation}\label{eqn:wein}
 \left\llbracket \tilde{\theta} \right\rrbracket (y) =  \left\llbracket \tilde{\theta} \right\rrbracket (x) + {\sf t} \cdot \left( y - x \right) \qquad \forall x,y \in S.
\end{equation}
Unlike the disclination-strength $\sf t$, \eqref{eqn:wein} shows that the \emph{dislocation-strength}, $\mathlarger{\int_l \, \tilde{k} \, dx}$, which equals the jump in the phase field across $S$, is not a well-defined topological constant (independent of the loop $l$) when the disclination-strength $\sf t \neq 0$. This is analogous to the statement for layered structures that $[\theta]_\gamma$ is not well-defined for a closed loop $\gamma$ unless the enclosed number of flips is even. When the disclination-strength ${\sf t} = 0$,  $\left\llbracket \tilde{\theta} \right\rrbracket$ is a constant on $S$ and the dislocation-strength is a well-defined topological constant. Another situation when this is so is when $S$ is a plane normal to $\sf t$. Of course,  it is possible in these situations for $\tilde{\theta}$ to be continuous as well.

The assumption that $\tilde{A}$ is a continuous field on $\Omega$ merits further discussion based on what is physically observed in elastic solids in contrast with nematics, smectics and patterns. In the presence of a dislocation in an elastic solid, i.e. considering a terminating surface $S$ on which a constant displacement discontinuity occurs, it can be seen that there is no jump in the limiting values of the displacement gradient field as the surface $S$ is approached. 
Consider now Frank's celebrated solutions for the entire family of straight, nematic wedge disclinations of integer multiples of $\frac{1}{2}$ strength (mathematically identical to the solution for the screw dislocation in a solid). In this case, writing $\tilde{k}(x) = cos \phi \, e_1 + sin \phi \, e_2$, where $(e_1,e_2)$ is a fixed orthonormal frame and $\phi$ is  a half integer multiple of the angle $\tilde{k}(x)$ makes with $e_1$ (plus a constant),  it can be checked that while across any surface $S$ whose trace on the $x_1-x_2$-plane is a straight radial ray from the origin there is no jump in the limiting values of $D\phi$, there is a jump in the limiting values of $D \tilde{k} = \partial_\phi \tilde{k} \otimes D\phi$ as $S$ is approached. For an  $S$ that is not necessarily planar,  e.g., in smectics and stripes where it can be chosen as a contour $\tilde{\theta} = m \pi$ and $\left\llbracket \tilde{k} \right\rrbracket = (1- e^{2 \pi i J}) \, \nu$ ({\em cf.} Eq.~\eqref{eq:twist}) is along $\nu$, the  direction of the local normal to the contour, $\tilde{A}$ has a nonzero distributional curl on $S$ (cf. Appendix \ref{app:angle_3D}). 

Consequently, we also consider the case when $\tilde{A}$ is symmetric, smooth, and curl-free in $\Omega \backslash S$. Repeating the above arguments, we now find that
\begin{equation}\label{eqn:A_discont}
\left\llbracket \tilde{k} \right \rrbracket (y) = \left \llbracket \tilde{k} \right \rrbracket (x) + \int^y_x \left \llbracket \tilde{A} \right \rrbracket \, dx \qquad \forall x,y \in S,
\end{equation}
and
\[
\left\llbracket \tilde{\theta} \right\rrbracket (y) =  \left\llbracket \tilde{\theta} \right\rrbracket (x) +  \int^y_x \left \llbracket \tilde{k} \right \rrbracket \, dx \qquad \forall x,y \in S.
\]
In case $\left\llbracket \tilde{A} \right\rrbracket$ is of the form $a \otimes \nu$ where $a$ is a vector field on $S$ and $\nu$ the unit normal field $S$, we note that the jump  $\left\llbracket \tilde{k} \right\rrbracket$ is still constant on $S$, even though $\tilde{A}$ is possibly discontinuous across $S$ and the jump in $\tilde{\theta}$ still satisfies \eqref{eqn:wein}. All the computational examples in this paper satisfy this condition which is, however, not a necessary feature of our theoretical or computational formalism.

Motivated by the simple arguments above, it is clear that in the class of defects that are 
integer multiples of $\frac{1}{2}$, 
our formalism is adapted to representing 
the $\pm \half$-strength defects as fundamental 
(based on kinematic grounds) and all others are necessarily represented as composites of these fundamental defects. We compute examples of the field of such composite defects in Secs. \ref{sec:+1} and \ref{sec:-3/2}. 

\subsection{Continuum mechanics of defects in scalar fields: natural patterns, smectics, and nematics (NPSN)} \label{sec:NPSN}

Let us denote the region of the interior hole (with boundary) in Fig.~\ref{fig:domain} as the `core' $C$. The considerations of \S\ref{sec:wein} show that the phase field is in general discontinuous in non simply connected domains or, alternatively, if the field $\tilde{A}$ was prescribed in the simply connected domain $\Omega \cup C$, but now with non-vanishing $curl$ supported in the core $C$. It can also be seen, by considering $\Omega$ to be a punctured domain (i.e., $C$ consists of a single point in 2-d or a curve in 3-d), that in such situations $\tilde{k} = D \tilde{\theta}$ is not in general a square-integrable field on $\Omega \cup C$ - for simplicity, consider the case when ${\sf t} = 0$ and the constant jump $\left\llbracket \tilde{\theta} \right\rrbracket \neq 0$. Even when $C$ is a set of full measure, $D \tilde{\theta}$ is not an integrable field on $\Omega$ and therefore its presence in any governing pde would be problematic in the presence of defects (characterized by non-$curl$-free $\tilde{A}$ and/or $\tilde{k}$ fields). 

Consequently, we think of allowing fields with at most (smoothed) bounded discontinuities and removing all (smoothed) concentrations, referred to as `singular' parts, from their gradients, these rehabilitated `gradients' being called `regular' parts. The intuition for this is as follow. Roughly speaking, we think of functions that are `smoothed analogs' of functions that belong to the space SBV in the following way: Consider, e.g., a piecewise-smooth function $k$ that has a bounded jump across a surface such as $S$ in Fig.~\ref{fig:domain} with the core $C$ shrunk to a point. Then the singular part of its gradient is concentrated on $S$. Suppose now we mollify this singular part; then this mollified field seems like one with a smooth concentration around $S$ - it is this smoothed field that, in  this instance, we think of as the `singular' field $B$. Finally subtracting this field from $Dk$, we have the `regular' part $A := Dk - B$, that is the analog of the absolutely continuous part of $Dk$. Since we envision a model where all of $A$, $B$, $Dk$ are going to be integrable functions, $Dk$ has no singular part in a strict sense. Therefore, \textit{we introduce an independent field $B$} in our model, which while being integrable, nevertheless has a smooth concentration around $S$ and a non-vanishing distributional $curl$ (with $curl \, A = - curl \,B)$, supported in a tube (with cross-section $C$ in Fig.~\ref{fig:domain}) rather than a curve. We are also interested in modeling possibly large collections of moving defects that interact and possibly intersect, and tracking the topology of the defected body with cores modeled as excluded `cylinders' is clearly impractical. Thus we seek a model that can be posed in simply connected domains, but nevertheless is descriptive of the topological properties of the line defects we are interested in. With this understanding, we consider a phase field $\theta$ with the regular part of its gradient denoted by $k$. The regular part of $D k$ in turn is denoted by $A$, with singular part by $B$ so that $A = D k - B$, as mentioned before. The terminology of `singular' is in the sense described above; when viewed at a microscopic scale these are smoothed concentrations on sets whose far-field identities are those of lower (than 3) dimensional objects, e.g. the field $B$ in the specific case discussed above would be supported on a set whose identity, on a spatial `zoom-out,'  is the 2-d surface $S$ and the field $-curl\,B =:\pi$ is supported on $C$, whose zoomed-out identity would be a 1-d curve. Finally, in classical governing equations for the phenomena of interest, developed for situations not containing defects, we admit the appearance of only the regular parts of fields (e.g. $A$ instead of $Dk$) and we supplement the model with an evolution equation for the new field $B$, arising from consideration of conservation of topological charge embodied in the field $\pi = - curl \, \pi$ (and the Second law of thermodynamics.

The energetic physics of the nematic director or the pattern phase gradient (far from onset of roll instabilities) is based on energetic cost of director gradients. The director has head-tail symmetry, and this is modeled by assigning null energetic cost to values of the field $B$ (the singular part of $D k$) that reflect local changes in director orientation by $180^\circ$ over a small coherence length, typically of the order of a linear dimension of a disclination core. The terminating `curve' of a `surface' on which $B$ is non-vanishing, say a constant, is a region where $- curl \, B = curl \, A =: \pi$ is supported, and such a region corresponds to a \emph{disclination} and we refer to the field $\pi$ as the disclination density.

With reference to Fig. \ref{fig:domain}, if the disclination density field $\pi$ were to be supported in the region $C$, then  a $curl$-free field $A$ and discontinuous fields $k$ and $\theta$ satisfying \eqref{eqn:gov_wein} without the $\tilde{}$ can certainly be defined. Moreover, if region $C$ were to contain two separate concentrations of the disclination density of opposite sign, i.e. a disclination dipole, such that $\int_s \pi \nu \, da = 0$ where $\nu$ is the unit normal field to any surface $s$ that transversally intersects $C$, then the field $\theta$ would have a constant jump on any admissible surface $S$. As well, if the field $A$ were to vanish for the moment and $curl \, k$ were to be supported in $C$, then again $\theta$ would have a constant jump on any $S$. And, of course, if a $\theta$ field had to be defined at least locally in some region, $curl\, k$ would have to vanish therein. When modeling smectics and natural patterns, we will energetically penalize $curl \, k$ strongly and refer to regions that contain concentrations of $curl\, k$ as a \emph{dislocation} and the field $-curl \,k =: \gamma$ as the dislocation density (this is slightly different from the definition for a solid). The considerations of \S\ref{sec:wein} suggest that a region containing a disclination dipole may also be considered as an effective dislocation in that the far-field topological identity of both cases, measured by integrating the director field along \emph{any} closed loop encircling the region, has to be a constant. 

The definition of the disclination density field as a $curl$, i.e., $\pi = curl \,A$, associates a `charge' with any closed curve $l$ in the body, given by $\int_s \pi \nu \, da$ where $s$ is any surface whose bounding curve is $l$, $\nu$ being the unit normal field on the surface. The flux of this charge across the bounding curve can also be kinematically characterized and is given by $- \pi \times V$, where $V$ is admitted (i.e. postulated to exist) as a velocity field, relative to the material, of the disclination density $\pi$, resulting in the conservation law 
\begin{equation}\label{eqn:pi_evol}
\dot{\pi} = - curl (\pi \times V)
\end{equation}
(see, e.g., \cite[Appendix B]{acharya2011microcanonical}, for a more physically detailed derivation of this conservation law, particularly the justification of the form of the flux). These considerations lead to the following kinematics of our model (here, $X$ represents the alternating tensor and the notation $A:X$ for $A$ a second order tensor is defined by $(A:X)_{j} = A_{ik}e_{ikj}$ for components w.r.t. an orthonormal basis):
\begin{align}\label{eqn:kin}
&\, k, B &&\text{fundamental kinematic fields} \notag\\
& A = D \, k - B && \text{regular part of director gradient} \notag\\
& \pi = curl\, A = -curl\, B &&  \text{disclination density} \notag\\
& \gamma = A :X + B:X = -curl\,k & & \text{dislocation density} \notag\\
& \dot{B} = curl \, B \times V  && \text{evolution of singular part of } grad\,k, 
\end{align}
where \eqref{eqn:kin}$_5$ follows from \eqref{eqn:pi_evol} up to a gradient. If `grain boundaries,' or thin regions with a 2-d skeleton on which $B$ has a concentration, are allowed to move transversely to themselves with velocity $V^\perp$, then \eqref{eqn:kin}$_5$ would be modified to read $ \dot{B} = curl \, B \times V + grad \left(BV^\perp\right)$ (without disturbing \eqref{eqn:pi_evol}). For simplicity, we do not consider this extra mechanism in this paper.

A sufficient condition for the construction of a scalar field $\theta$, corresponding to the fields $k, B$, is that it be possible to remove the support of the disclination density and dislocation density fields from the body and the resulting body (say $\Omega$) be amenable to being rendered simply connected by the removal of a connected surface in it. In that case, a field $\theta$ may constructed satisfying $D \theta = k$ in $\Omega$ which is generally discontinuous. In general, it is unclear if there is a single connected surface whose removal will permit the construction of a single valued  phase $\theta$ on the complement ({\em cf}. the random stripe pattern in Fig.~\ref{fig:random}). Even if $\Omega$ can be rendered simply connected by removing a surface, it is natural to expect that there would be more than one surface with the same property and each such surface would correspond to a different $\theta$ field on $\Omega$. Thus, when $\theta$ can be constructed, on an appropriately `reduced' domain, in the presence of dislocations and disclinations, it can be expected to be `massively' non-unique. This non-uniqueness of $\theta$ is a price one has to pay in going from a microscopic model for the system, which resolves behaviors on the scale of the underlying periodicity, to a macroscopic phase description.

Let $\xi$ be a length-scale corresponding to the linear dimension of a disclination core. A typical example of a free-energy density function for the model (which possesses all qualitative properties that we would like embodied in the description of a nematic/smectic system, and one we use for all examples solved in this paper) is
\begin{equation}\label{eqn:energy}
\psi = P_1 \left( |k| - 1\right)^2 + P_2 |curl\, k|^2 + \alpha K^* f( |B|) + K |D k - B|^2 + \veps |\pi|^2.
\end{equation}
The vector field $k$ is physically non-dimensional. The material constant $K$ characterizes the elasticity of director gradients, $\frac{P_1 \xi^2}{K} \gg 1$, $\frac{P_2}{K} \gg 1$ are penalizing constants, $f$ is a nondimensional, nonconvex function of $|B|$ of the type defined below, $\alpha > 0$ is a nondimensional number that tunes the strength of the nonconvexity of $f$, $\frac{K^* \xi^2}{K} \approx 1$, and  $\frac{\varepsilon}{K \xi^2} \approx 1$. The function $f$ is used to model the head-tail symmetry of the director through an energetic penalty in our model. This is achieved by assigning approximately vanishing elastic cost for pointwise values of the director gradient of the type $grad \, k \approx \frac{n - (-n)}{a \xi} \otimes l$, where $0 < a \leq 1$ and $n, l$ are unit vectors, the latter representing the direction along which the jump of $n$ occurs. This implies that the two wells of $f$ should be at $|B| = 0, \frac{2}{a\xi}$ (based on rough energy-minimization arguments disregarding constraints of compatibility on $D k$). Finally, to model pure nematics one sets $P_2 = 0$.

In what follows, we refer to a simply-connected domain or body within which the mechanics of interest takes place as $\Omega$.

Beyond the specification of the energy density of the system, the disclination velocity $V$ has to be specified. Guidelines for that specification arises from demanding that, up to contributions from the boundary of the body, the evolution of $B$ results in a non-increasing free energy $\int_\Omega \psi \, dv$ evolution, i.e. $\frac{d}{dt}\int_\Omega \psi \, dv \leq 0$, which is a simplified embodiment of the Second law of Thermodynamics. We consider a free energy density with the following dependencies:
\begin{equation}\label{eqn:psi_npsn}
\psi(k, D k, B, \pi),
\end{equation}
noting that $-curl\,k = grad\, k : X$. Then
\begin{align}\label{eqn:diss}
\dot{\overline{\int_\Omega \psi \, dv \, }} & = \int_\Omega \left ( \partial_k \psi - div \left( \partial_{\,D k} \psi \right) \right) \cdot \dot{k} \, dv \notag\\
& \ \  + \int_\Omega \left( \partial_B \psi + curl \, \partial_\pi \psi \right) : \left( curl \, B \times V \right)\, dv  + \text{boundary terms}.
\end{align}
Consequently, requiring
\begin{align}\label{eqn:B_evol}
M_B^{-1}\, \dot{B} & = - \,curl \, B \times \left[ X \left\{ \left(\partial_B \psi + curl \, \partial_\pi \psi \right)^T curl\, B \right\} \right] \\
M_k^{-1}\, \dot{k} &= - \left ( \partial_k \psi - div \left( \partial_{\,D k} \psi \right) \right) \label{eqn:ang_mom_bal}
\end{align}
with $M_B, M_k$ being positive, scalar, mobility constants, is sufficient for the contribution to the rate of change of the total free-energy of the body to be non-positive due to the evolution of the fields $B$ and $k$, up to contributions from the boundary. An energy density specification with the dependencies as in \eqref{eqn:psi_npsn} along with \eqref{eqn:B_evol}-\eqref{eqn:ang_mom_bal} constitute the closed set of statements defining the model. The objects $B,k$ are its fundamental fields. It is to be noted that the `Hamilton-Jacobi \textit{system}' structure of \eqref{eqn:B_evol} is a direct consequence of the conservation law of topological charge given by \eqref{eqn:pi_evol} along with the Second law of Thermodynamics; in particular, and unlike a gradient-flow, the knowledge of an energy density alone does not suffice to specify it.

For nematics and smectics, \eqref{eqn:ang_mom_bal} in the limit of large mobility $M_k$ corresponds to the balance law of angular momentum \cite{stewart2004static} (here, we are considering no material motion for simplicity; its consideration would lead to accounting for balance of linear momentum, along with viscous dissipation and material inertia \cite{leslie1992continuum,stewart2004static}). Convection patterns, in contrast are driven by organized, collective motion of materials, and $k$ is not directly associated with a conserved quantity, i.e. mass, momentum or energy. Nonetheless,   the late stages in the evolution of a convection pattern can be written as the gradient flow for the reduced Cross-Newell energy~\eqref{eq:rcn} that can be expressed in terms of $k$. When the primary concern is to understand dynamics close to local minima of the system free-energy, it therefore suffices to consider, in addition to \eqref{eqn:B_evol}, the `gradient flow' \eqref{eqn:ang_mom_bal} governing $k$.

\subsection{Continuum mechanics of defects in vector and tensor fields: elastic solids}\label{sec:solids}
~For defects in scalar fields, it sufficed to consider $k$ as a fundamental field and consider only the regular and defect parts of $D k$. This was primarily dictated by the nature of the energy density function for such systems, in particular elasticity arising due to director gradients, with a non-convex contribution accommodating energetically preferred states of $D k$.

To understand the similarities and differences between defects in elasticity of solids and NPSN, it is useful to first consider `anti-plane' deformations of elastic solids, i.e., a body undergoing displacement in the out-of-plane direction as a function of in-plane coordinates. Then the displacement vector field has one non-trivial component, analogous to the phase field $\theta$. A primary difference, however, arises, from the energetics. In elastic solids, the primary elasticity arises from displacement gradients and it is important to consider regular and defect parts of the displacement gradient. Moreover, crystal periodicity dictates energetically preferred displacement gradient states. Hence, the kinematics of defected `anti-plane' elastic solids is given in Table \ref{table:kin_solid}.

\begin{table}
\begin{tabular}{l l}
\centering
$ \theta, p, B$ & fundamental kinematic fields \\
$p$ & defect part of displacement gradient =: plastic distortion\\
$k = D \, \theta - p$ & regular part of displacement gradient =: elastic distortion \\
$B$ & defect part of $Dk$ =: eigenwall field \\
$A = D k - B = D^2 \theta -D  p - B$ & regular part of elastic distortion gradient \\
$\pi = curl\, A = -curl\, B$ & g.disclination density\\
$\gamma = A :X + B:X = curl \, p$ & dislocation density (customarily defined as $A:X$) \\
$\stackrel{.}{p} = - curl\, p \times V^\gamma$ & evolution of plastic distortion\\
$\dot{B} = curl \, B \times V^\pi$ & evolution of eigenwall field
\end{tabular}
\label{table:kin_solid}
\caption{\emph{Governing fields and equations for defect dynamics (the fields $\theta, p$ are slaved to $k$ for NPSN)}. }
\end{table}
A typical energy density function for defects in `anti-plane' elastic solids (screw dislocations with Burgers vector and line direction in the out-of-plane direction, and twist disclinations with axis and rotation vector in the in-plane directions) looks like
\begin{equation}\label{energy_solids}
\psi = \mu |D \theta - p|^2 + g_1( p) + \veps_1 |curl\, p|^2 + K |D k - B|^2 + g_2(B) + \veps_2 |curl \, B|^2,
\end{equation}
where $\mu$ is the elastic shear modulus, $K$ is a modulus related to couple-stress elasticity, $g_1$ is a non-convex function reflecting preferred strain states due to lattice periodicity, $g_2$ represents a non-convex grain boundary energy reflecting preferred lattice misorientations, $\veps_1$ is a material parameter characterizing dislocation core energy, and $\veps_2$ is a constant characterizing (g.)disclination \cite{zhang2018relevance, zhang2018finite} core energy. The \textit{balance laws of linear and angular momentum (involving second-derivatives in time)} provide the governing equations for the evolution of $\theta$, and Table \ref{table:kin_solid}$_{8,9}$ represent the evolution of the fields $p,B$. Constitutive guidance for $V^\gamma, V^\pi$ for closing the model are deduced from thermodynamic arguments following similar argument as in deriving \eqref{eqn:B_evol} \cite{acharya2015continuum}. A \textit{primary difference between elastic solids and NPSN} is reflected in the scaling $\mu \gg K  \xi^{-2}$, where $\xi$ is assumed to be a typical linear dimension of a core for NPSN defects. The elastic modulus $\mu$ depending on $p$ allows the modeling of \textit{earthquake dynamics}  \cite{zhang2015single}. 

In 3D elasticity, all fields in Table \ref{table:kin_solid} are tensors of one higher order than for the anti-plane case (and the elastic modulus is a  $4^{th}$-order tensor). Nonlinear elasticity requires the energy density to depend on  $k^T k$ (where $k$ now is the elastic distortion field) \cite{zhang2018finite,arora_acharya_ijss,arora2020unification,arora2020finite}, and it can be a smooth function which is at least rank-one convex. And, of course, elasticity with defects in solids is fundamentally about material deformation and motion and singularities (at a macroscopic scale) in such deformation.

Finally, we note that if $P(t)$ is the power supplied to, and $K(t)$ the kinetic energy of, the body at time $t$, then our thermodynamic formalism ensures that $P \geq \ \dot{K} + \dot{E}$ for all $t$, so that $\dot{E}$ is not necessarily $\leq 0$, allowing for externally driven, strongly out-of-equilibrium phenomena involving rapid material motion in our model.

That the type of model discussed above is realistic for elastic solids and earthquake rupture dynamics, even at the level of being robustly computable in dealing with objects that are macroscopically viewed as nasty singularities is demonstrated in \cite{zhang2016non, zhang2018relevance, zhang2018finite,zhang2015single, acharya2015continuum,garg2015study}. Connections of such models to NPSN are shown in \cite{zhang2016non} and alluded to in \cite{newell2017elastic}. Similar models applicable to NPSN, with intriguing analogies to cosmology and the  Standard Model of particle physics are discussed in \cite{newell2012pattern,newell2017elastic}. 
\section{Illustration of theory} \label{sec:numerics}
In this section we demonstrate salient \emph{equilibrium} features of the theory developed above through particular examples. Dynamical aspects of our framework will be discussed elsewhere.

\subsection{Nondimensional gradient flow dynamics}\label{sec:gf_derivation}
To non-dimensionalize Equations \eqref{eqn:ang_mom_bal}, \eqref{eqn:B_evol}, and \eqref{eqn:energy}, we introduce the following dimensionless variables,
\begin{eqnarray*}
\tilde{x_i} = \frac{1}{\xi}x_i;\quad \tilde{s} = K M_2 t; \quad \tilde{P_1} = \frac{\xi^2}{K}P_1; \quad \tilde{P_2} = \frac{1}{K}P_2; \quad \tilde{K^*} = \frac{\epsilon^2}{K}K^*; \quad  \tilde{\epsilon} = \frac{1}{K \xi^2} \epsilon; \quad \tilde{B} = \xi B.
\end{eqnarray*}
and assume $ M_1 = M_2 \xi^2$ without loss of generality here, since the gradient flow equation for $k$ can be treated as simply a device to get equilibrium of $k$ with $B$ fixed. The non-dimensionalized gradient flow equations of the energy \eqref{eqn:energy} read as:
\begin{eqnarray*}
\begin{aligned}
\frac{\partial k_i}{\partial \tilde{s}} = (k_{i,j} - \tilde{B}_{ij})_{,\,j} - \tilde{P_1} (|k| - 1) \frac{k_i}{|k|} + \tilde{P_2} e_{sjk} e_{sri} k_{k,jr} \\ 
\frac{\partial \tilde{B}_{ij}}{\partial \tilde{s}} = (D k - \tilde{B})_{ij} - \alpha \tilde{K^*} \frac{\partial f}{\partial \tilde{B}_{ij}} + \tilde{\epsilon} ( e_{tmn}e_{tsj} \tilde{B}_{in,ms}).
\end{aligned}
\end{eqnarray*}
\emph{For convenience, we remove all tildes in remaining work} and use the following nondimensional evolution equations in the rest of the paper:
\begin{eqnarray} \label{eqn:grad_dimen}
\left.
\begin{aligned}\label{eqn:grad_flow_gov}
\frac{\partial k_i}{\partial s} = (k_{i,j} - B_{ij})_{,\,j} - P_1 (|k| - 1) \frac{k_i}{|k|} + P_2 \, e_{sjk} e_{sri} k_{k,jr} \\ 
\frac{\partial B_{ij}}{\partial s} = (D k - B)_{ij} - \alpha K^* \frac{\partial f}{\partial B_{ij}} + \epsilon ( e_{tmn}e_{tsj} B_{in,ms})
\end{aligned}
\right\} \text{in the body $B$}.
\end{eqnarray}

Note that the evolution equation \eqref{eqn:grad_dimen}  for $B$ is different from \eqref{eqn:B_evol}. It is shown in \cite{zhang2016non} that while the $L^2$-gradient flow dynamics \eqref{eqn:grad_dimen}  for the energy density  \eqref{eqn:energy} describes defect equilibria well, it is not able to adequately describe defect interaction and evolution in important situations, e.g. the elastic interaction and annihilation of a pair of positive and negative half-strength disclination. On the other hand, \eqref{eqn:B_evol}, a dynamics based on kinematics of topological charge conservation and thermodynamics, succeeds in this task, as demonstrated in \cite{zhang2016non}. While a theoretical explanation for this inadequacy of the gradient flow dynamics for these co-dimension 2 defects remains to be addressed (speculation is provided in \cite{zhang2016non}), with possible relation to similar phenomena for the equal well-depth case for the co-dimension 1 case analyzed in \cite{rubinstein1989fast} (recognizing that the mutual elastic interaction of co-dimension 2 defects is much stronger than for co-dimension 1), in this paper we simply rely on the gradient flow dynamics to predict approximate equilibria, and reach physical conclusions based simply on comparisons of total energy content of various defect configurations.

\subsection{Computational Examples} \label{sec:computations2d}

In this Section, we assume $a=1$, $\xi=0.1$ and the size of the domain to be $20\xi \times 20\xi$. Unless specified otherwise, for all results pertaining to modeling nematics, we use the following default values for the (non-dimensional) material constants: $P_1=100$, $P_2=0$, $\alpha=10$, $K^*=5$, and $\epsilon=1$.

In the following computed examples, $k$ is specified at one point (that eliminates rigid translation in pure statics), along with the natural boundary condition corresponding to \eqref{eqn:grad_flow_gov}$_1$, $((D k - B) \cdot n + P_2 \, curl \, k \times n)  = 0$, where $n$ is the outward normal to boundary of the domain. Also, the natural boundary condition for  \eqref{eqn:grad_flow_gov}$_2$, $ curl \, B \times n = 0$ is applied. The one-point specification of $k$ allows the prediction of distinct director patterns for defects with identical magnitude of strength, e.g., the `target' and the `source' for the strength $+1$ defect, utilizing identical $B$ fields, as well as the $\pm \half$ defects which involve initial conditions on $B$ with differing sign. In the following calculations, results from the gradient flow \eqref{eqn:grad_flow_gov} with both $k$ and $B$ evolving are referred as \textit{equilibrium}, and results of evolving $k$ with specified $B$ are referred as \textit{constrained equilibrium}. In the constrained equilibrium calculations, $B$ is not evolved from its specified initial condition. Our Finite Element method based algorithm is presented in Appendix \ref{app:numerical_derivation}. The acceptance criterion for a (constrained) local equilibrium state for all calculations is $\frac{|E_{s} - E_{s-1}|}{E_{s-1} \Delta s} < 10^{-5}$, where $E_s$ is the total energy at discrete time $s$, and $\Delta s$ is the time step at time $s$.

\subsubsection{Strength $\pm \half$ defects} \label{sec:half_defects}
As mentioned in \S\ref{sec:model}, $f(|B|)$ has two wells at $0, \frac{2}{a\xi}$. In this section, we prescribe initial conditions for the gradient flow calculations for the $B$ field as non-zero within a layer. For a positive half strength disclination,
\begin{equation}\label{eqn:B_positive}
B(x, y) = \begin{cases}
-\frac{2}{a \xi}\bfe_1 \otimes \bfe_2, & \text{ if $|y|<{\frac{a\xi}{2}}$  and $x < 0$} \\
0, & otherwise.
\end{cases}
\end{equation}
For a negative half strength disclination, 
\begin{equation}\label{eqn:B_negative}
B(x,y) = \begin{cases}
\frac{2}{a \xi}\bfe_1 \otimes \bfe_2, & \text{ if $|y|<{\frac{a\xi}{2}}$  and $x < 0$} \\
0, & otherwise.
\end{cases}
\end{equation}

\begin{figure}
\centering
\subfigure[Prescription of $B$ of positive half disclination. $B$ is non-zero inside the layer, with $B_{12}$ being non-zero component.]{
\includegraphics[width=0.4\linewidth]{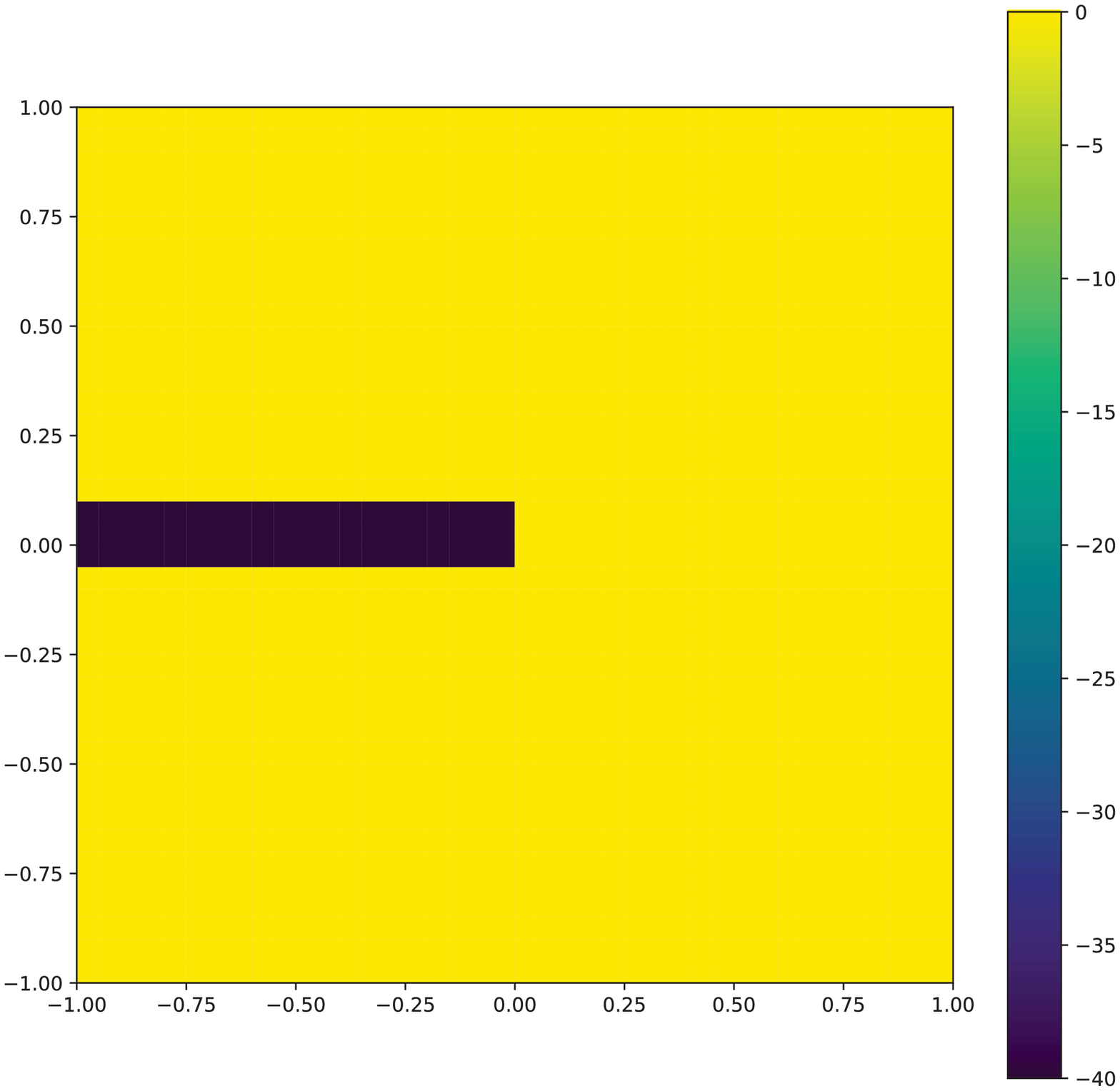}
\label{fig:positive_half_B}}\qquad
\subfigure[Prescription of $B$ of negative half disclination. $B$ is non-zero inside the layer, with $B_{12}$ being non-zero component with opposite sign of the positive case.]{
\includegraphics[width=0.4\linewidth]{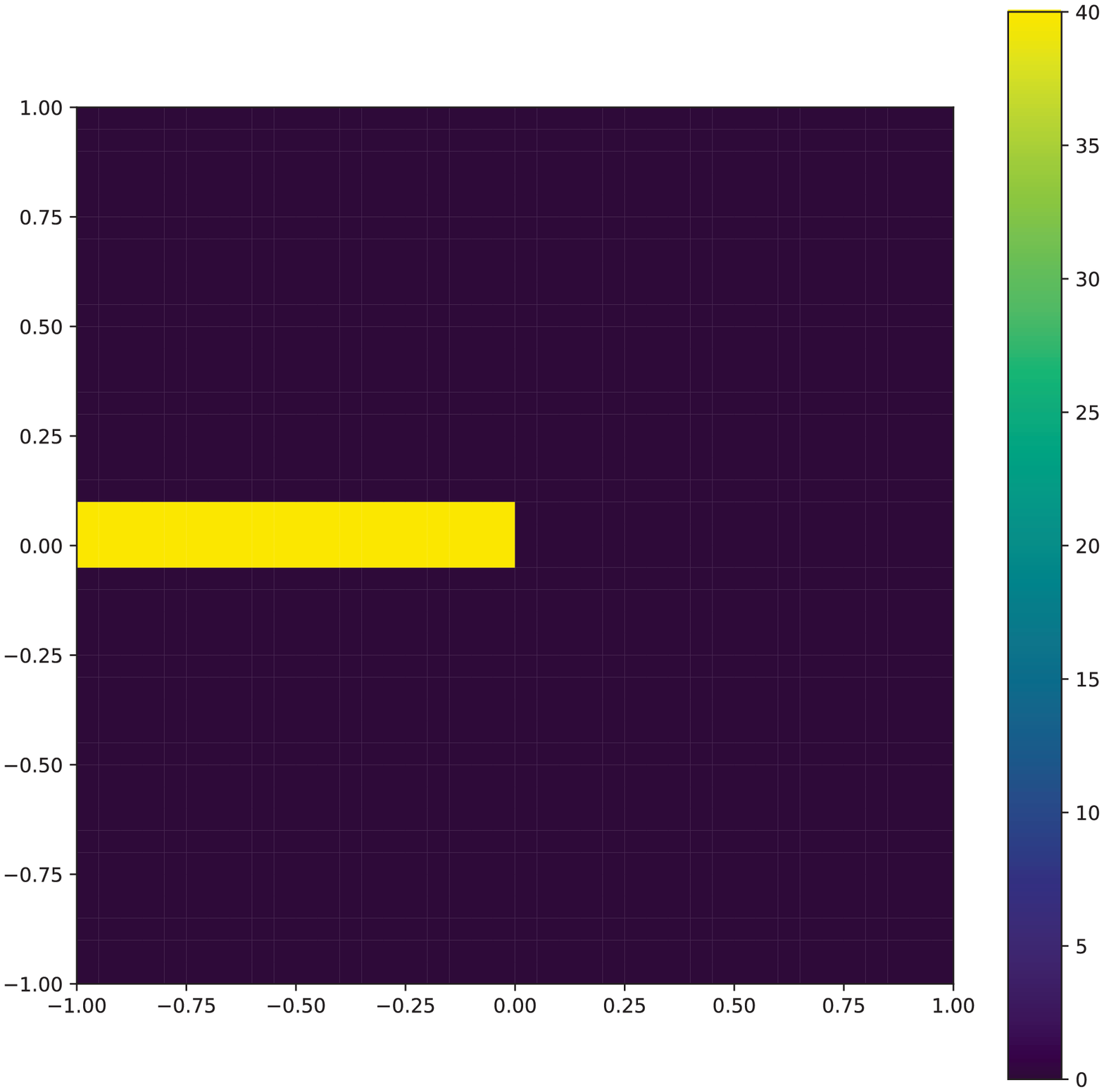}
\label{fig:negative_half_B}}\qquad
\caption{Prescription of $B$ for both positive half and negative half disclinations.}\label{fig:half_B}
\end {figure}

Fig \ref{fig:positive_half_B} and \ref{fig:negative_half_B} show the prescription of $B(x,y)$ for both a positive half strength and  a negative half strength disclination. These specifications of initial conditions correspond to being at the minima of the function $f$, pointwise.

\begin{figure}
\centering
\subfigure[Equilibrium of $k$ of positive half disclination.]{
\includegraphics[width=0.4\linewidth]{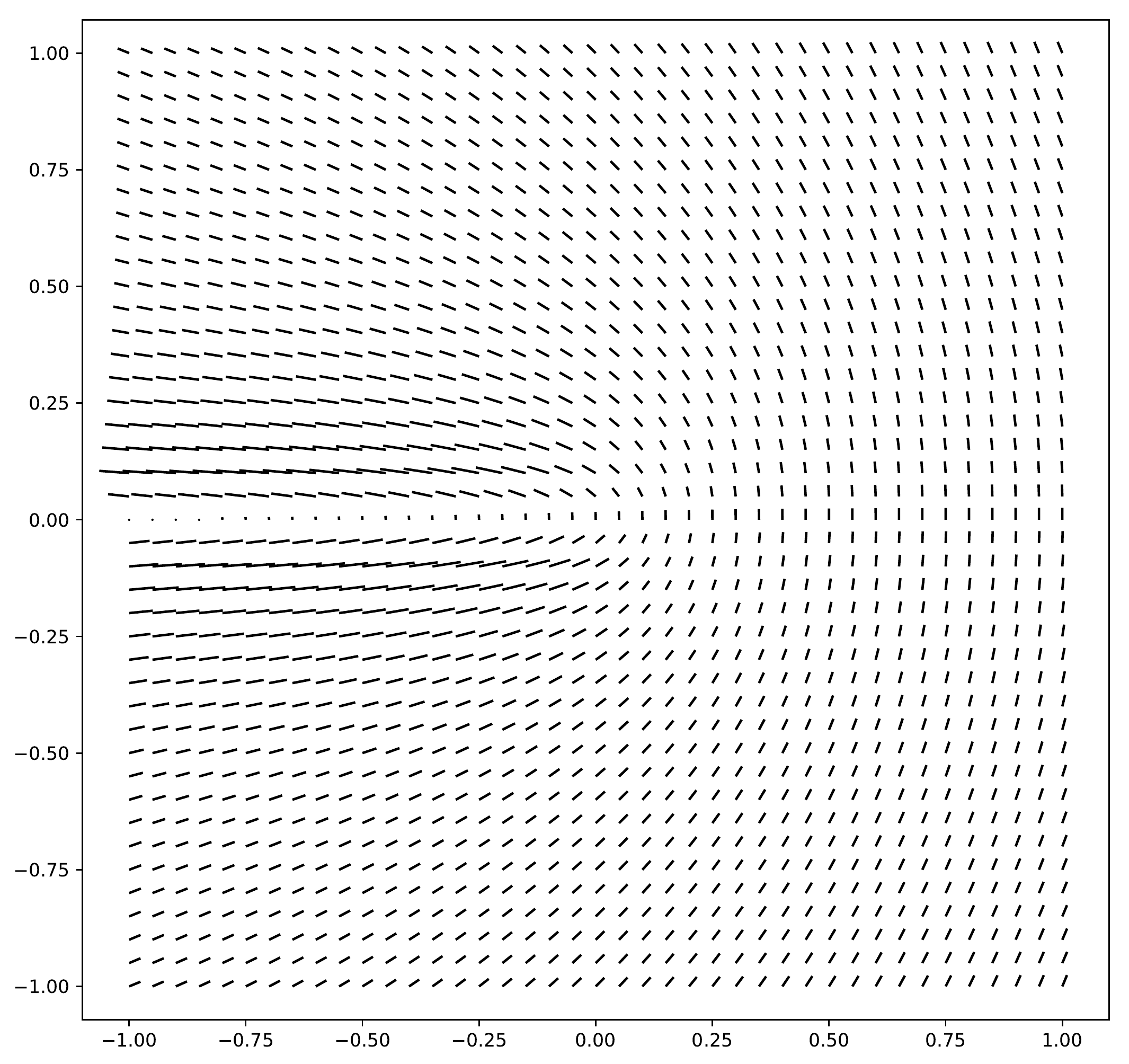}
\label{fig:positive_half_k}}\qquad
\subfigure[Equilibrium of $k$ of negative half disclination.]{
\includegraphics[width=0.4\linewidth]{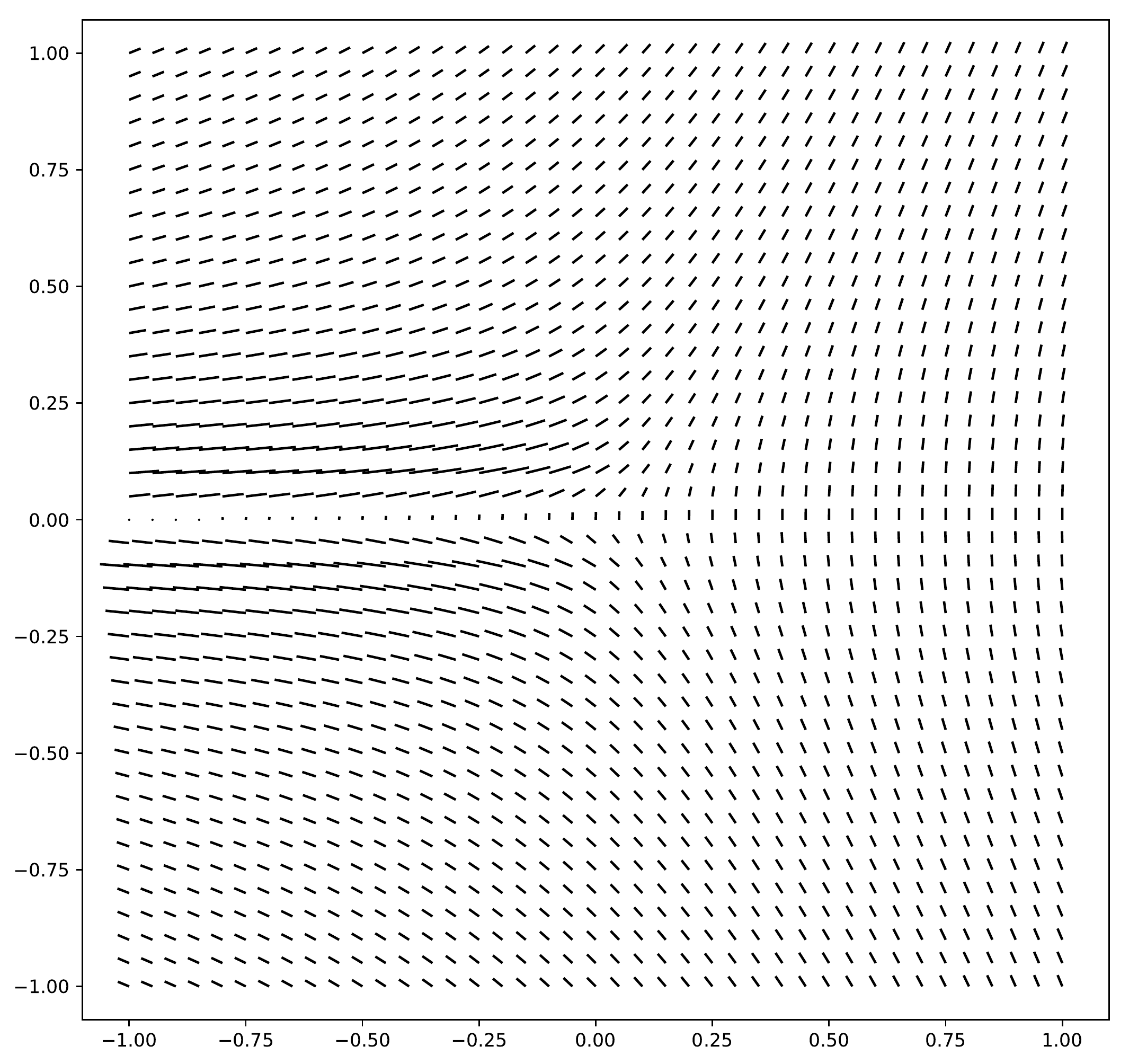}
\label{fig:negative_half_k}}\qquad
\caption{Result of $k$ for both positive half and negative half disclinations. $k$ dramatically changes within the layer.}\label{fig:half_k}
\end {figure}

\begin{figure}
\centering
\subfigure[Energy density of positive half disclination.]{
\includegraphics[width=0.4\linewidth]{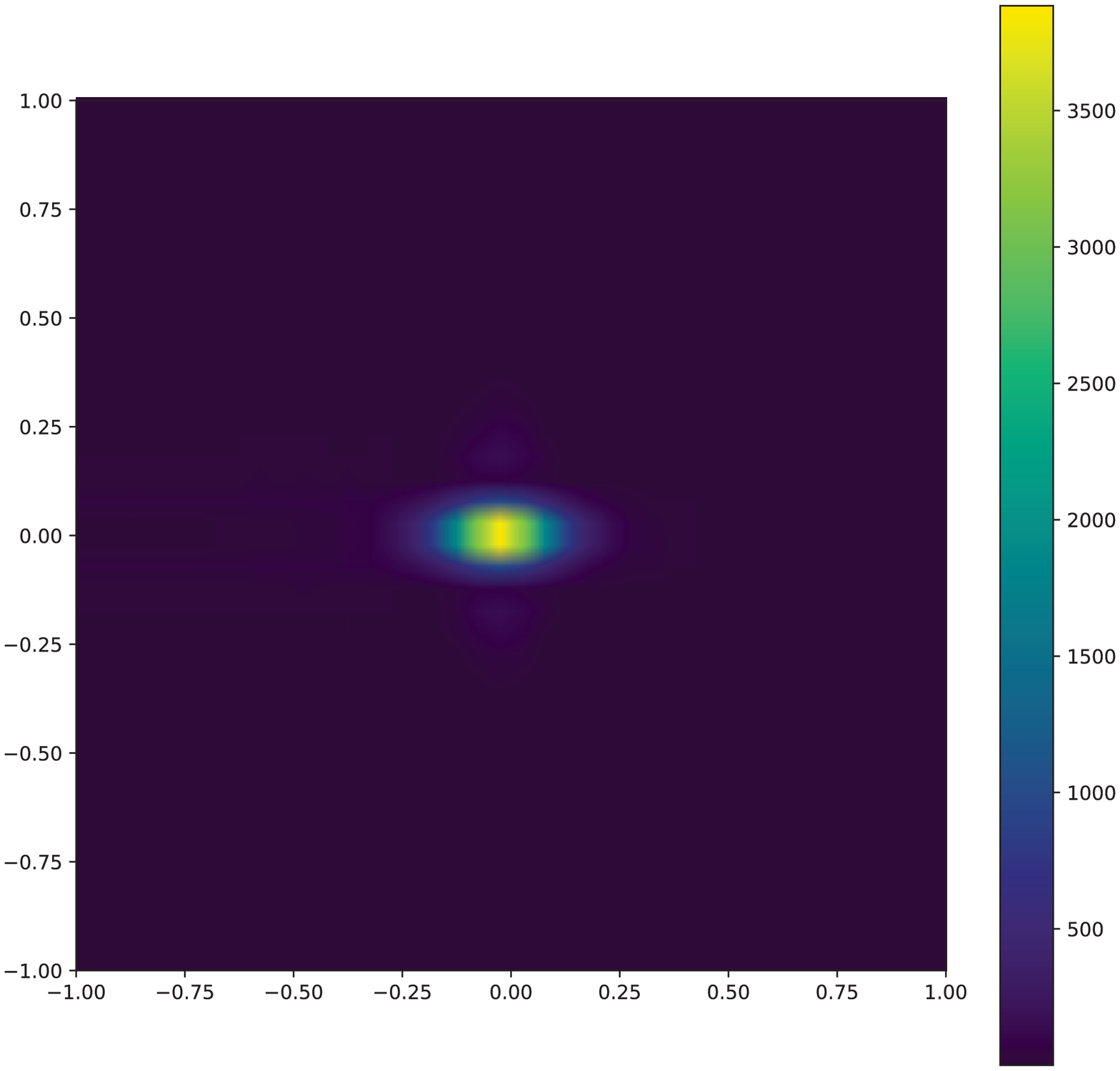}
\label{fig:positive_half_ed}}\qquad
\subfigure[Energy density of negative half disclination.]{
\includegraphics[width=0.4\linewidth]{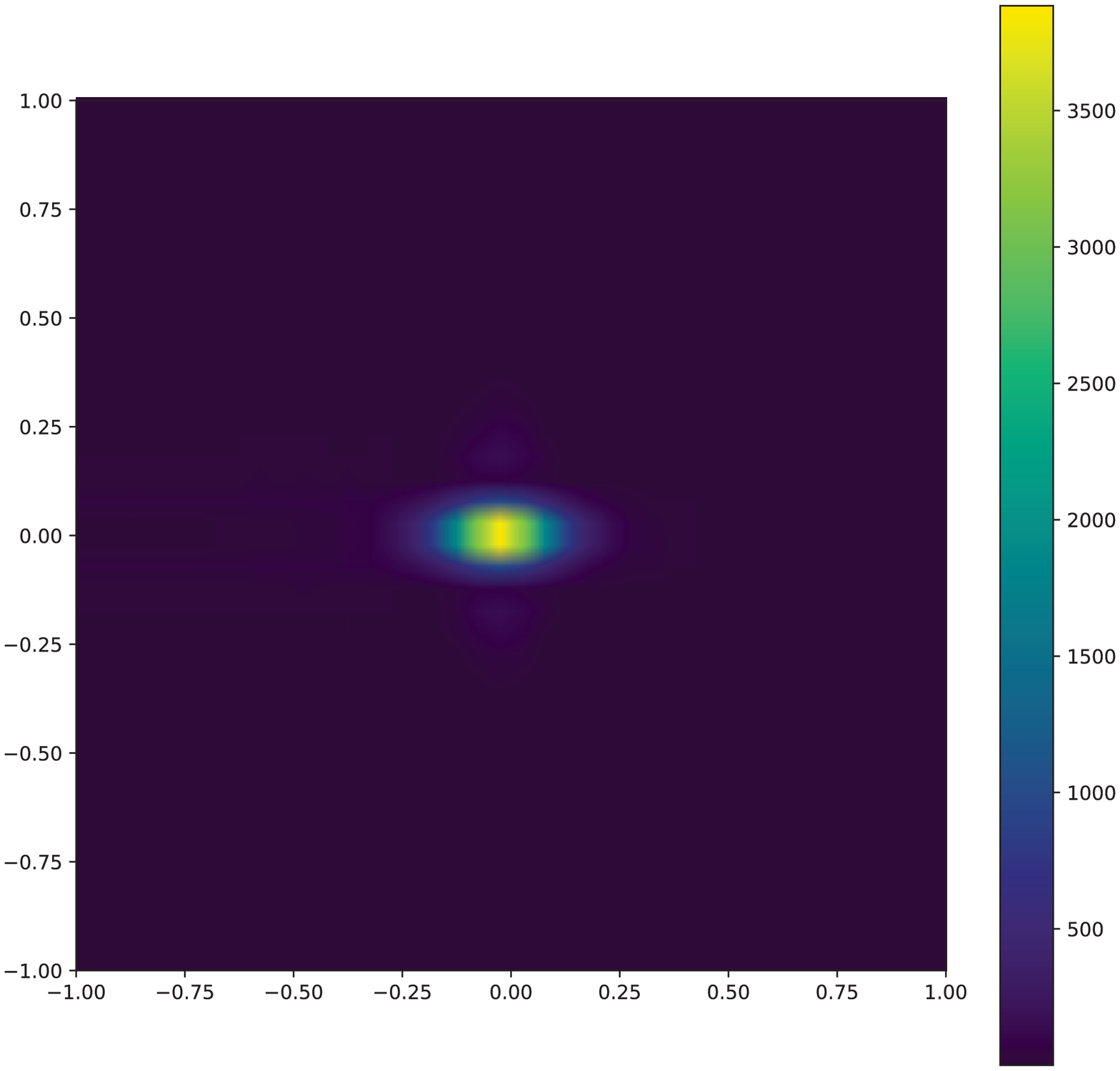}
\label{fig:negative_half_ed}}\qquad
\caption{Energy density plots for both positive half and negative half disclinations. Energy is localized at disclination cores. The energy densities of positive and negative disclinations are similar.}\label{fig:half_ed}
\end {figure}

In addition to different initial conditions for $B$, the director field $k$ is specified, for all times, at one point on the top of the boundary of the layer. Fig \ref{fig:half_k} shows numerically computed equilibria of $k$ obtained from the gradient flow equations \eqref{eqn:grad_dimen} and Fig \ref{fig:half_ed} shows energy density plots, for both the positive half and negative half disclinations respectively. Noteworthy is the fact that although $k$ dramatically changes both direction and length within the layer, the energy density localizes only around cores. 

\begin{figure}
\centering
\subfigure[Frank Energy density of positive half disclination.]{
\includegraphics[width=0.4\linewidth]{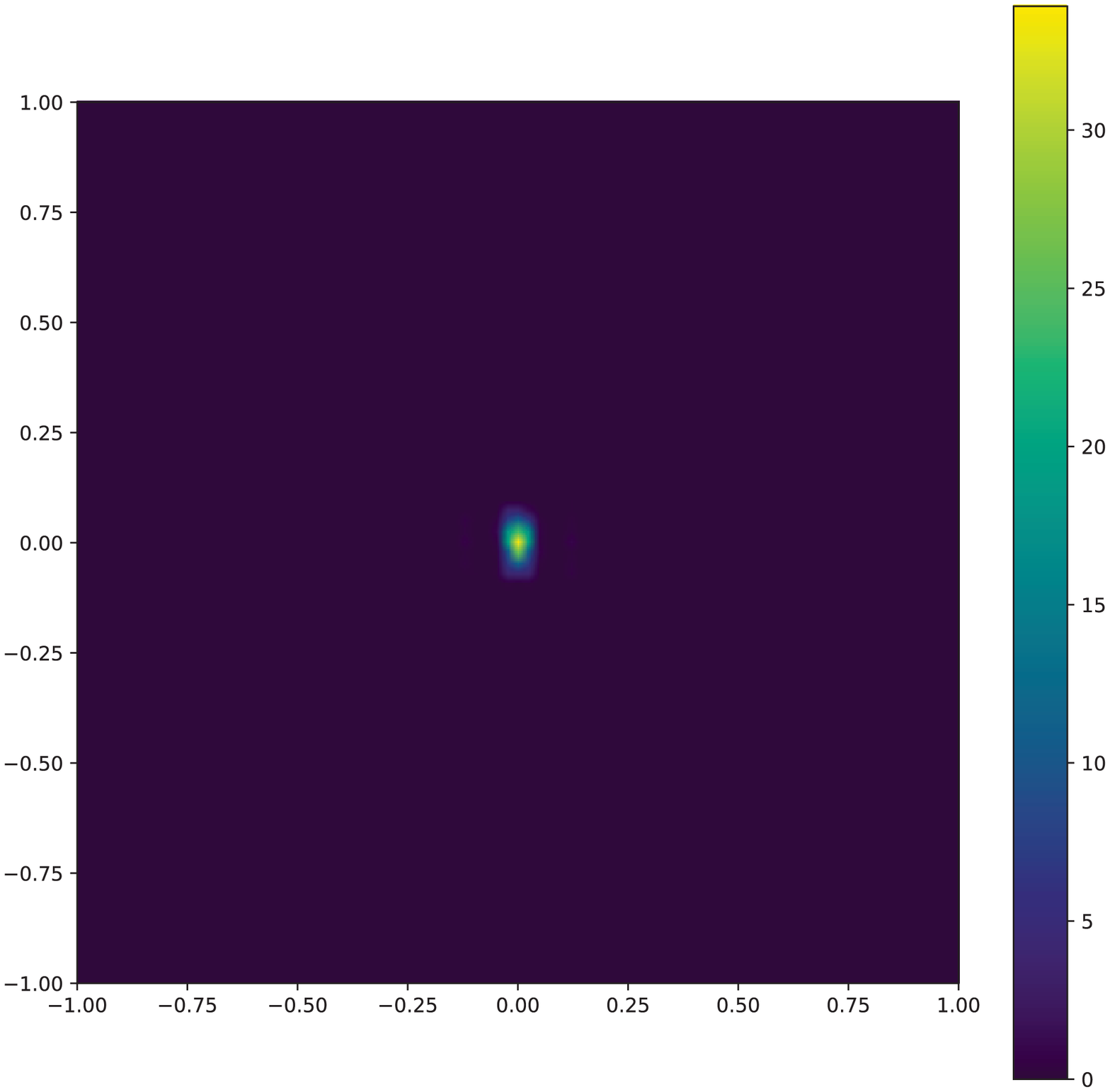}
\label{fig:positive_half_frank_ed}}\qquad
\subfigure[Frank energy density along the middle of the layer.]{
\includegraphics[width=0.4\linewidth]{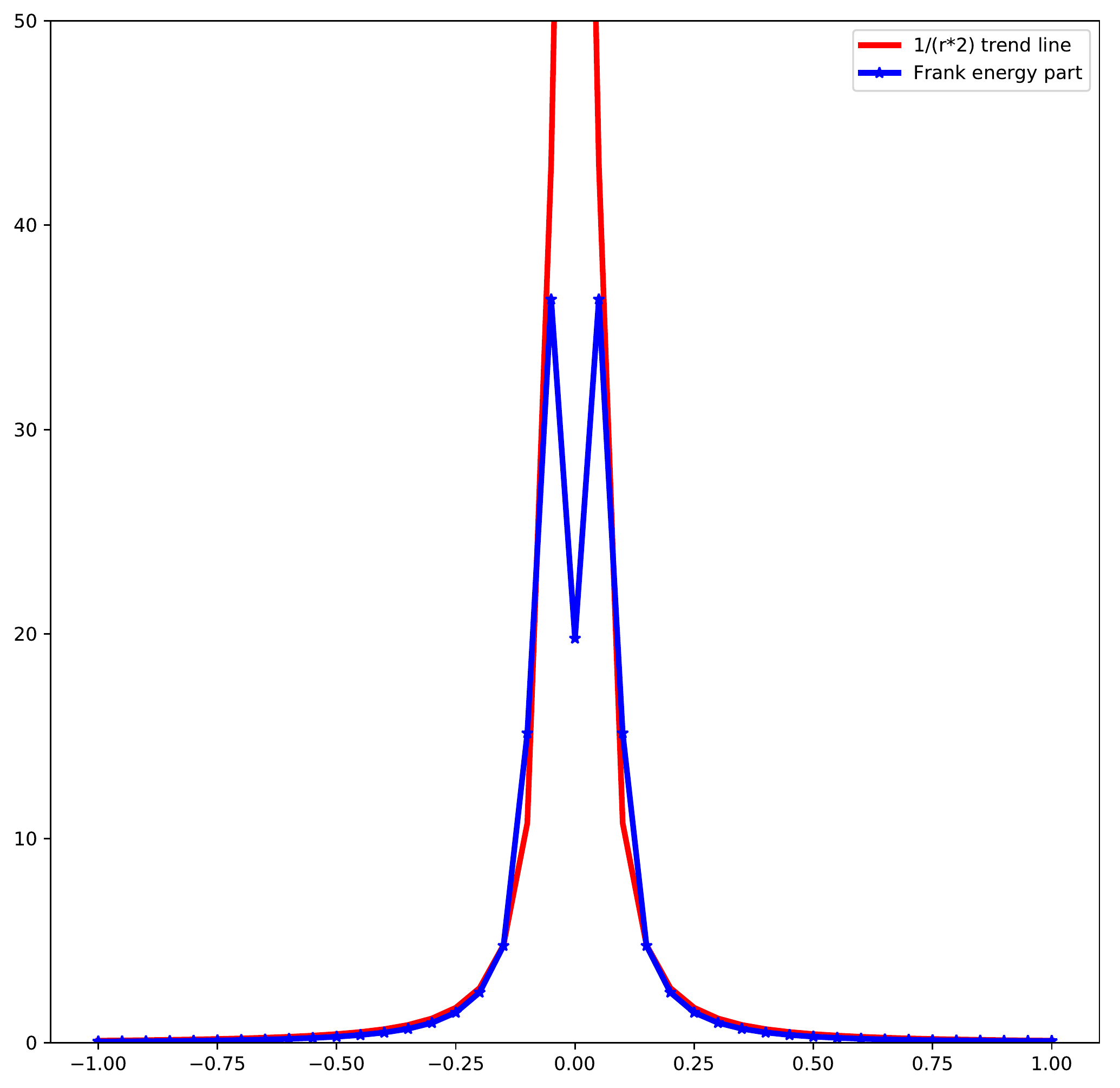}
\label{fig:positive_half__midlayer_ed}}\qquad
\caption{Frank energy density for positive half strength disclinations. Frank energy density is localized around disclination core. The Frank energy matches $1/r^2$ decaying rate outside the core and yields finite energy density inside the core.}\label{fig:half_ed_frank}
\end {figure}

Figure \ref{fig:half_ed_frank} shows the Frank energy contribution ($K|D k - B|^2$) for the positive half disclination and the comparison between the Frank energy density along $x_2=0$ with the function $1/r^2$, where $r$ is the distance of a point from the origin. The Frank energy density is also localized around the core, and it follows the $1/r^2$ decay rate outside the core, while yielding finite energy density inside the core.

To illustrate the effect of the prescription of $B$, we model $-\frac{1}{2}$ defect with a different layer field $B$ as follows,
\begin{equation}\label{eqn:B_negative_alt}
B(x, y) = \begin{cases}
-\frac{\sqrt{2}}{2a \xi}\bfe_1 \otimes \bfe_1 + \frac{\sqrt{2}}{2a \xi}\bfe_2 \otimes \bfe_1, & \text{ if $|x|<{\frac{a\xi}{2}}$  and $y < 0$} \\
0, & otherwise.
\end{cases}
\end{equation}

\begin{figure}
\centering
\subfigure[Prescription of $B$ of negative half disclination with vertical layer.]{
\includegraphics[width=0.3\linewidth]{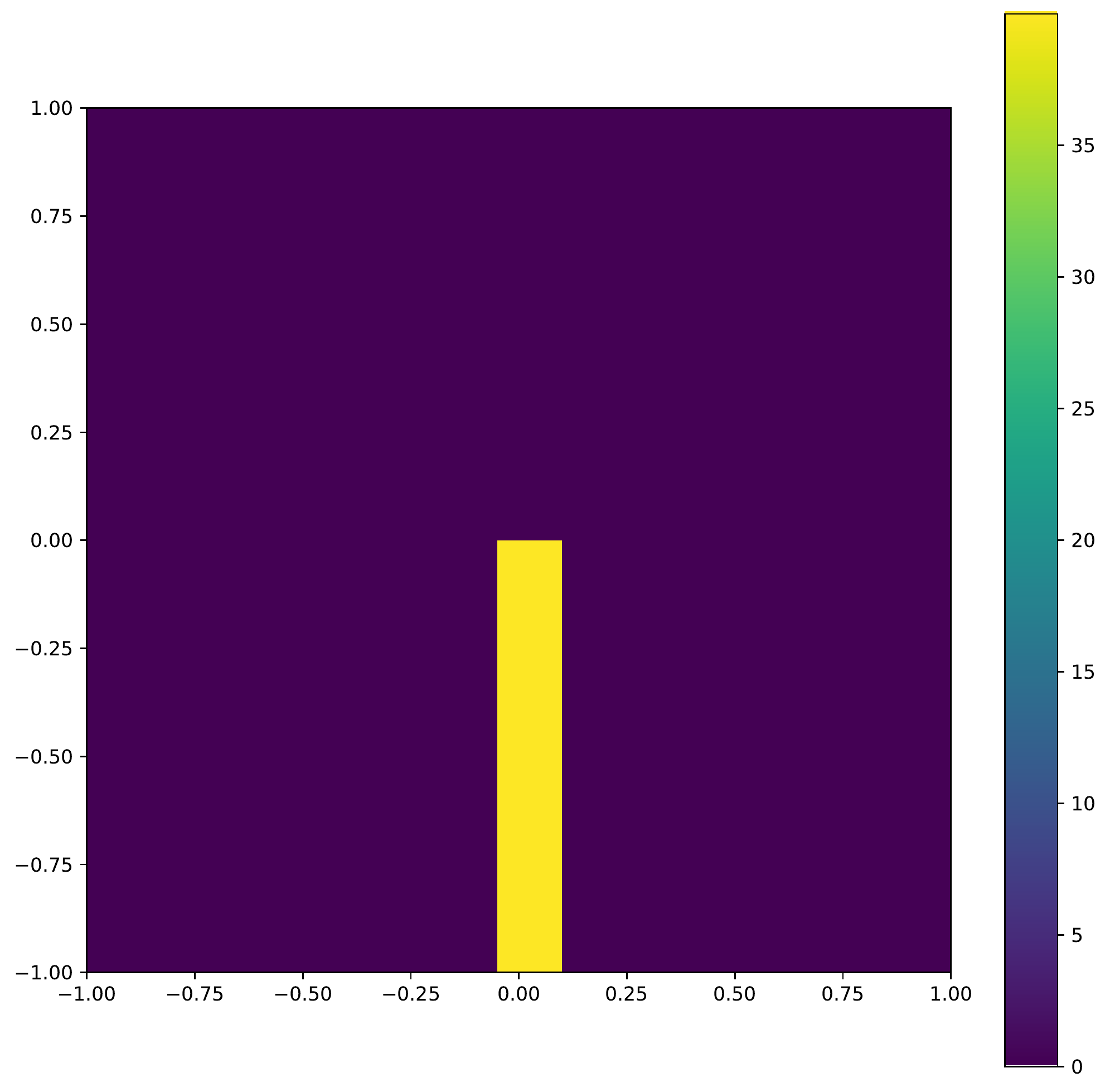}
\label{fig:half_neg_vertival_Bini}}\quad
\subfigure[Equilibrium of $k$ of negative half disclination modeled with vertical layer.]{
\includegraphics[width=0.3\linewidth]{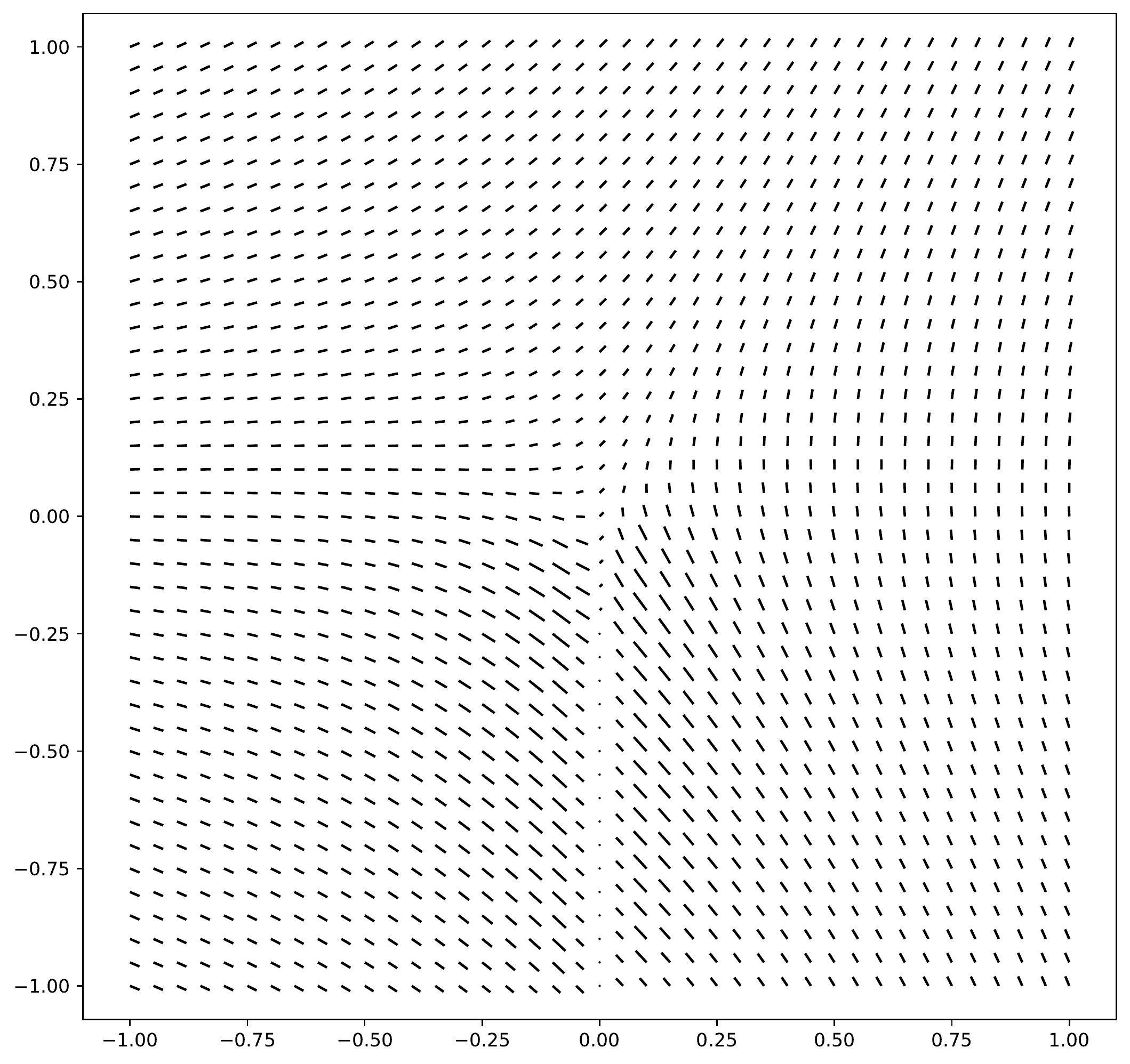}
\label{fig:half_neg_vertical_kres}}\quad
\subfigure[Energy density of negative half disclination modeled with vertical layer.]{
\includegraphics[width=0.3\linewidth]{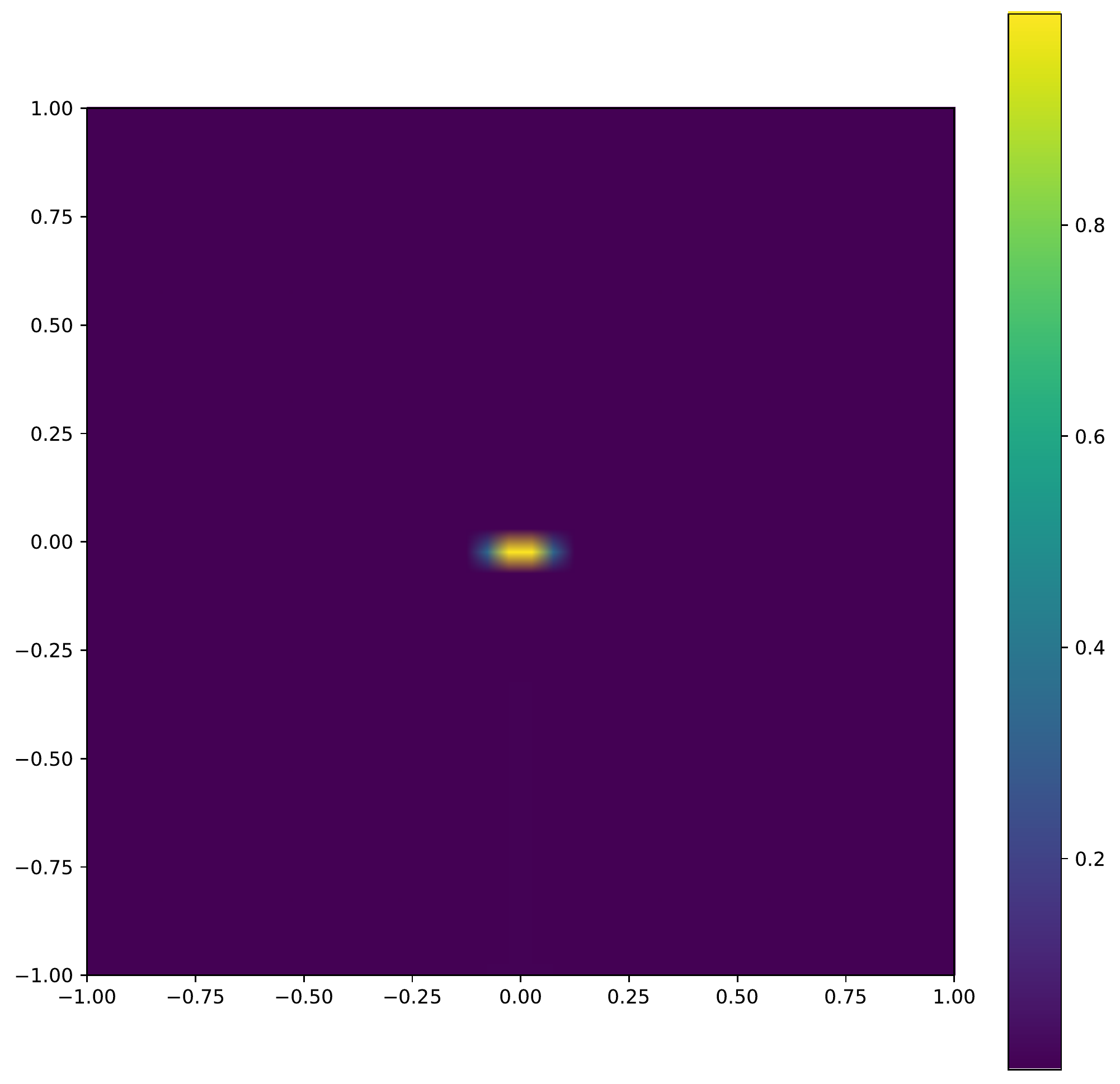}
\label{fig:half_neg_vertical_edres}}\quad
\caption{Prescription of $B$, Equilibrium of $k$, and energy density of negative half disclination modeled with vertical layer.}\label{fig:half_neg_vertical}
\end {figure}

Fig.~\ref{fig:half_neg_vertival_Bini} shows the prescription of $B$. Fig.~\ref{fig:half_neg_vertical_kres} and Fig.~\ref{fig:half_neg_vertical_edres} are the static results for director $k$ and energy density field respectively. Although the prescription of $B$ is very different compared to Fig.~\ref{fig:negative_half_B}, the static equilibria of $k$ and energy density outside cores and the energy density core shapes are similar.

\subsubsection{Strength $+1$ defect}\label{sec:+1}

\begin{figure}
\centering
\subfigure[Initialization of $B$ for $+1$ defect.]{
\includegraphics[width=0.4\linewidth]{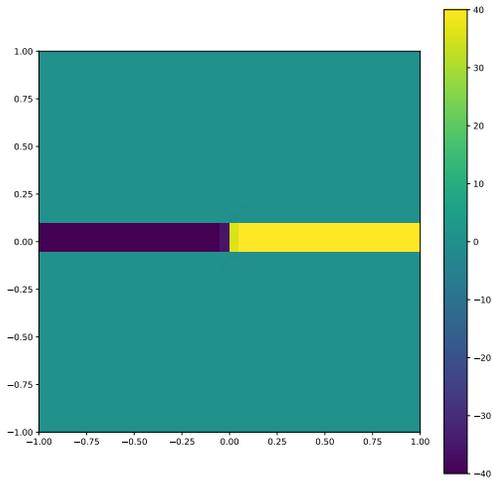}
\label{fig:positive_one_B_ini}}\qquad
\subfigure[Initialization of $k$ for $+1$ defect.]{
\includegraphics[width=0.4\linewidth]{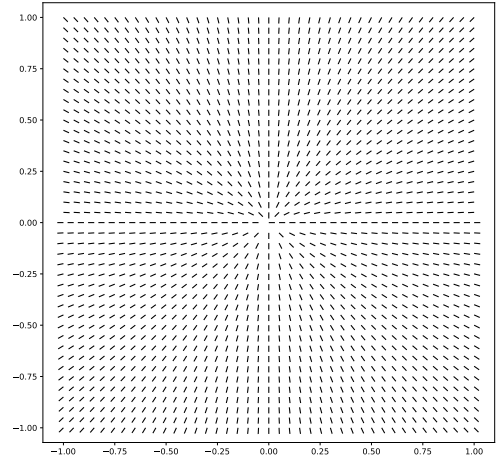}
\label{fig:positive_one_k_ini}}\qquad
\caption{Initialization of $B$ and $k$ for positive one disclination.}
\end{figure}

\begin{figure}
\centering
\subfigure[Constrained equilibrium of $k$ for $+1$ `source' defect.]{
\includegraphics[width=0.4\linewidth]{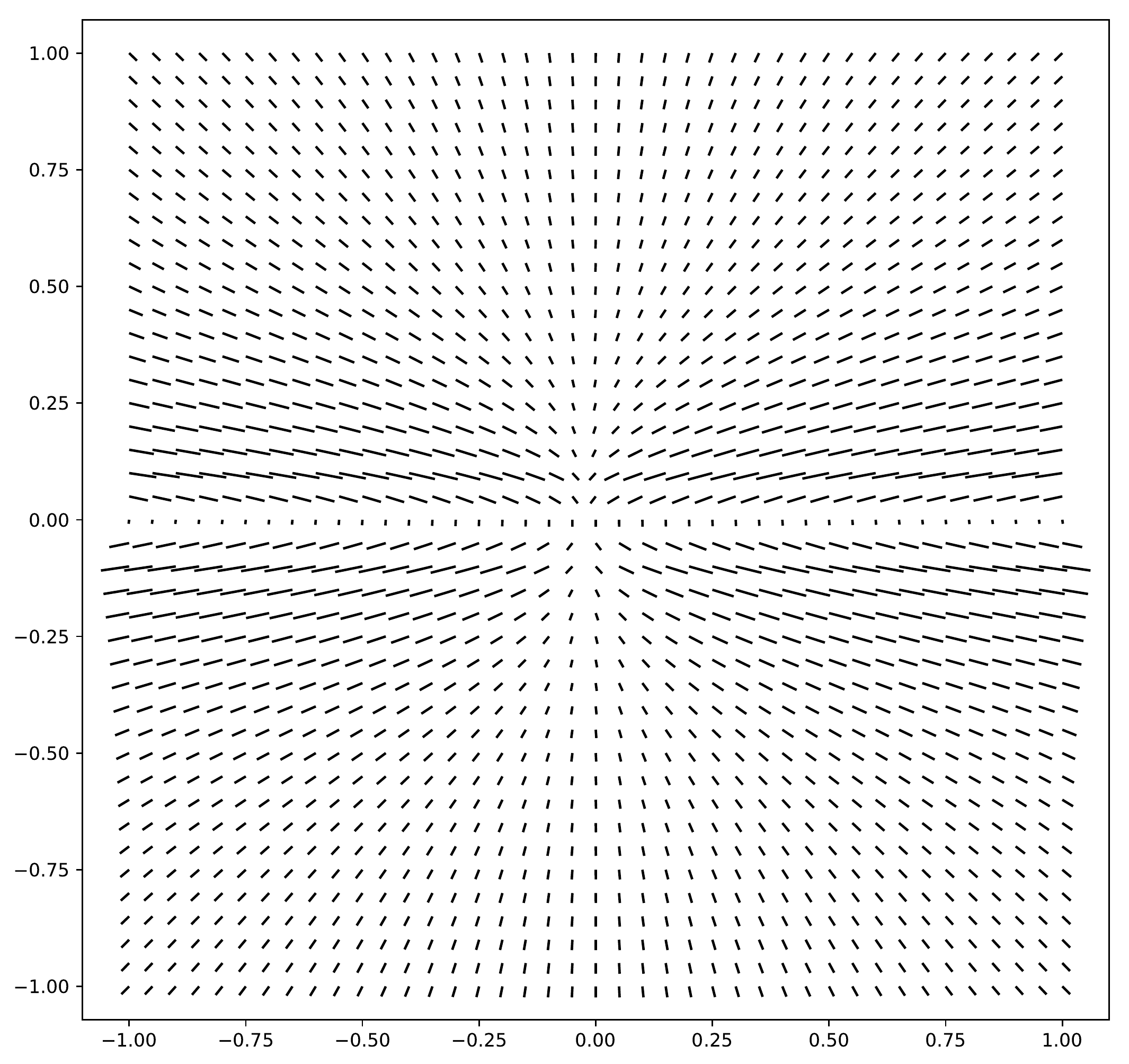}
\label{fig:positive_one_k_result}}\qquad
\subfigure[Energy density.]{
\includegraphics[width=0.4\linewidth]{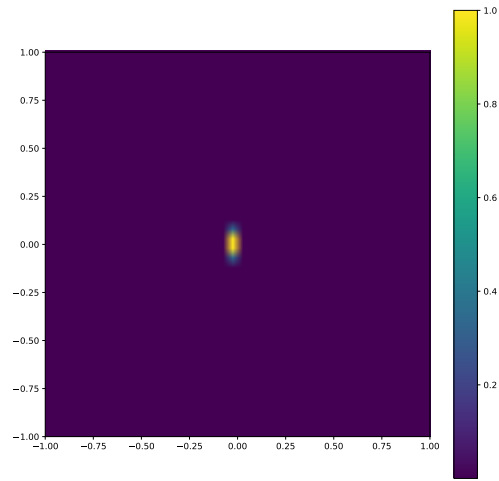}
\label{fig:positive_one_k_ed}}\qquad
\caption{Constrained equilibrium of director field $k$ and energy density for $+1$ `source' defect.}
\label{fig:positive_one_result}
\end{figure}

\begin{figure}
\centering
\subfigure[Equilibrium of $k$ of opposite half-strength defects at small distance.]{
\includegraphics[width=0.4\linewidth]{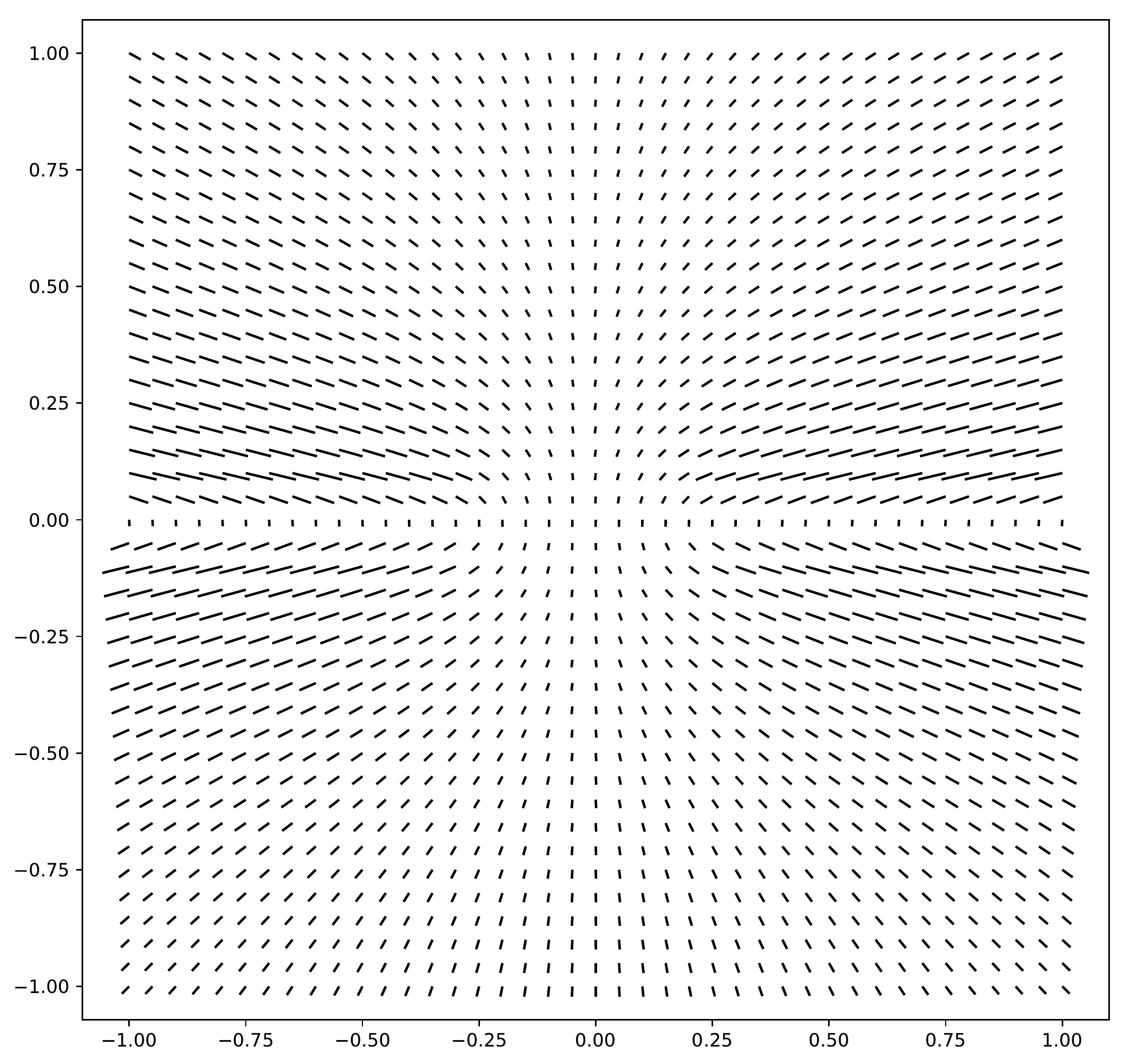}
\label{fig:positive_one_k_split_small}}\qquad
\subfigure[Equilibrium of $k$ of opposite half-strength defects at large distance.]{
\includegraphics[width=0.4\linewidth]{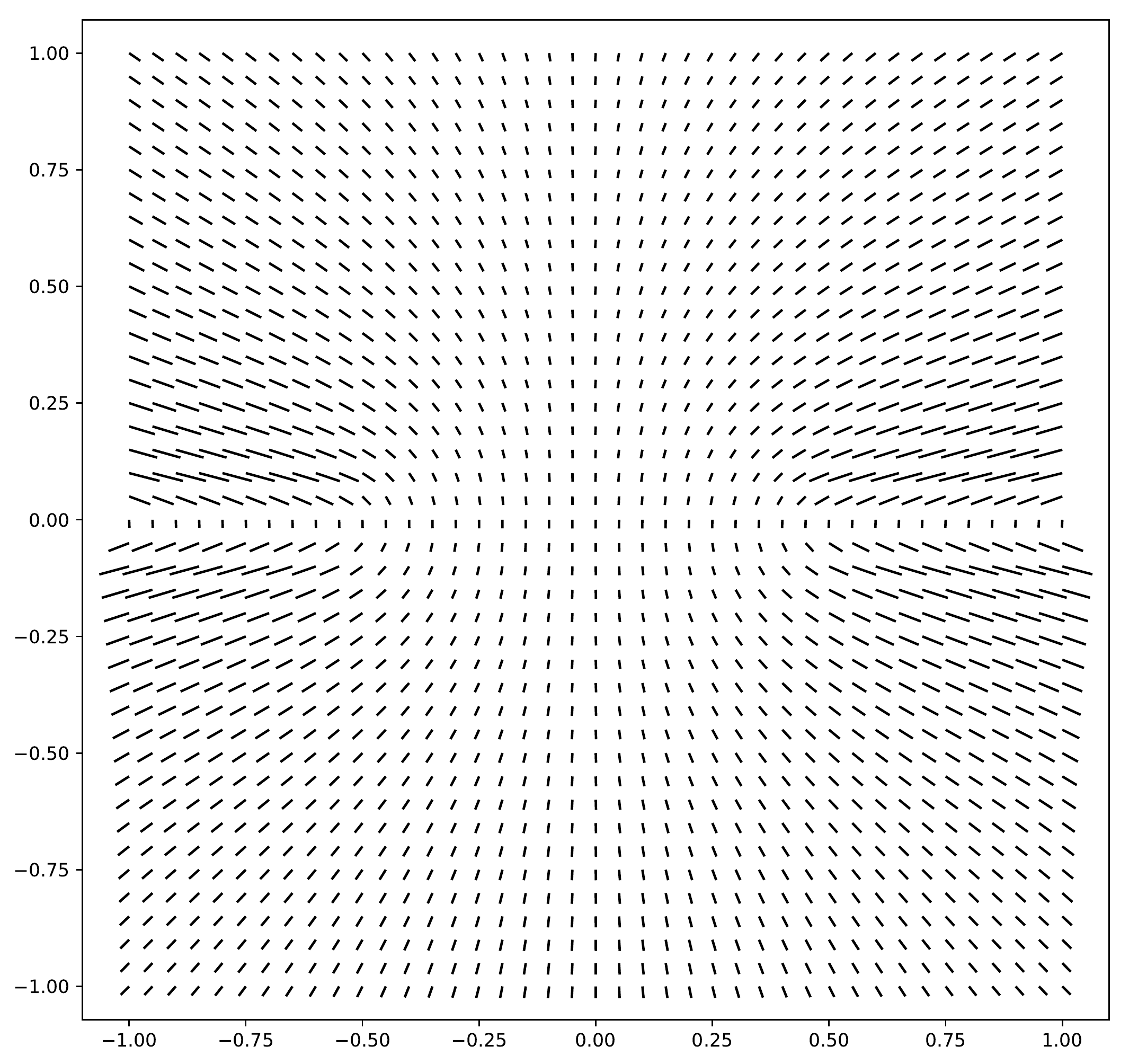}
\label{fig:positive_one_k_split_large}}\qquad
\subfigure[Equilibrium of energy density of opposite half-strength defects at small distance.]{
\includegraphics[width=0.4\linewidth]{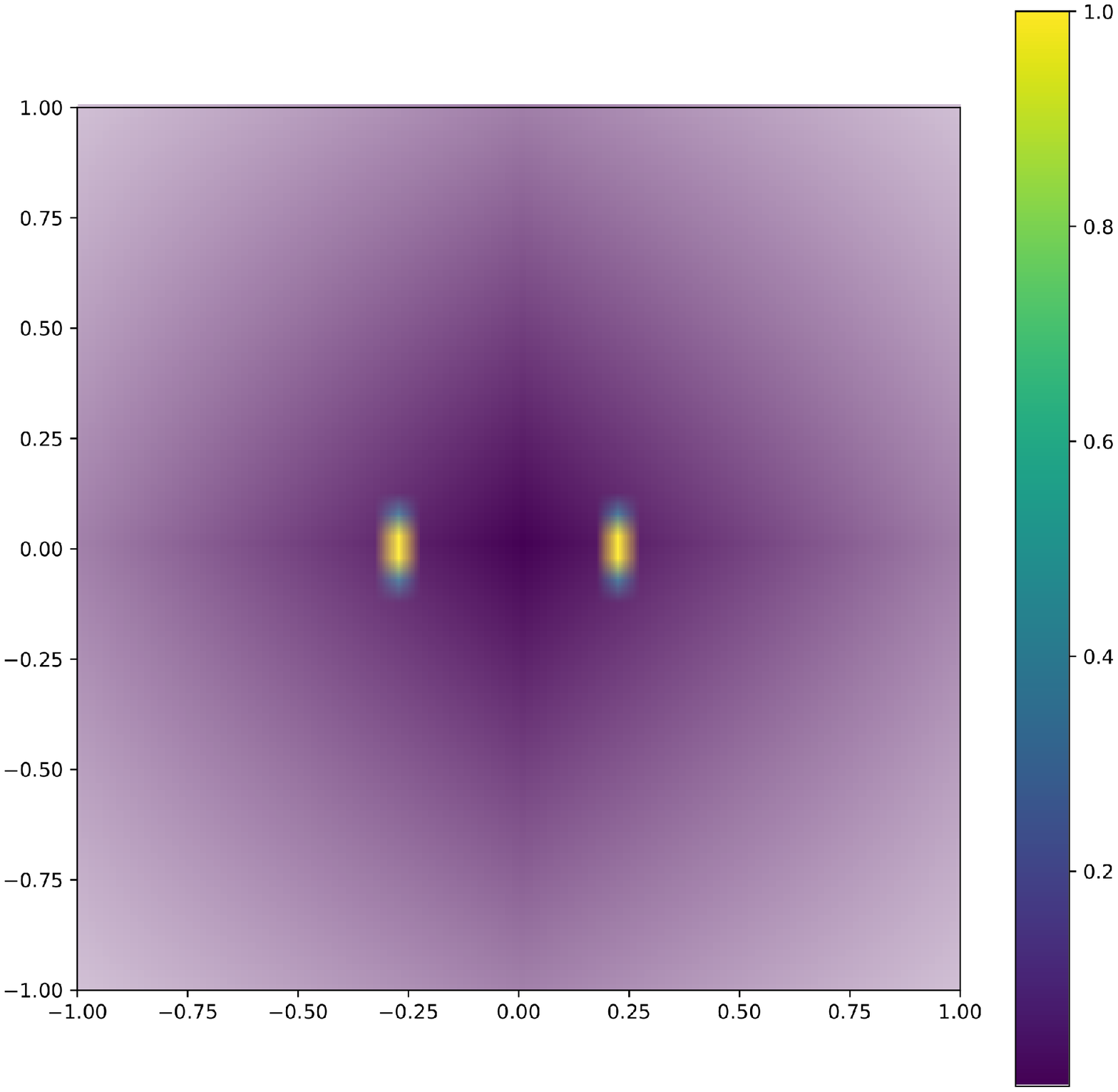}
\label{fig:positive_one_ed_split_small}}\qquad
\subfigure[Equilibrium of energy density of opposite half-strength defects at large distance.]{
\includegraphics[width=0.4\linewidth]{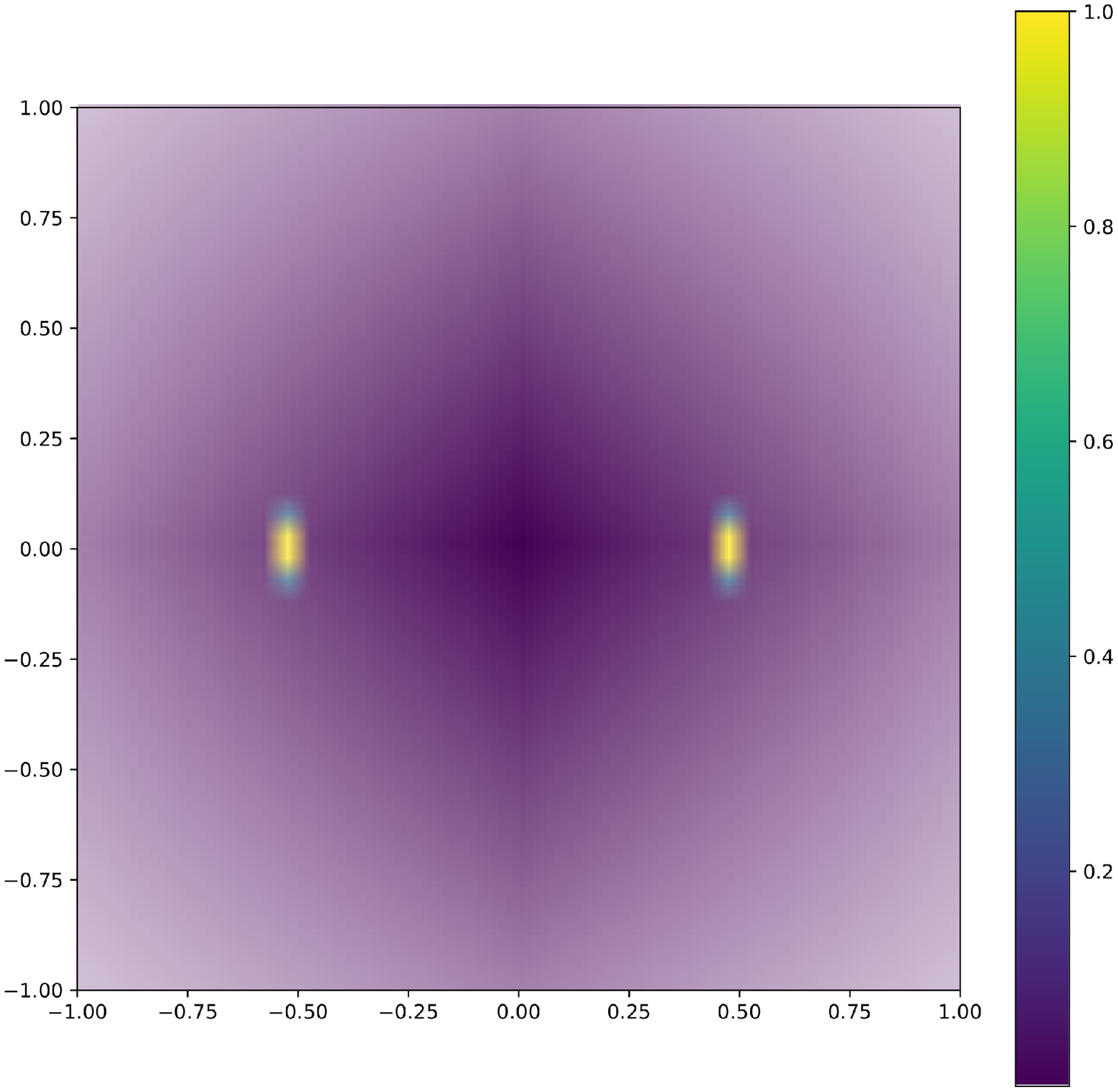}
\label{fig:positive_one_ed_split_large}}\qquad
\caption{Equilibrium of director field $k$ and energy density for a split pair of $+\frac{1}{2}$ defects.}
\label{fig:positive_one_split_result}
\end{figure}

A strength ($+1$) defect can be represented as a \textit{composite} defect in our model by putting two $+\frac{1}{2}$ defects close together, as shown in Fig \ref{fig:positive_one_B_ini}. The initialization of the $k$ field is shown in Fig \ref{fig:positive_one_k_ini}. Since the strength one defect is energetically unstable, we increase $\alpha$ to 50 in this example to constrain the diffusion of $B$. Both $B$ and $k$ are evolved following the gradient flow dynamics \eqref{eqn:grad_dimen}. The equilibrium of the director field $k$ and the energy density are shown in Figure \ref{fig:positive_one_result}. In this section the color legends of all energy density plots are normalized by their maximum values. Since the strength one defect is energetically unstable, it tends to split into two half-strength defects with opposite signs which, subsequently, repel each other. As mentioned in Section \ref{sec:gf_derivation}, the gradient flow dynamics \eqref{eqn:grad_dimen} is not capable of capturing this  behavior. Thus, we calculate the equilibrium configurations and total energies corresponding to two strength-half defects at different separation distances to approximate the strength-one defect splitting process. In Fig \ref{fig:positive_one_k_split_small} and \ref{fig:positive_one_k_split_large}, we show equilibria of the $k$ field corresponding to two opposite half-strength defects at specified distances. Fig \ref{fig:positive_one_ed_split_small} and \ref{fig:positive_one_ed_split_large} are their corresponding energy density fields. The total non-dimensionalized energies for the positive one strength defect is $1.812 \times 10^5$. After being normalized by the positive one strength defect's total energy, the total energies for the two opposite half-strength defect configurations at small and large separation distances are $0.508$ and $0.498$, respectively. Thus, a pair of + half-strength defects are energetically preferable in comparison to a single strength +1 defect, and the elements of the pair repel each other.

\begin{figure}
\centering
\subfigure[Constrained equilibrium of $k$ for 'target' pattern of $+1$ disclination.]{
\includegraphics[width=0.4\linewidth]{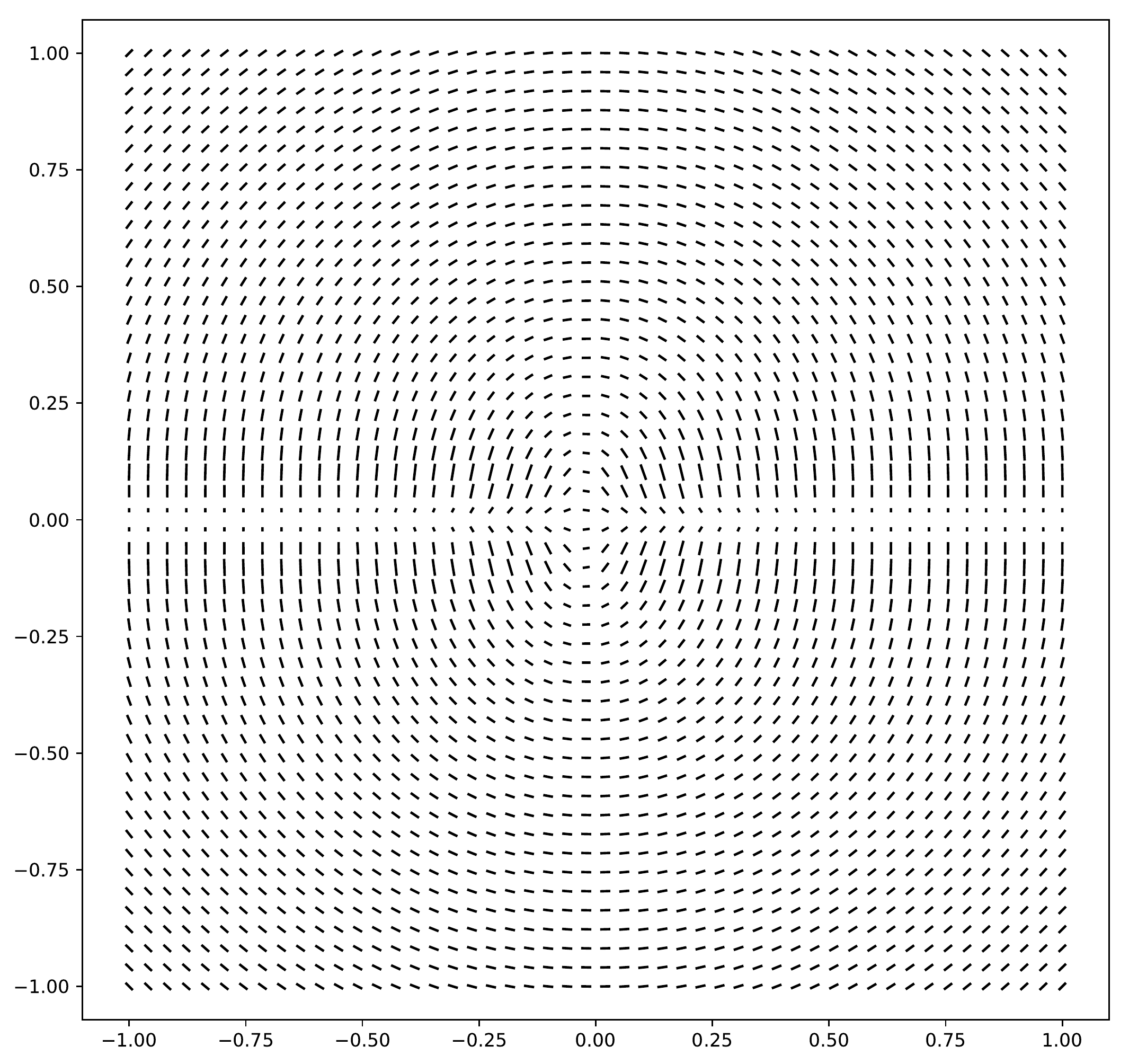}
\label{fig:positive_one_target_k}}\qquad
\subfigure[Energy density for 'target' pattern of $+1$ disclination.]{
\includegraphics[width=0.4\linewidth]{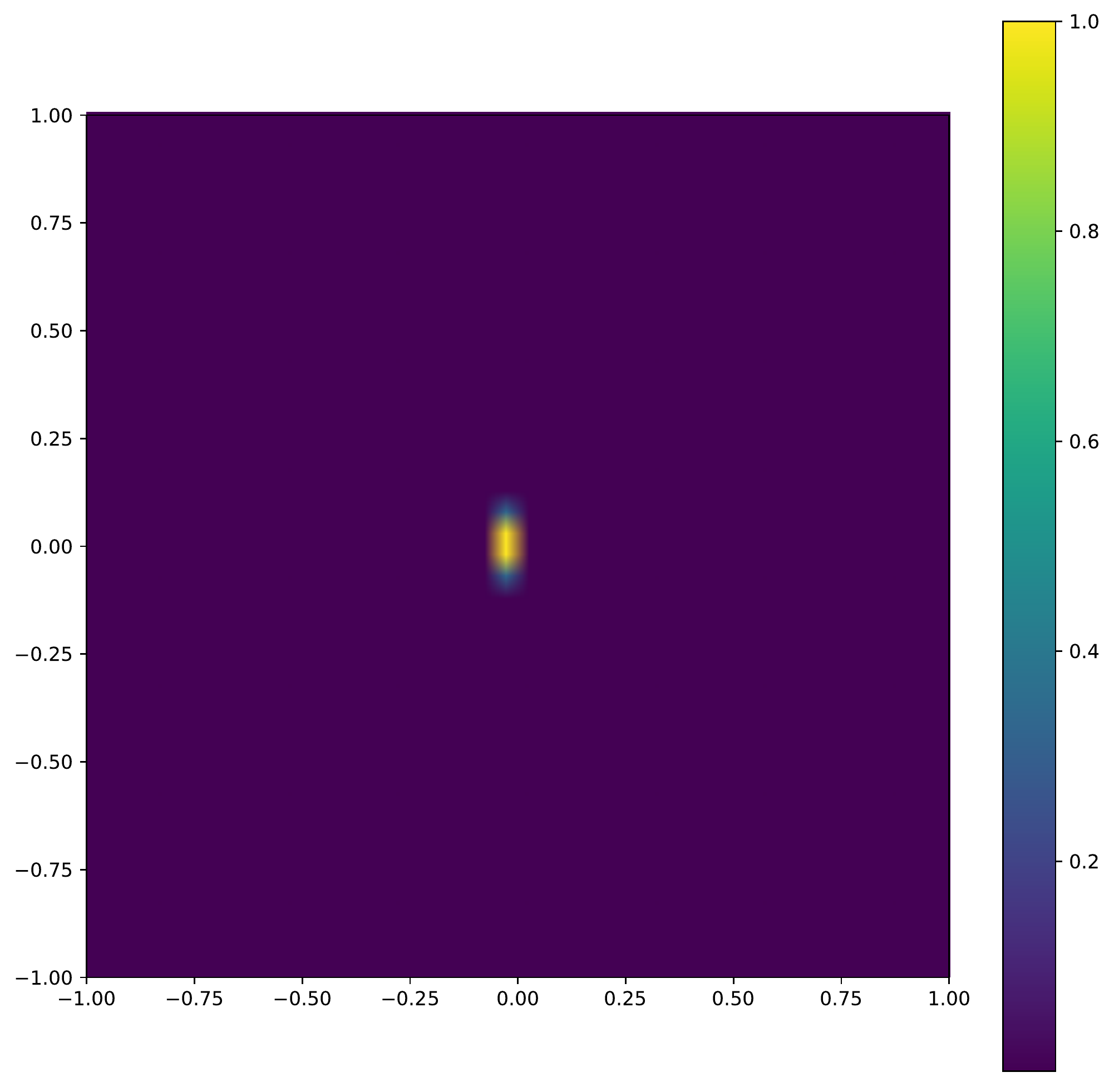}
\label{fig:positive_one_target_energy}}\qquad
\caption{Constrained equilibrium of $k$ and energy density for 'target' pattern of $+1$ disclination.}
\label{fig:positive_one_target}
\end{figure}

In addition to the `source' pattern of the $+1$ disclination, we also calculate the `target' pattern of the same strength $+1$ disclination, whose director field $k$ and energy density are shown in Figure \ref{fig:positive_one_target}. For the `star' pattern, $k$ flips horizontally across the layer, and $B_{12}$ is nonzero within the layer. For the `target' pattern, $k$ flips vertically across the layer, and $B_{22}$ is nonzero within layer.

We note that \textit{that in all the examples solved in this paper, the width of the layer(s) for the specification of $B$ through initial conditions can be made arbitrarily small without affecting the qualitative properties of the solutions.}

\subsubsection{Strength $-3/2$ defect}\label{sec:-3/2}

\begin{figure}
\centering
\subfigure[Initialization of $B$ for $-3/2$ defect.]{
\includegraphics[width=0.4\linewidth]{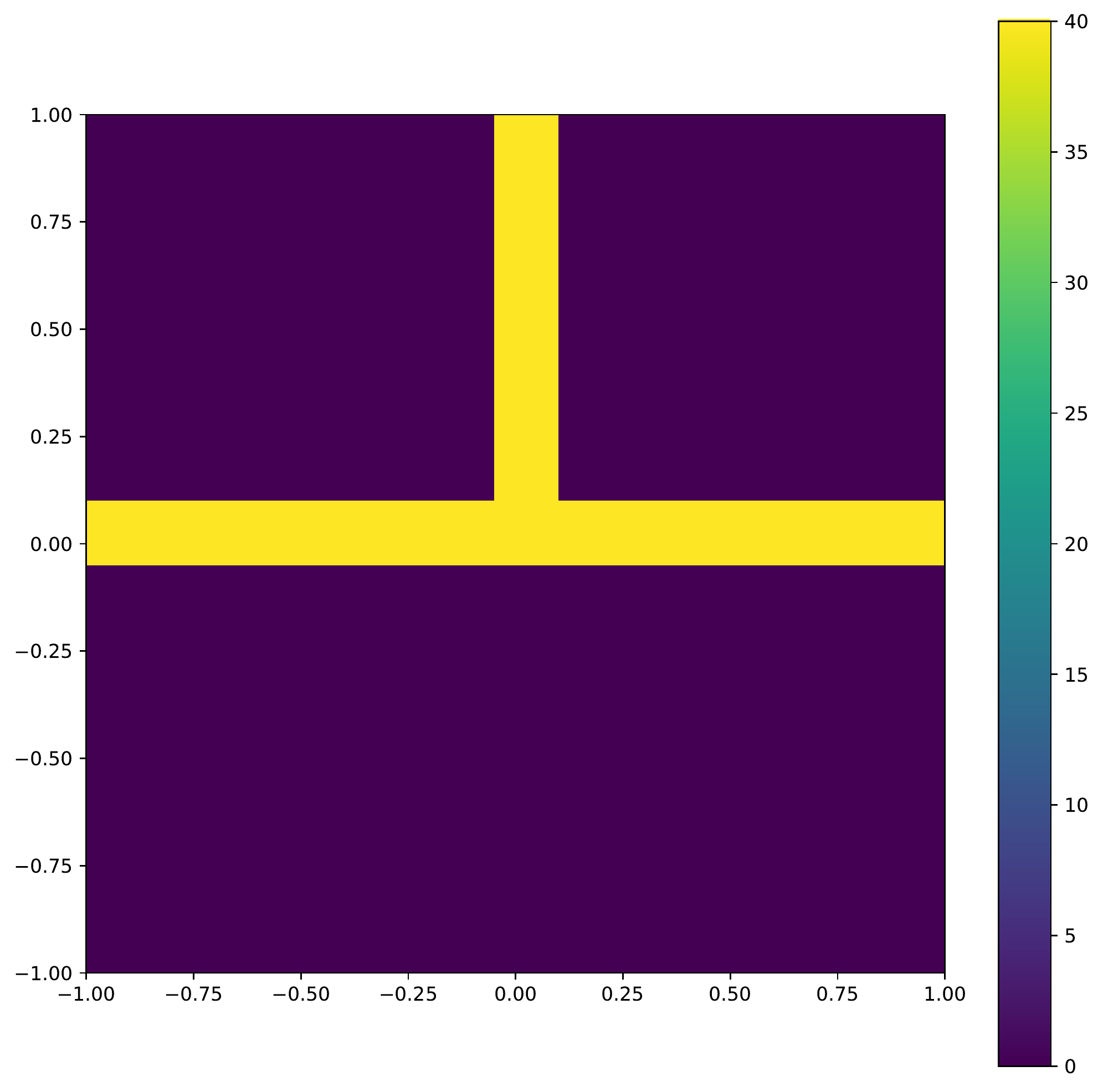}
\label{fig:negative_onehalf_B}}\qquad
\subfigure[Constrained equilibrium of $k$ for $-3/2$ defect. A superposed contour explains the evaluation of the strength of the defect.]{
\includegraphics[width=0.4\linewidth]{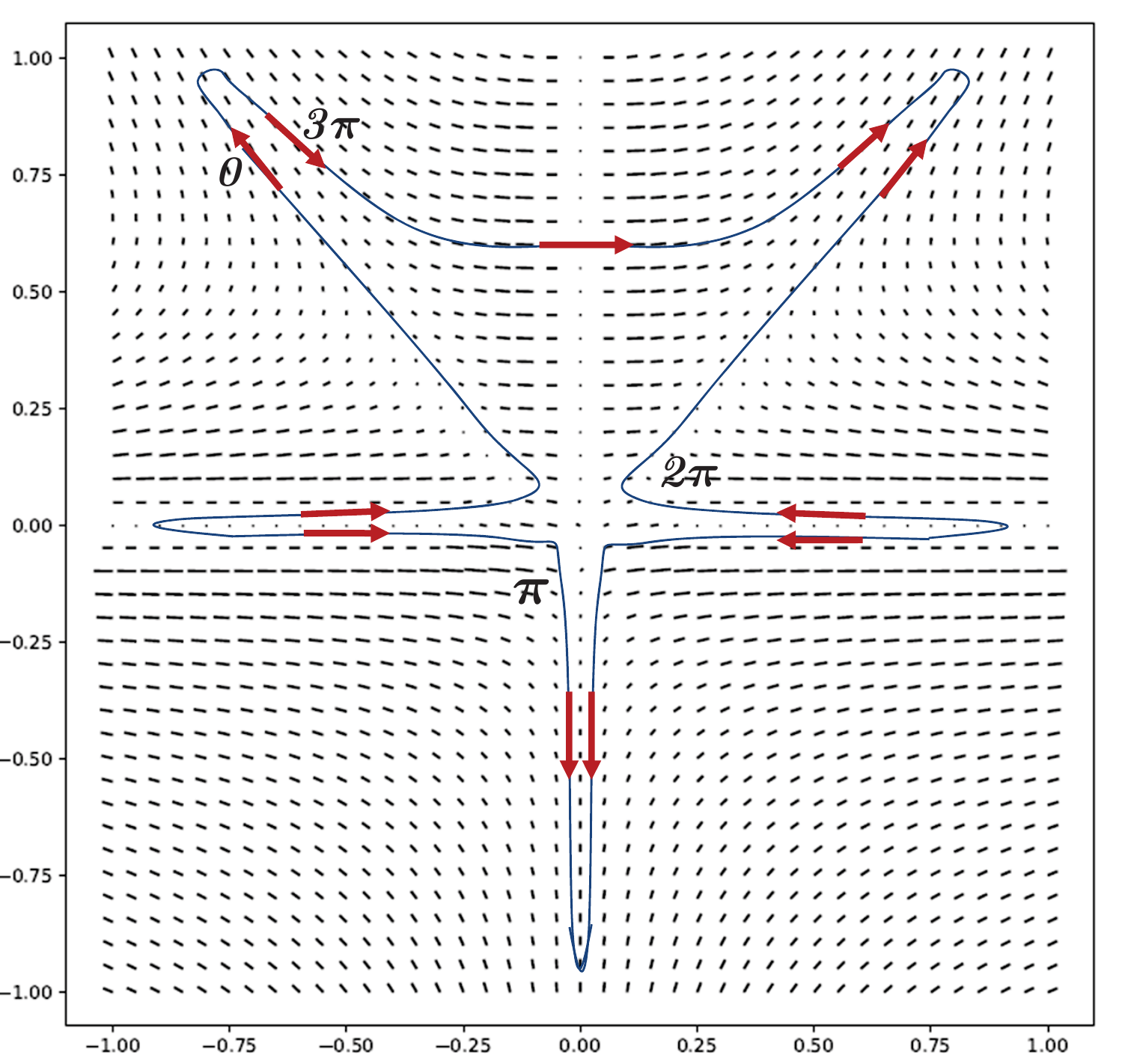}
\label{fig:negative_onehalf_k}}\qquad
\caption{Constrained equilibrium of $k$ for $-3/2$ disclination.}
\end{figure}

\begin{figure}
    \centering
    \subfigure[Energy density for $-3/2$ disclination. The color legend is normalized by its maximum.]{
    \includegraphics[width=0.4\linewidth]{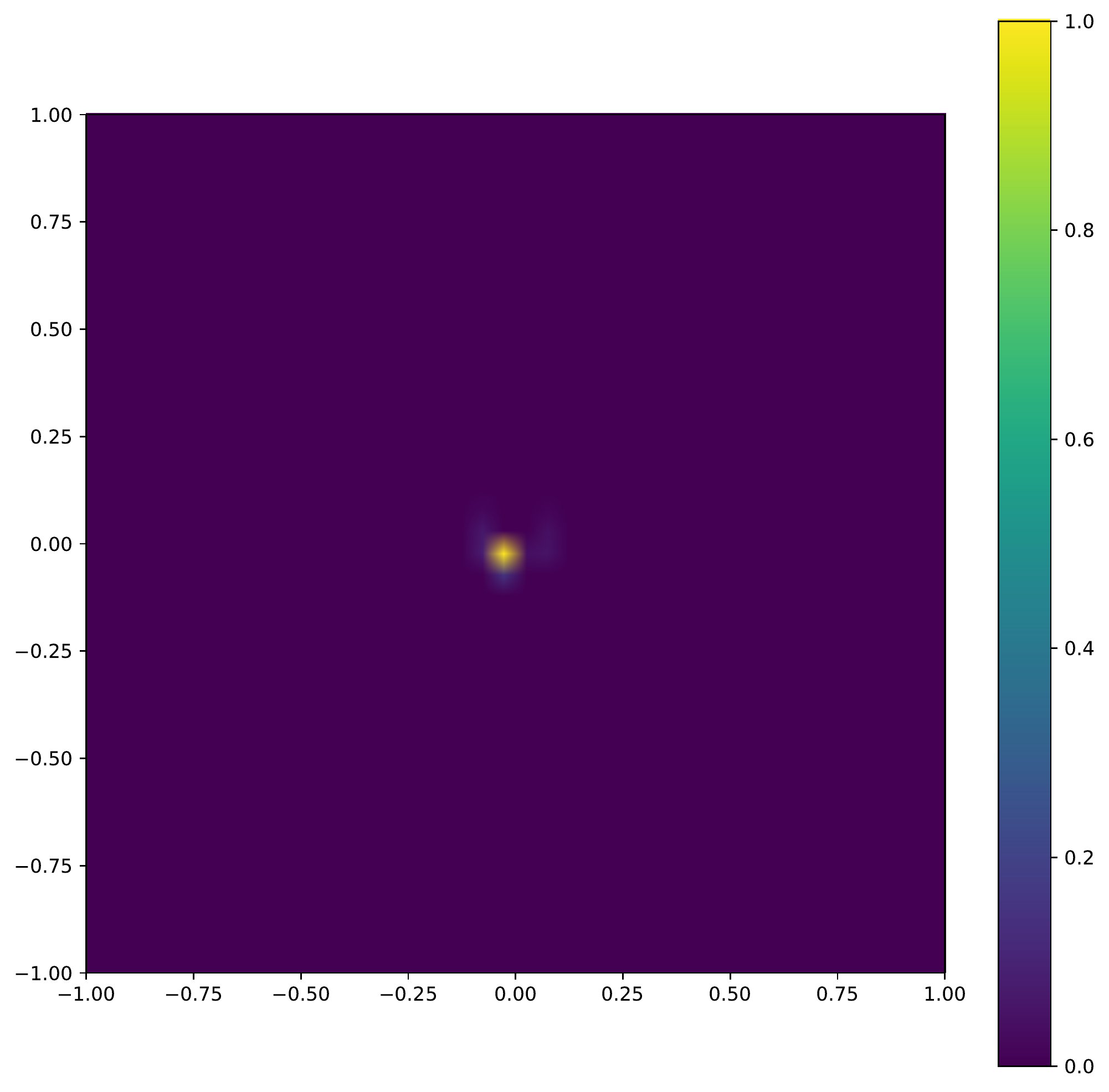}
    \label{fig:negative_onehalf_ed}}\qquad
    \subfigure[Normalized energy density in logarithmic scale zoomed in near the core. ]{
    \includegraphics[width=0.4\linewidth]{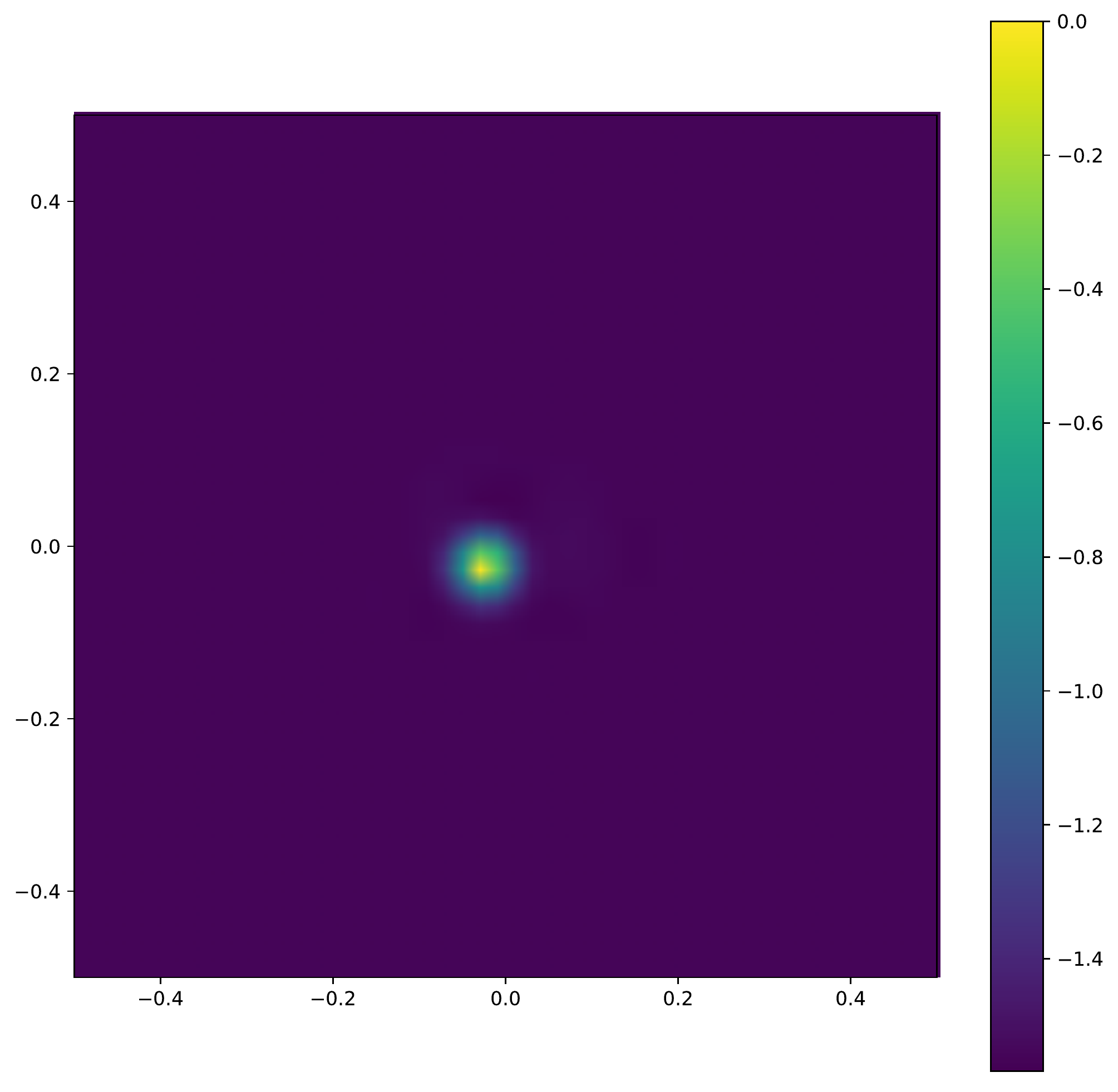}
    \label{fig:negative_onehalf_ed_logzoom}}\qquad
    \caption{Energy densities for a negative one  and a half strength defect.}
\end{figure}

Here we demonstrate a $-3/2$ strength defect as another interesting example of the capability of  our theory in modeling composite defects of higher strength. Figure \ref{fig:negative_onehalf_B} shows the $|B|$ field for the initial condition $B(x, y)$. The initial condition is a piecewise-constant field with three different constant values of $B$ in the layers, all with $|B| = 2$. In this calculation, $B$ is not allowed to evolve from its initial conditions (this is an energetically unstable defect, and we are simply interested in demonstrating a negative-strength composite here), and $k$ evolves following the gradient flow equations \eqref{eqn:grad_dimen}$_1$. Figure \ref{fig:negative_onehalf_k} shows the constrained equilibrium of the director field. Figure \ref{fig:negative_onehalf_ed} shows the non-dimensionalized energy density, where the layers are completely invisible and the energy density shows a strong, non-singular (by design) concentration at the core.

\subsubsection{Defect loop in 3D} \label{sec:squared_loop}
\begin{figure}
    \centering
    \subfigure[Prescription of $B$ of a squared loop defect.]{
    \includegraphics[trim={50 0 50 0}, clip, width=0.4\linewidth]{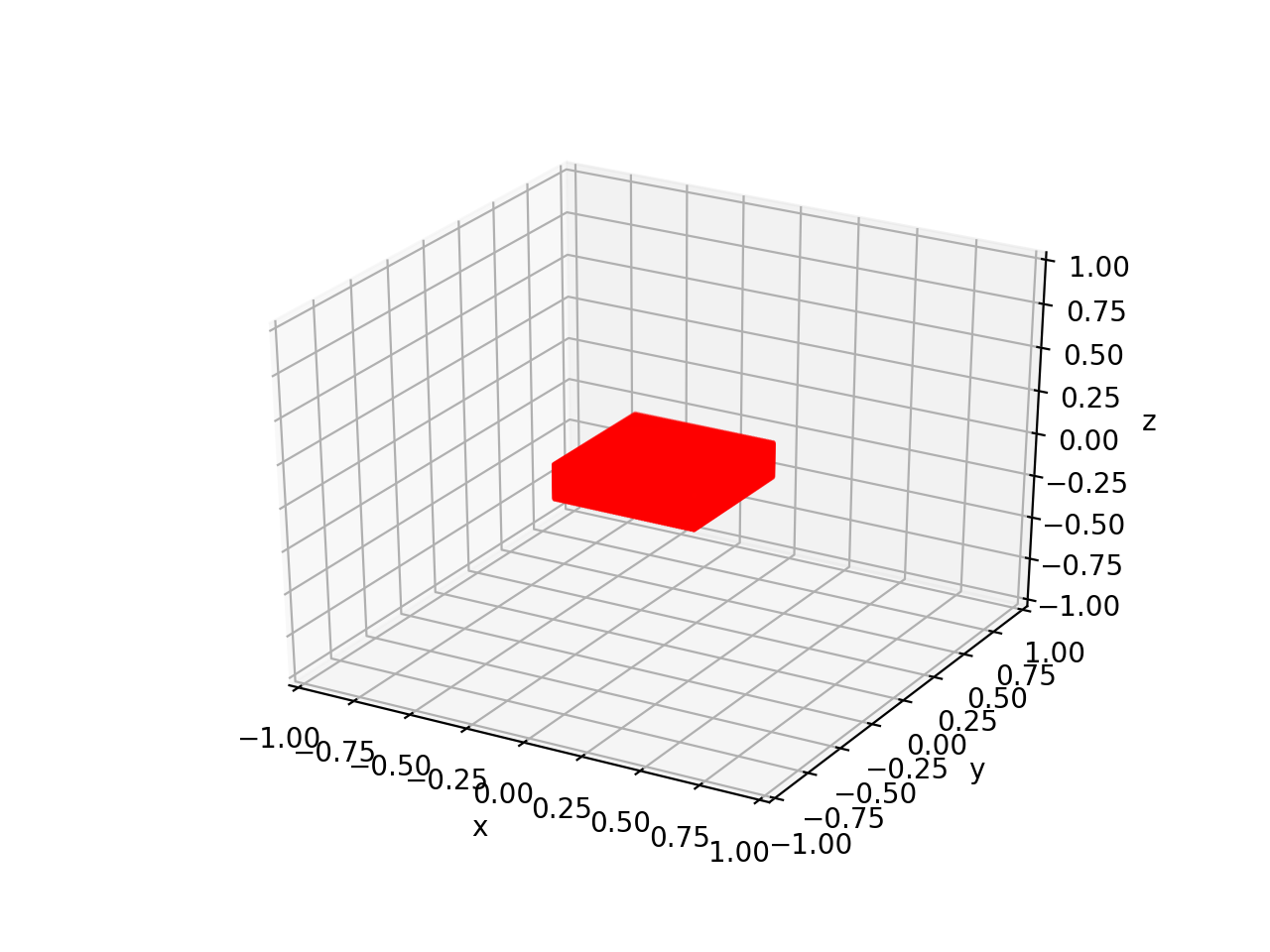}
    \label{fig:loop_ini_B}}\qquad
    \subfigure[Prescription of $\pi$ of a square loop defect.]{
    \includegraphics[trim={50 0 50 0}, clip, width=0.4\linewidth]{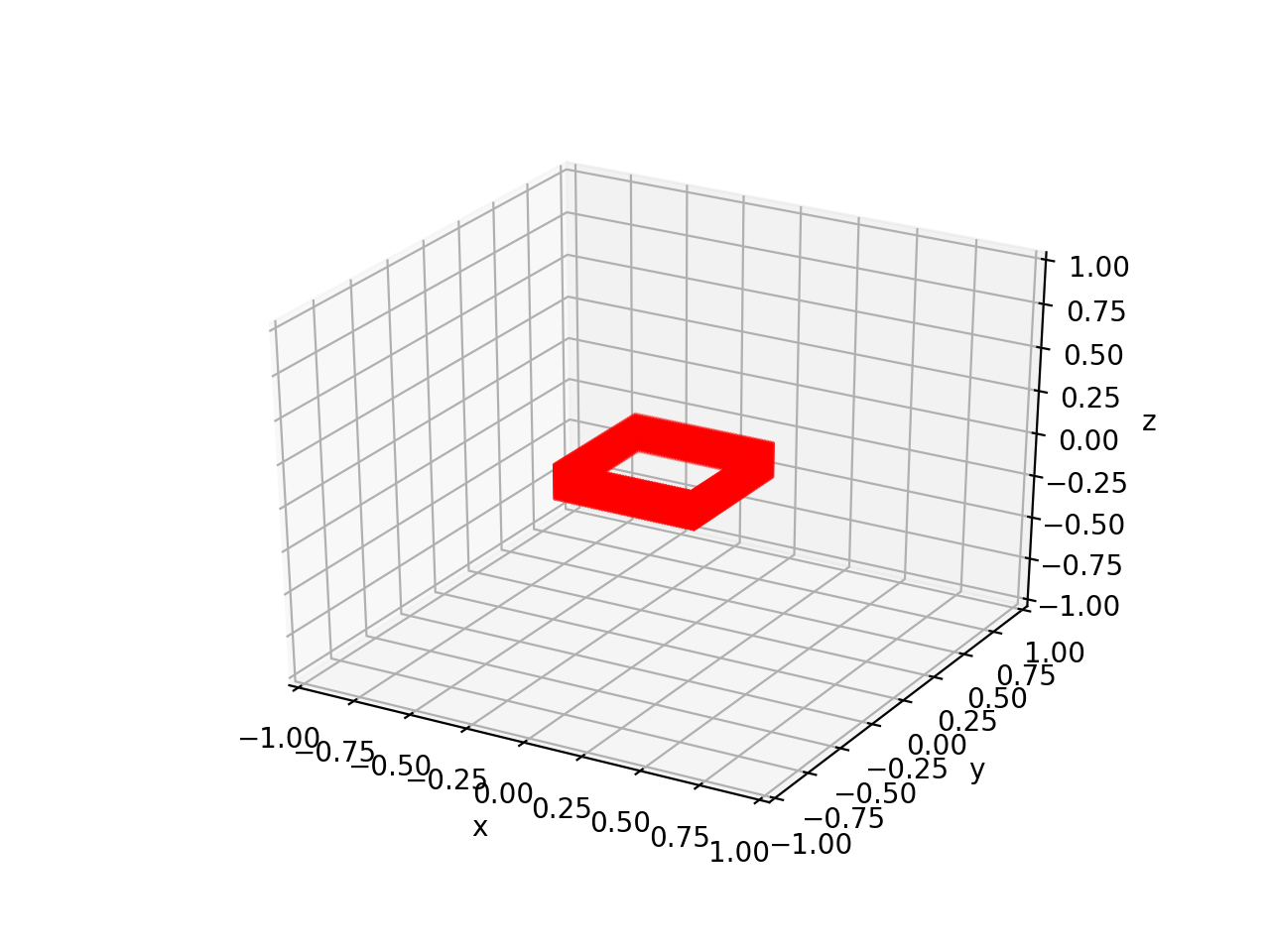}
    \label{fig:loop_ini_pi}}
    \caption{Initial prescriptions of $B$ and $\pi$ of squared loop defect. Red areas indicate where $B$ or $\pi$ is nonzero.}
\end{figure}

A square half strength defect loop in 3D case is demonstrated in this part. Fig.~\ref{fig:loop_ini_B} shows the prescription of $B$ given as follows,
\begin{equation}\label{eqn:B_loop}
B(x,y,z) = \begin{cases}
\frac{2}{2a \xi}\bfe_1 \otimes \bfe_3, & \text{ if $|z| \le {\frac{a\xi}{2}}$, $|x|\le d$, and $|y| \le d$} \\
0, & otherwise,
\end{cases}
\end{equation}
where $d$ represents the half length of defect square side. And Fig.~\ref{fig:loop_ini_pi} shows the corresponding prescription of $\pi$ field. The one-point specification of $k$ is applied at $(-1, -1, -1)$ in addition to zero moment boundary condition. The initial prescription of $k$ is given as 
\begin{equation}\label{eqn:k_loop}
k(x,y,z) = \begin{cases}
\cos(\frac{\arctan (z, x-d) + \pi}{2}) \bfe_1 + \sin(\frac{\arctan (z, x-d) + \pi}{2}) \bfe_3, & \text{ if $|y| \le d, x > d, z > |\frac{a\xi}{2}|$} \\
\cos(\frac{\arctan (z, -x-d) + \pi}{2}) \bfe_1 + \sin(\frac{\arctan (z, -x-d) + \pi}{2}) \bfe_3, & \text{ if $|y| \le d, x < -d, z > |\frac{a\xi}{2}|$} \\
\cos(\frac{\arctan (z, y-d) + \pi}{2}) \bfe_1 + \sin(\frac{\arctan (z, y-d) + \pi}{2}) \bfe_3, & \text{ if $|x| \le d, y > d, z > |\frac{a\xi}{2}|$} \\
\cos(\frac{\arctan (z, -y-d) + \pi}{2}) \bfe_1 + \sin(\frac{\arctan (z, -y-d) + \pi}{2}) \bfe_3, & \text{ if $|x| \le d, y < -d, z > |\frac{a\xi}{2}|$} \\
sign(z)\bfe_1, & \text{if $|x| \le d, |y| \le d, |z| = \frac{a\xi}{2}$} \\
0 & \text{if $|x| \le d, |y| \le d, z < |\frac{a\xi}{2}|$} \\
\bfe_3, & otherwise.
\end{cases}
\end{equation}

\begin{figure}
    \centering
    \subfigure[Director constrained equilibrium of a square loop defect.]{
    \includegraphics[trim={150 80 150 80}, clip, width=0.4\linewidth]{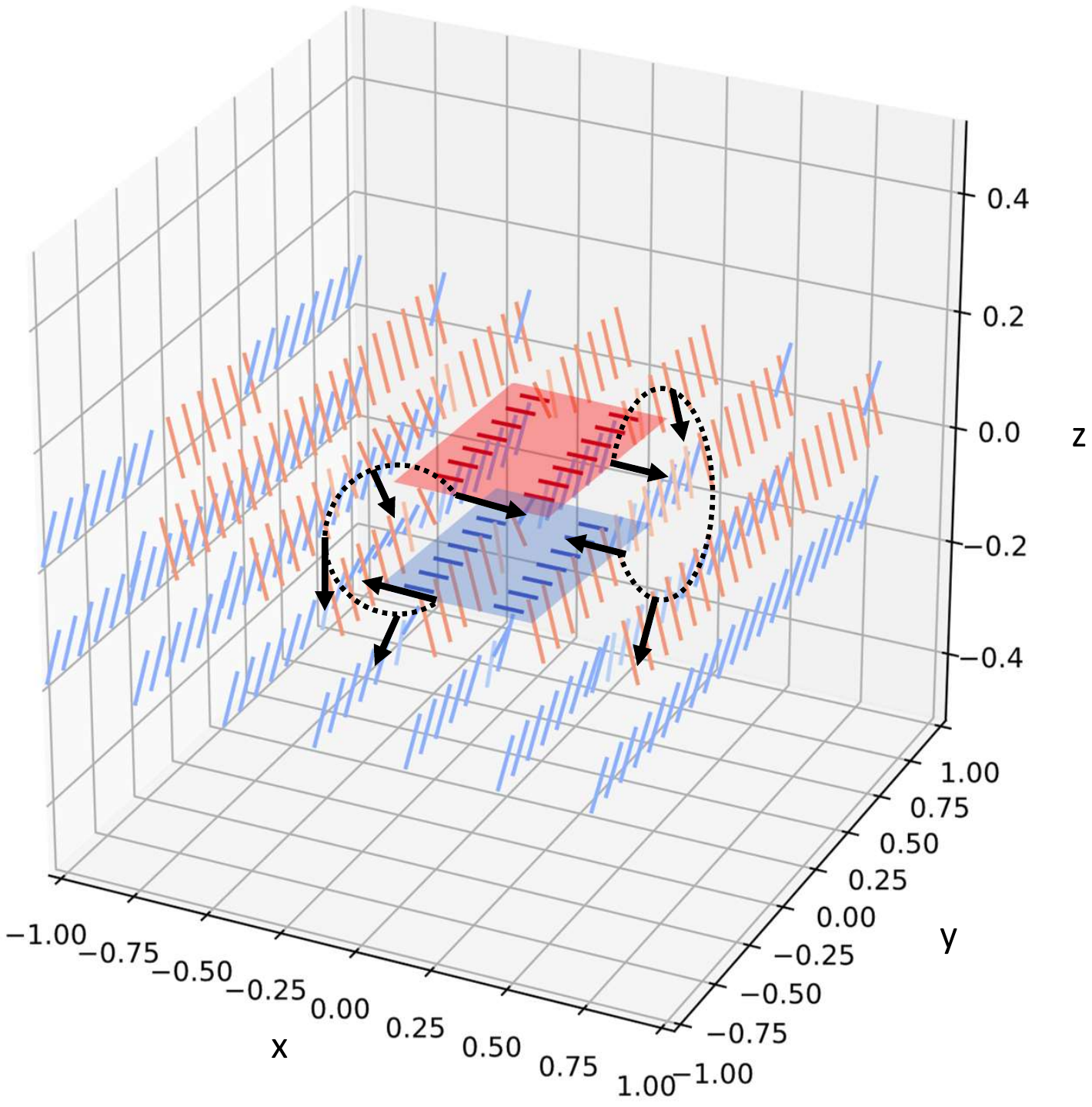}
    \label{fig:loop_res_k}} \qquad
    \subfigure[Director constrained equilibrium in $x$-$z$ cross sections.]{
   \includegraphics[height=0.35\linewidth, width=0.4\linewidth]{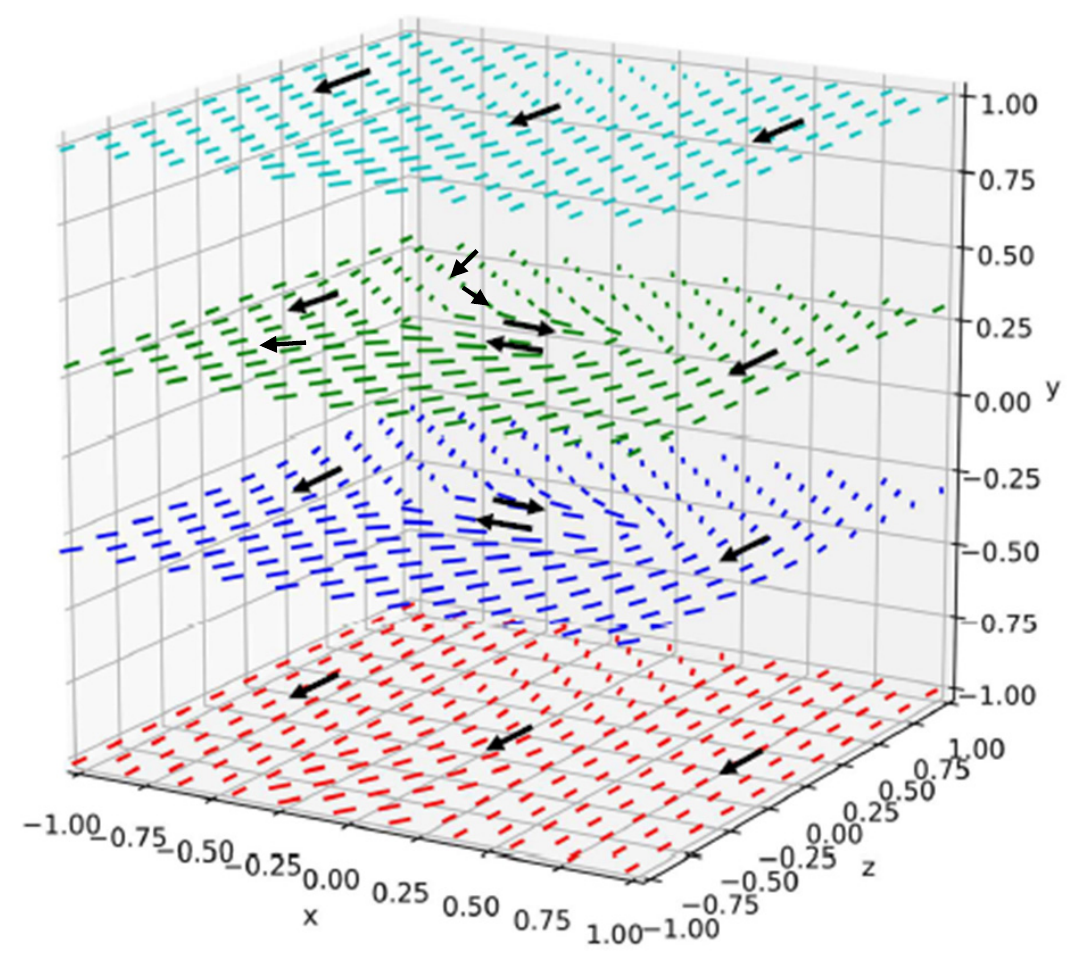}
    \label{fig:loop_res_k_twist}} \qquad
    \subfigure[Director constrained equilibrium on $x$-$z$ cross section.]{
    \includegraphics[trim={180 50 180 50}, clip, width=0.4\linewidth]{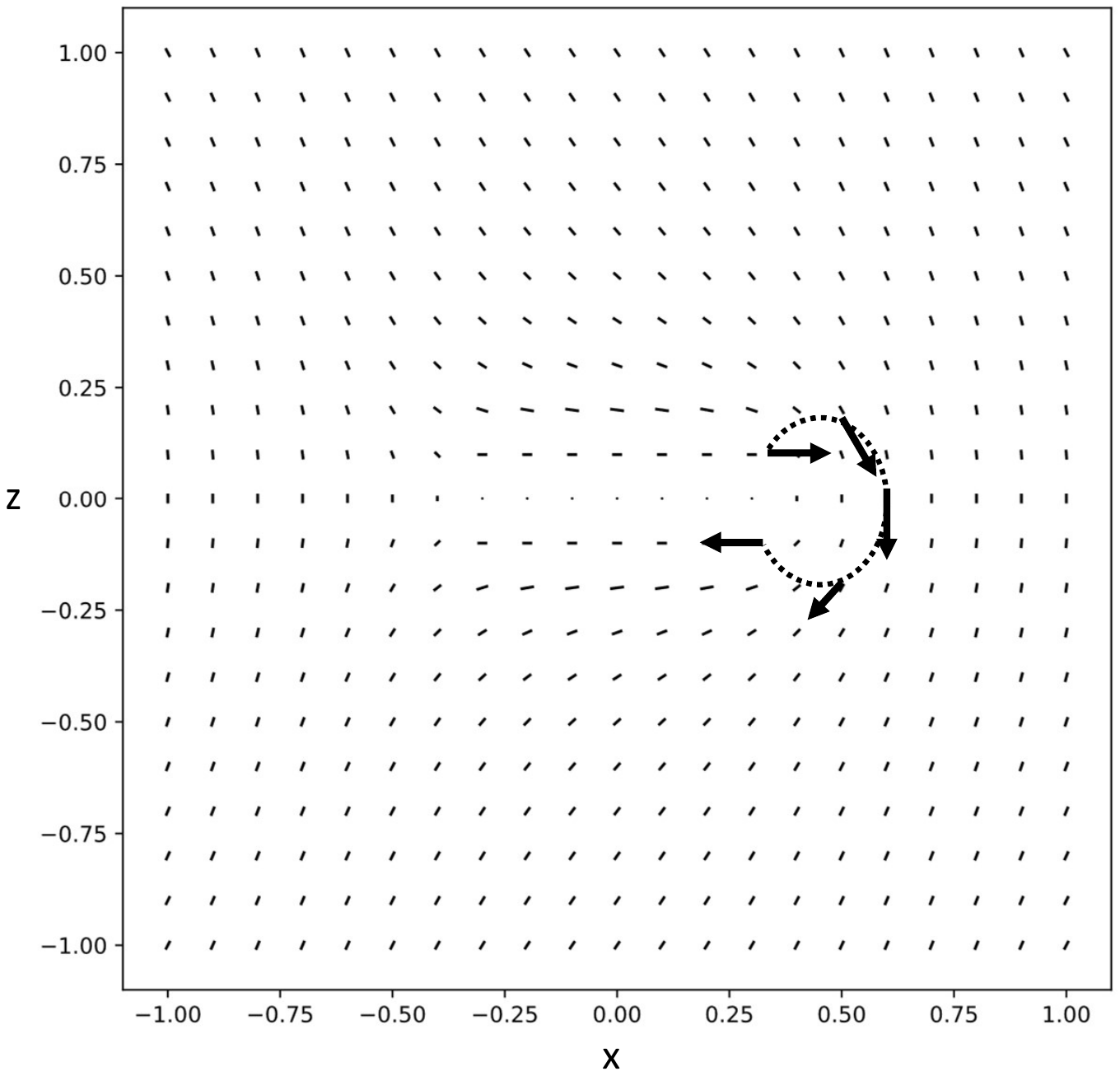}
    \label{fig:loop_res_k_xz}} \qquad
    \subfigure[Director constrained equilibrium on $y$-$z$ cross section.]{
    \includegraphics[trim={180 50 180 50}, clip, width=0.4\linewidth]{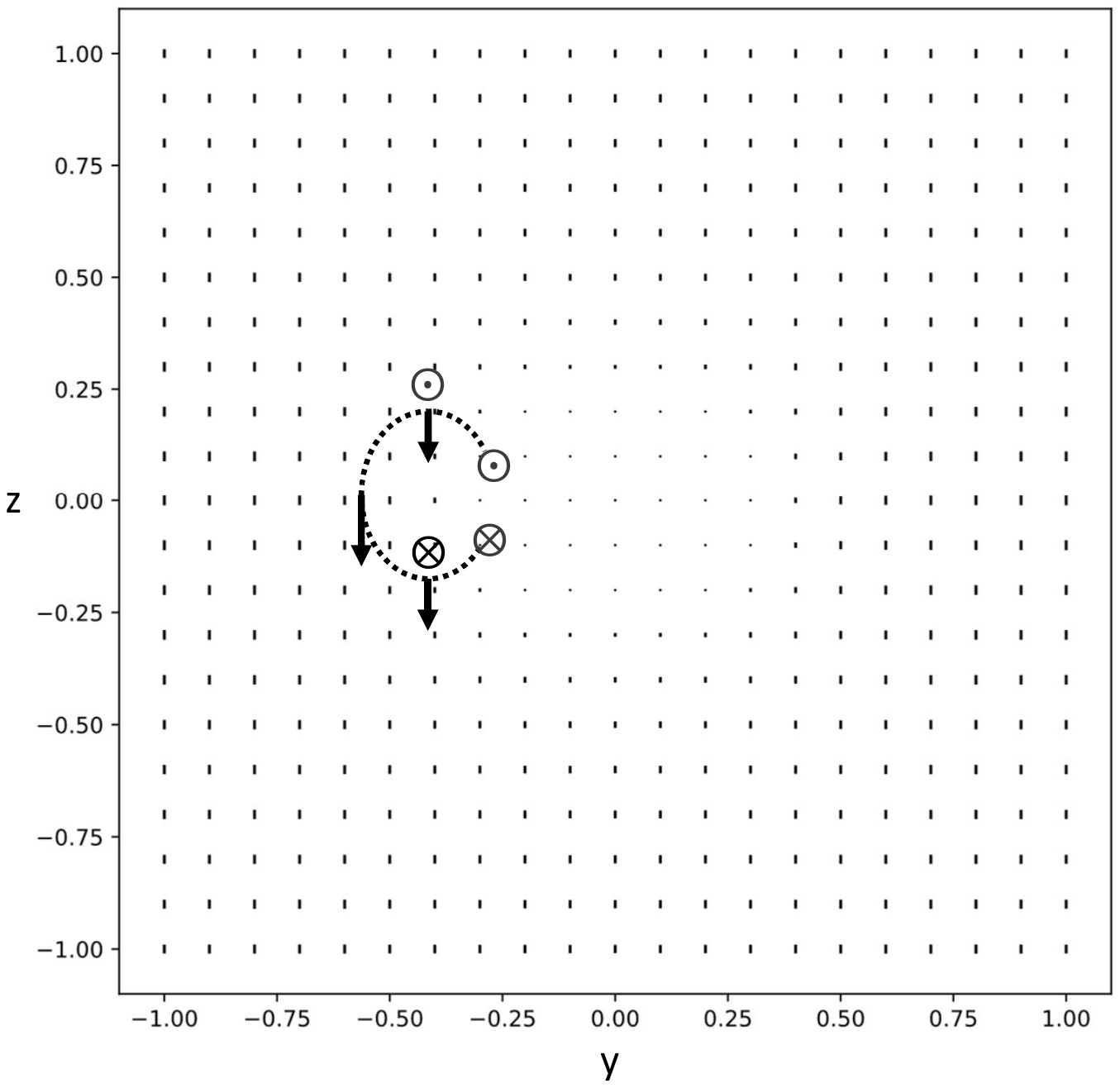}
    \label{fig:loop_res_k_yz}}
    \caption{Illustrations of director transition for square loop defect.}
\end{figure}

\begin{figure}
    \centering
    \subfigure[Energy densities at $y$-$z$ section.]{
    \includegraphics[width=0.4\linewidth]{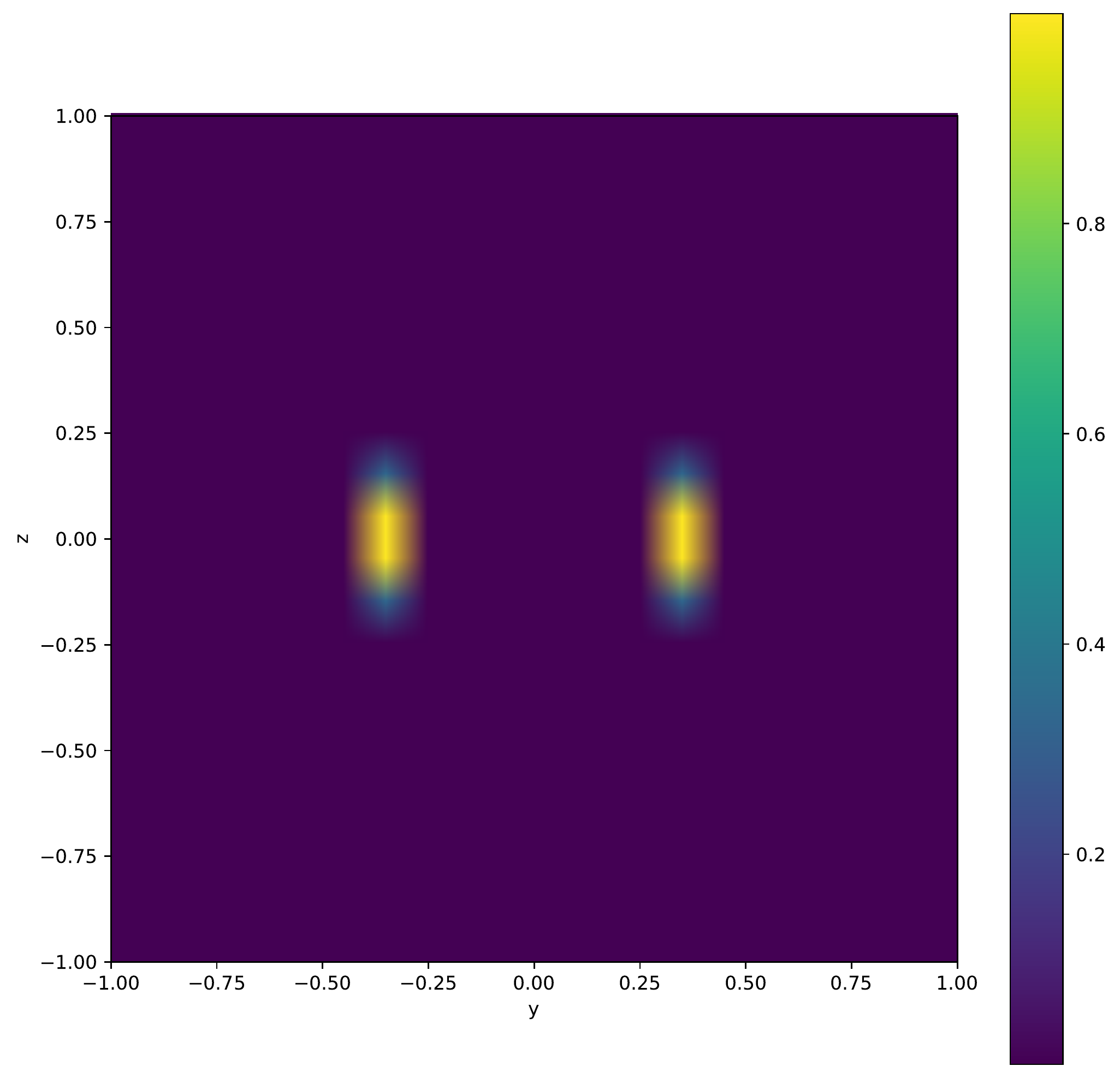}
    \label{fig:loop_energy_yz}}\qquad
    \subfigure[Energy densities at $x$-$z$ section.]{
    \includegraphics[width=0.4\linewidth]{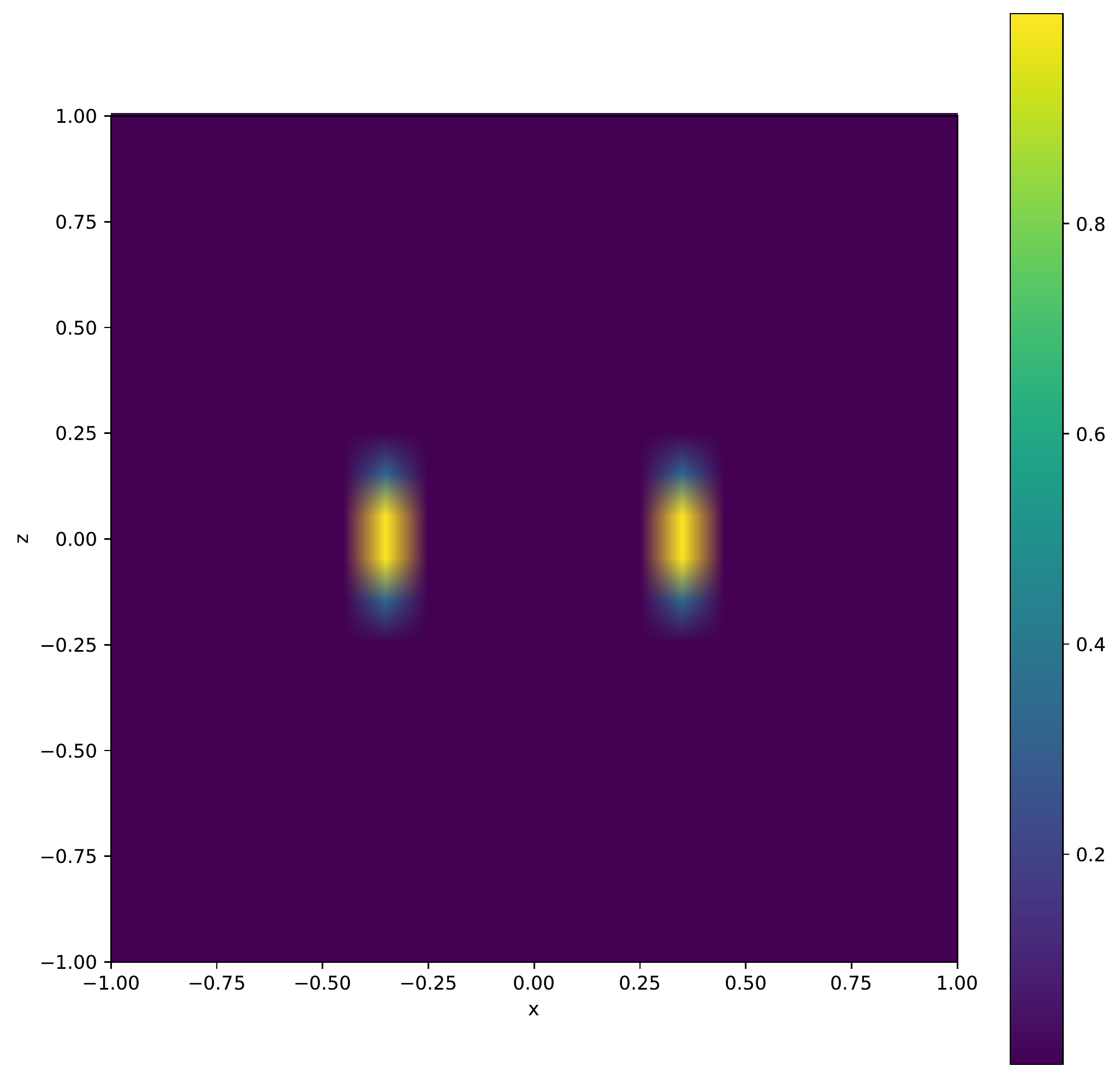}
    \label{fig:loop_energy_xz}}
    \caption{Energy densities of a square loop defect.}
    \label{fig:loop_energy}
\end{figure}

\begin{figure}
    \centering
    \subfigure[Norm of disclination density $\pi$ at $y$-$z$ section.]{
    \includegraphics[width=0.4\linewidth]{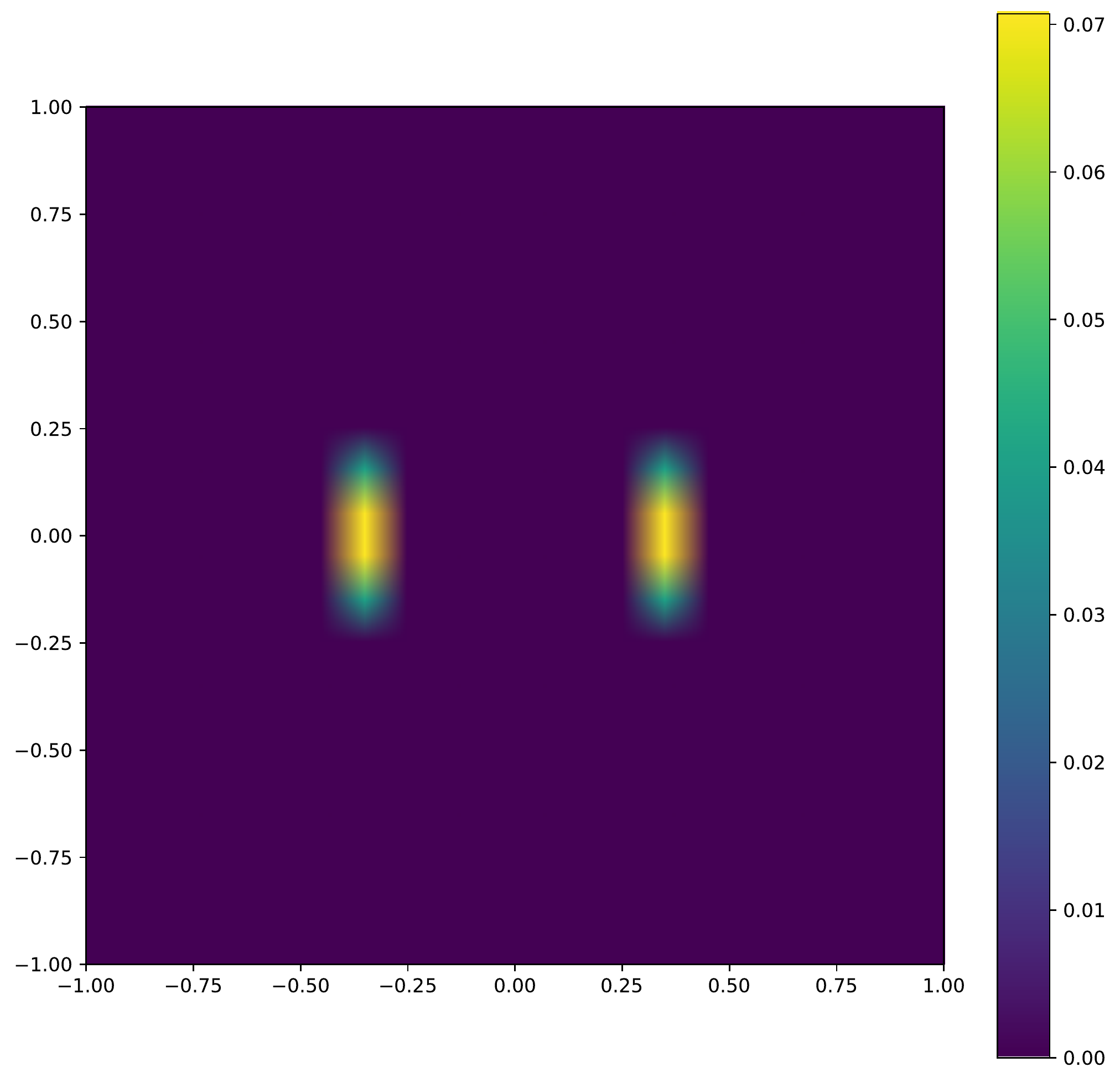}
    \label{fig:loop_disclination_density_yz}}\qquad
    \subfigure[Norm of disclination density $\pi$ at $x$-$z$ section.]{
    \includegraphics[width=0.4\linewidth]{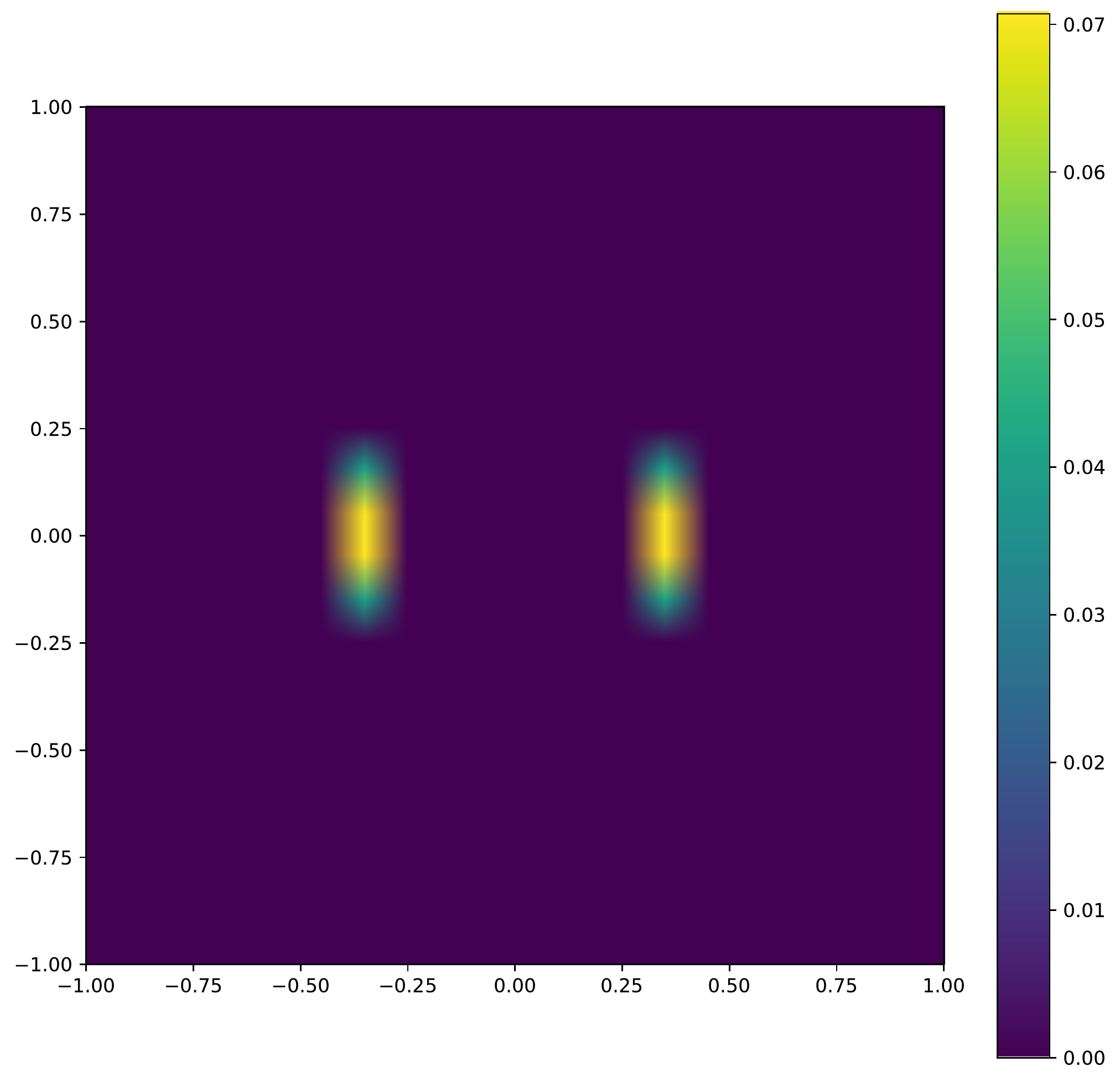}
    \label{fig:loop_disclination_density_xz}}
    \caption{Disclination densities of a square loop defect.}
    \label{fig:loop_disclination_density}
\end{figure}

Fig.~\ref{fig:loop_res_k} shows the constrained equilibrium of director $k$ nearby defect layer. Director colors represent the norm of projection on $x$ axis, where red means $k_1>0$ and blue means $k_1<0$. In this case, directors on the upper surface are $\bfe_1$ (red) while the ones on the bottom are $-\bfe_1$ (blue). Fig.~\ref{fig:loop_res_k_twist} shows an illustration of director constrained equilibrium on multiple $x$-$z$ cross sections. In Fig.~\ref{fig:loop_res_k}, we draw two circuits corresponding to ones in Fig.~\ref{fig:loop_res_k_xz} and in Fig.~\ref{fig:loop_res_k_yz} respectively. Fig.~\ref{fig:loop_res_k_xz} shows the director projection on $x$-$z$ section ($y=0$) and Fig.~\ref{fig:loop_res_k_yz} shows the director projection on $y$-$z$ section ($x=0$). To make visualization cleaner, we draw black solid arrows along the circuit in each figure, zooming in their director arrows at each point. If we follow a circuit from upper surface to bottom surface in $x$-$z$ plane, the directors transit from $\bfe_1$ to $-\bfe_1$ by varying director and size in plane. On the other hand, the director transition happens out of plane if following a circuit in $y$-$z$ plane. For example, in Fig.~\ref{fig:loop_res_k_yz}, black arrows represent director projections on $y$-$z$ cross section with $\odot$ meaning director pointing out and $\otimes$ meaning director pointing in. Fig.~\ref{fig:loop_energy} shows the energy densities at $y$-$z$ section ($x=0$) and at $x$-$z$ section ($y=0$), indicating that the energy is localized around core where $\pi \neq 0$. Fig.~\ref{fig:loop_disclination_density} shows the disclination density norms at $y$-$z$ section and $x$-$z$ section, which demonstrates that the disclination density is localized along defect core loop.

In the disclination loop being considered, the segments parallel to the $x$-axis in Fig. \ref{fig:loop_ini_pi} are of twist character, (with axis of rotation along the $y$-axis) and the segments parallel to the $y$-axis are of wedge character. The solution demonstrates that such a loop produces minimal far-field director distortion away from the loop (this is similar to what happens for the displacement field corresponding to dislocation loops in solids). Closer examination also reveals that the director field is planar - in the $x$-$z$ plane - almost everywhere except very close to the loop, with spatial variation in all three directions. It is instructive to compare this almost planar director field with the field of the loop computed in \cite{pourmat2015} which is of twist character everywhere and comprises a (strictly) planar director field with 3-d spatial variation. As shown there, such a loop with the same character everywhere induces significant far-field director variations, implying significantly more total energy content than the wedge-twist loop computed in this paper.

\subsubsection{Smectic boundary}

\begin{figure}
\centering
\subfigure[Configuration of $B$ with small defect dipole distance.]{
\includegraphics[width=0.4\linewidth]{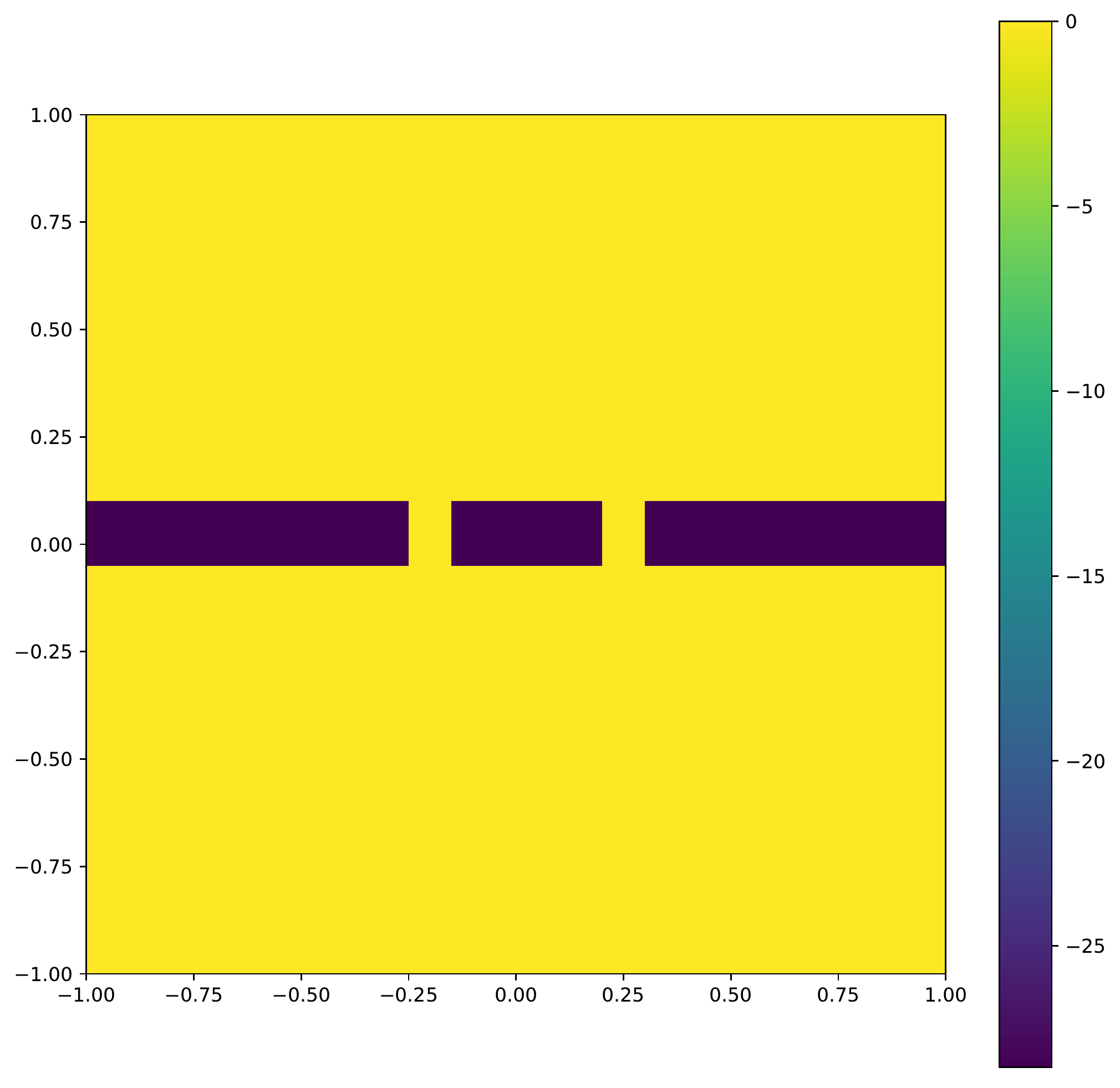}
\label{fig:smectic_B_small}} \qquad
\subfigure[Configuration of $k$ with small defect dipole distance.]{
\includegraphics[width=0.4\linewidth]{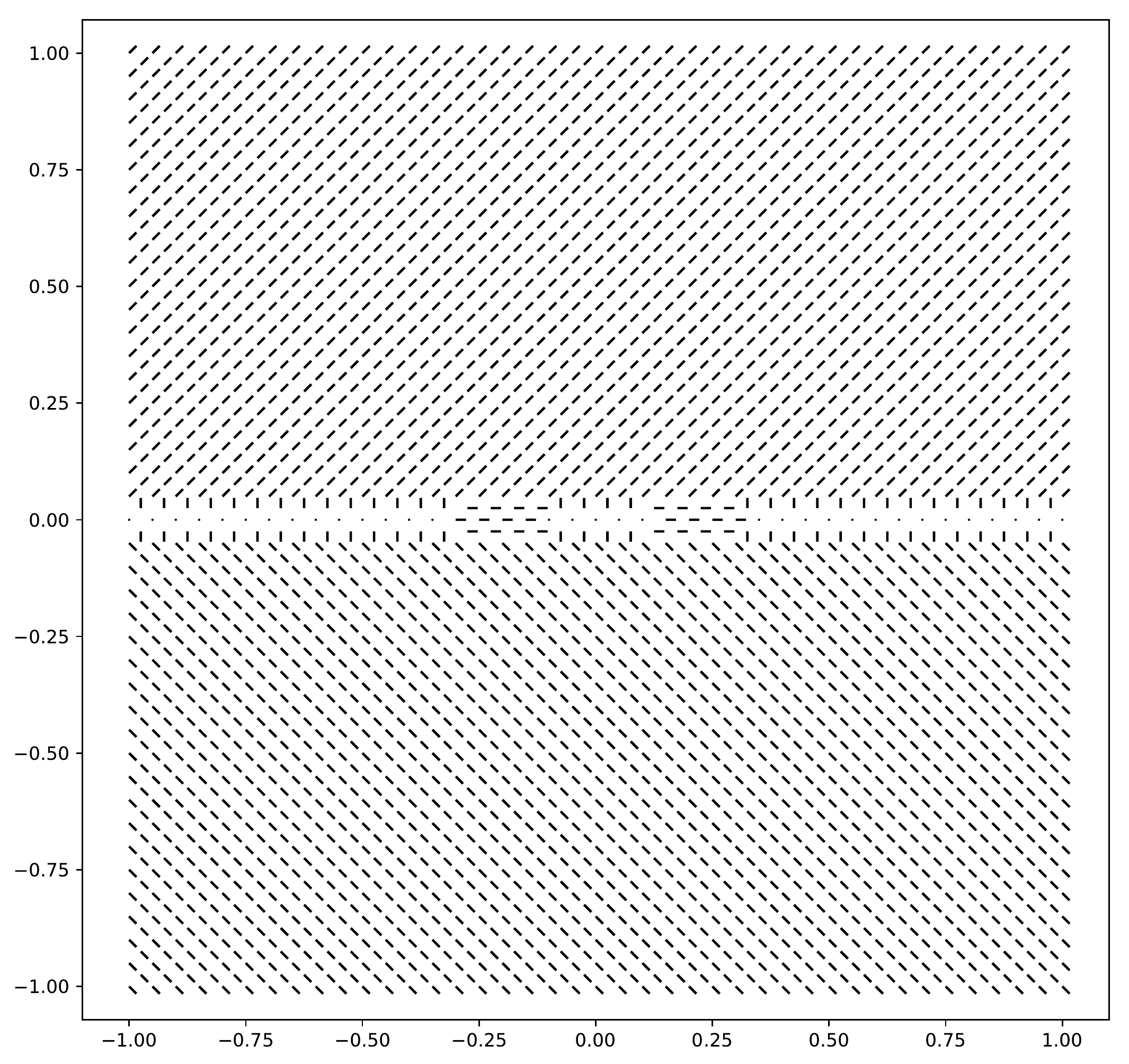}
\label{fig:smectic_k_small}} \qquad
\subfigure[Configuration of $B$ with mid defect dipole distance.]{
\includegraphics[width=0.4\linewidth]{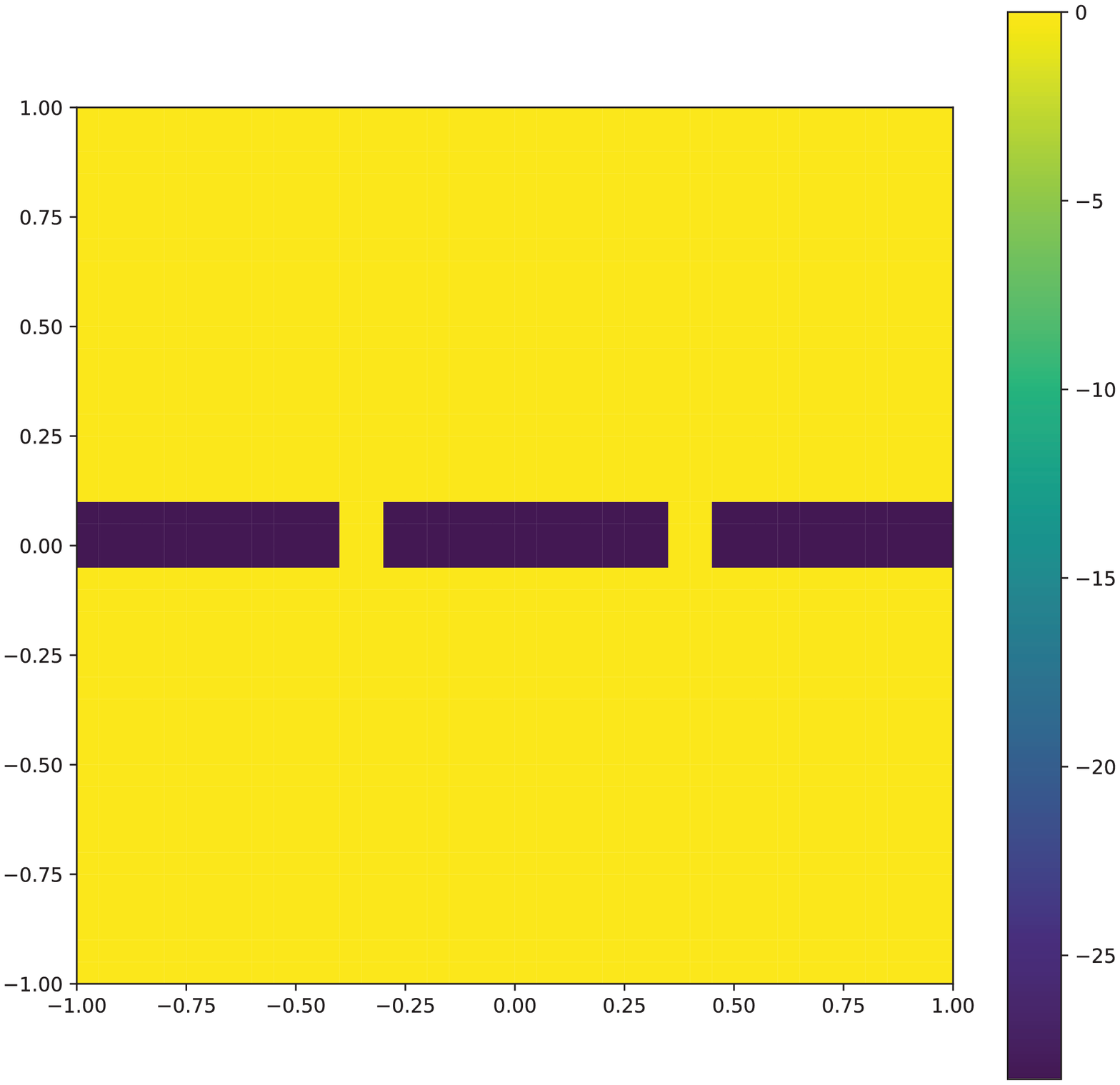}
\label{fig:smectic_B_mid}} \qquad
\subfigure[Configuration of $k$ with mid defect dipole distance.]{
\includegraphics[width=0.4\linewidth]{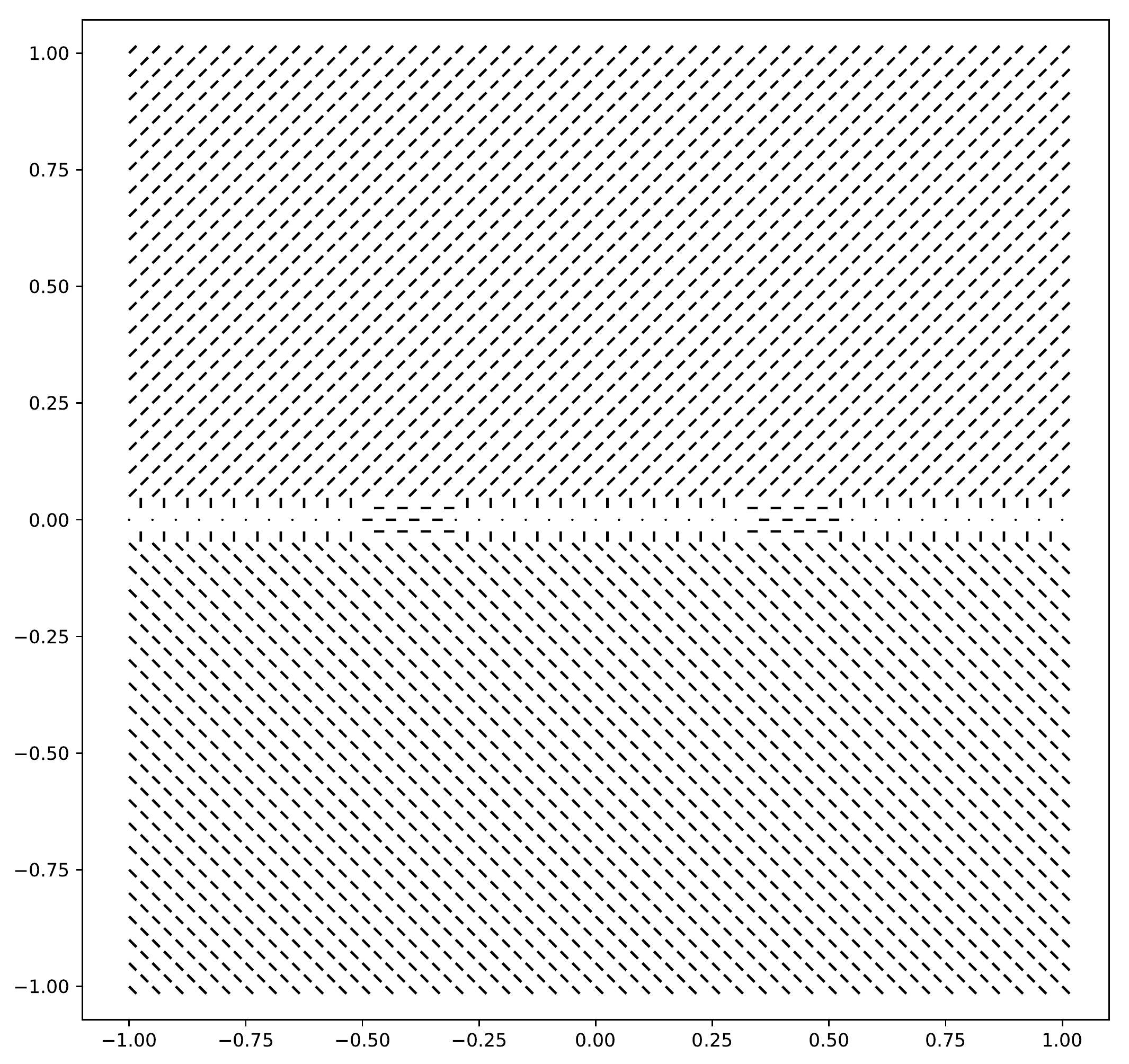}
\label{fig:smectic_k_mid}} \qquad
\subfigure[Configuration of $B$ with large defect dipole distance.]{
\includegraphics[width=0.4\linewidth]{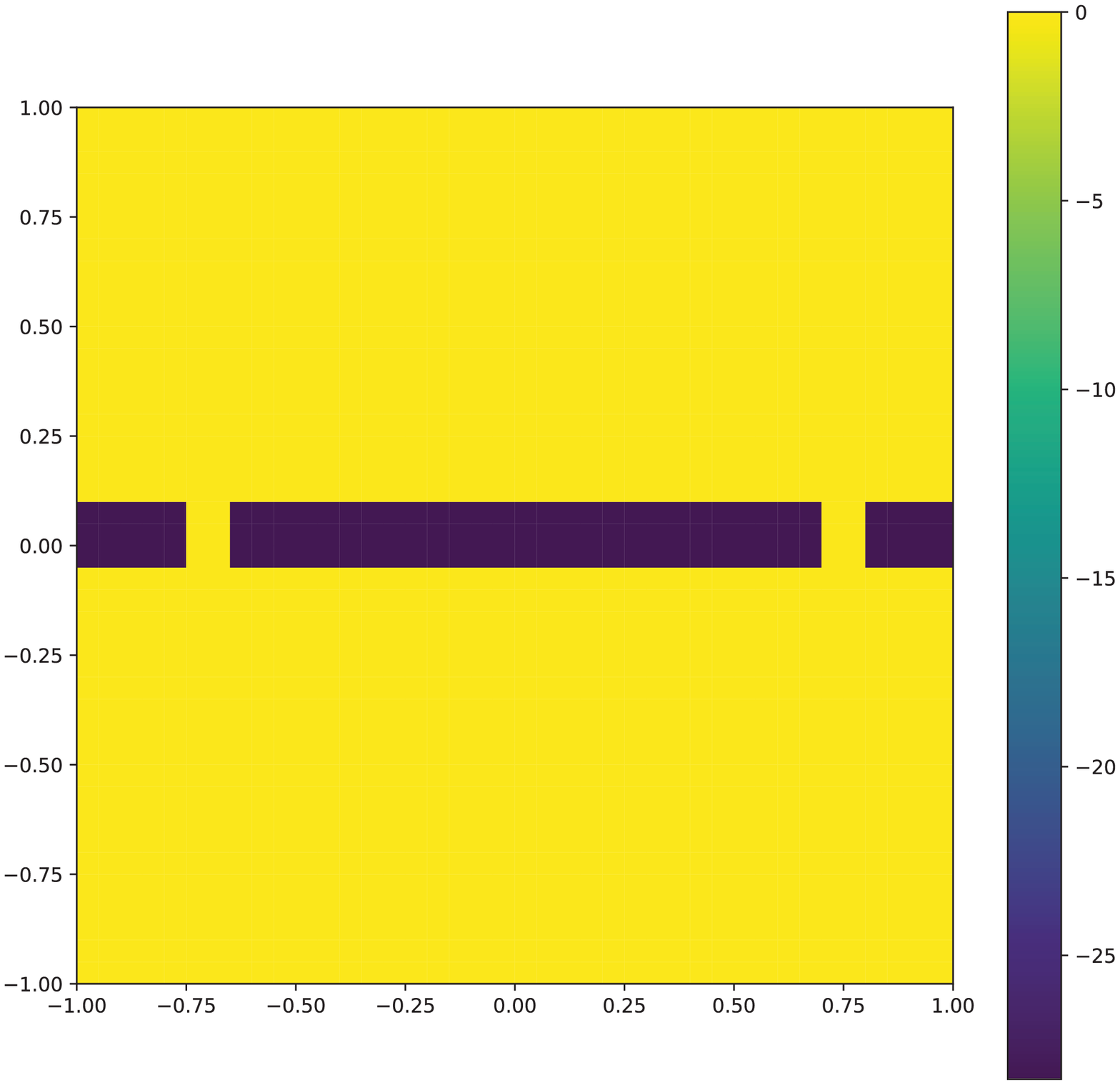}
\label{fig:smectic_B_large}} \qquad
\subfigure[Configuration of $k$ with large defect dipole distance.]{
\includegraphics[width=0.4\linewidth]{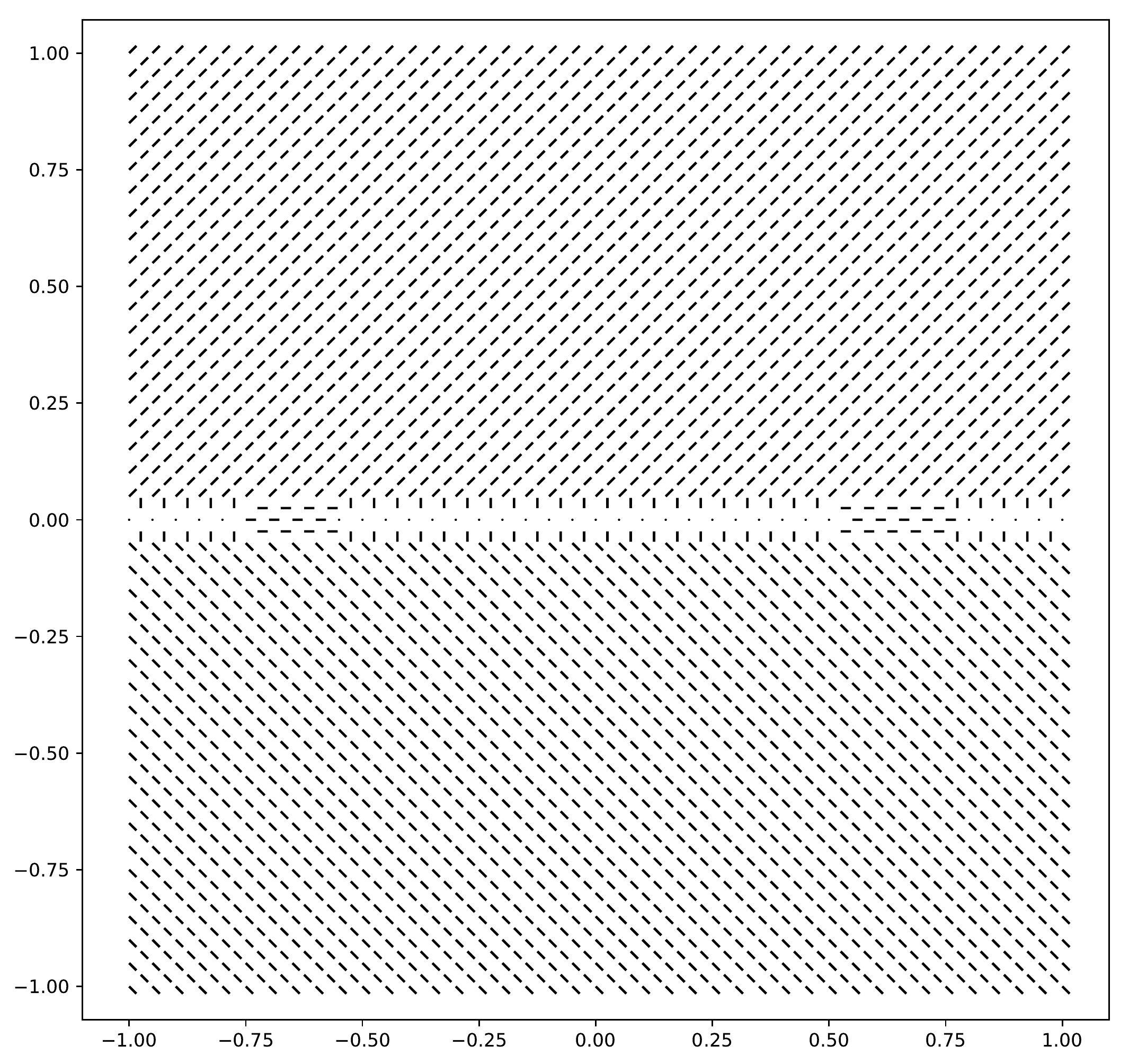}
\label{fig:smectic_k_large}} \qquad
\caption{Configurations of $B$ and $k$ representing smectic boundary with different defect dipole distances.}
\label{fig:smectic_B_k_init}
\end {figure}

\begin{figure}
\centering
\subfigure[Equilibrium of $k$ with small defect dipole distance.]{
\includegraphics[width=0.4\linewidth]{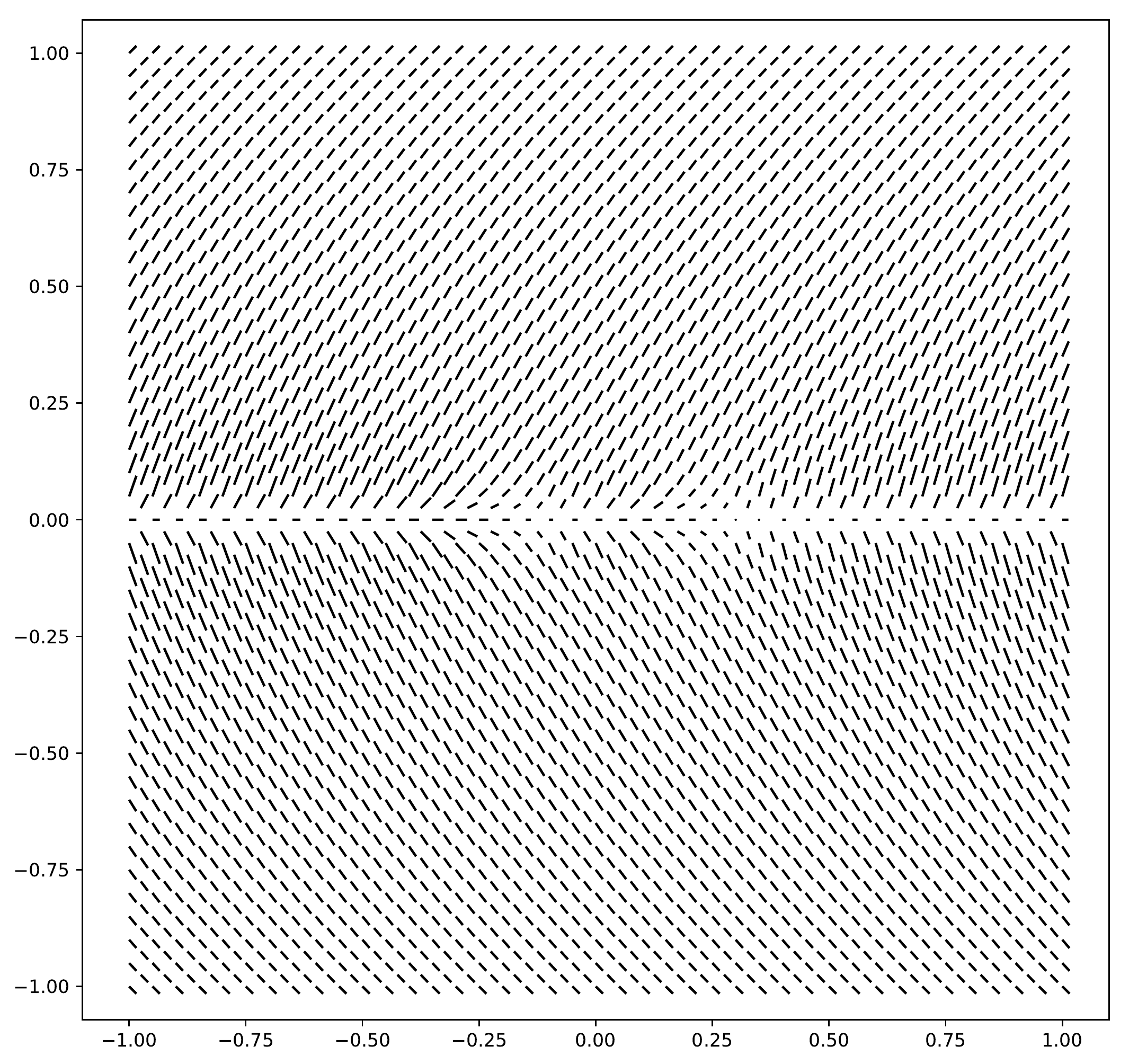}
\label{fig:smectic_k_small_res}} \qquad
\subfigure[Equilibrium of $k$ with mid defect dipole distance.]{
\includegraphics[width=0.4\linewidth]{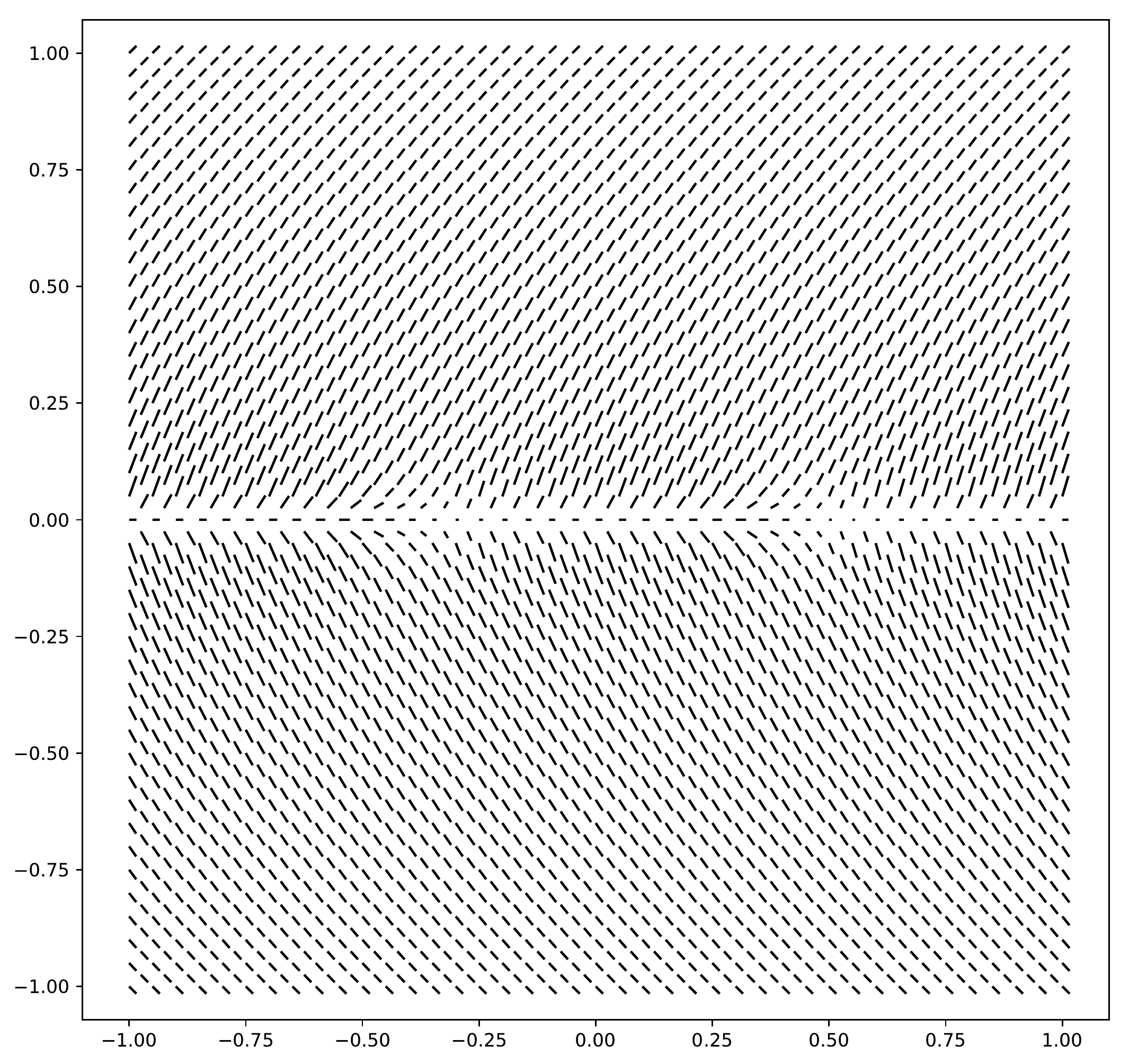}
\label{fig:smectic_k_mid_res}} \qquad
\subfigure[Equilibrium of $k$ with large defect dipole distance.]{
\includegraphics[width=0.4\linewidth]{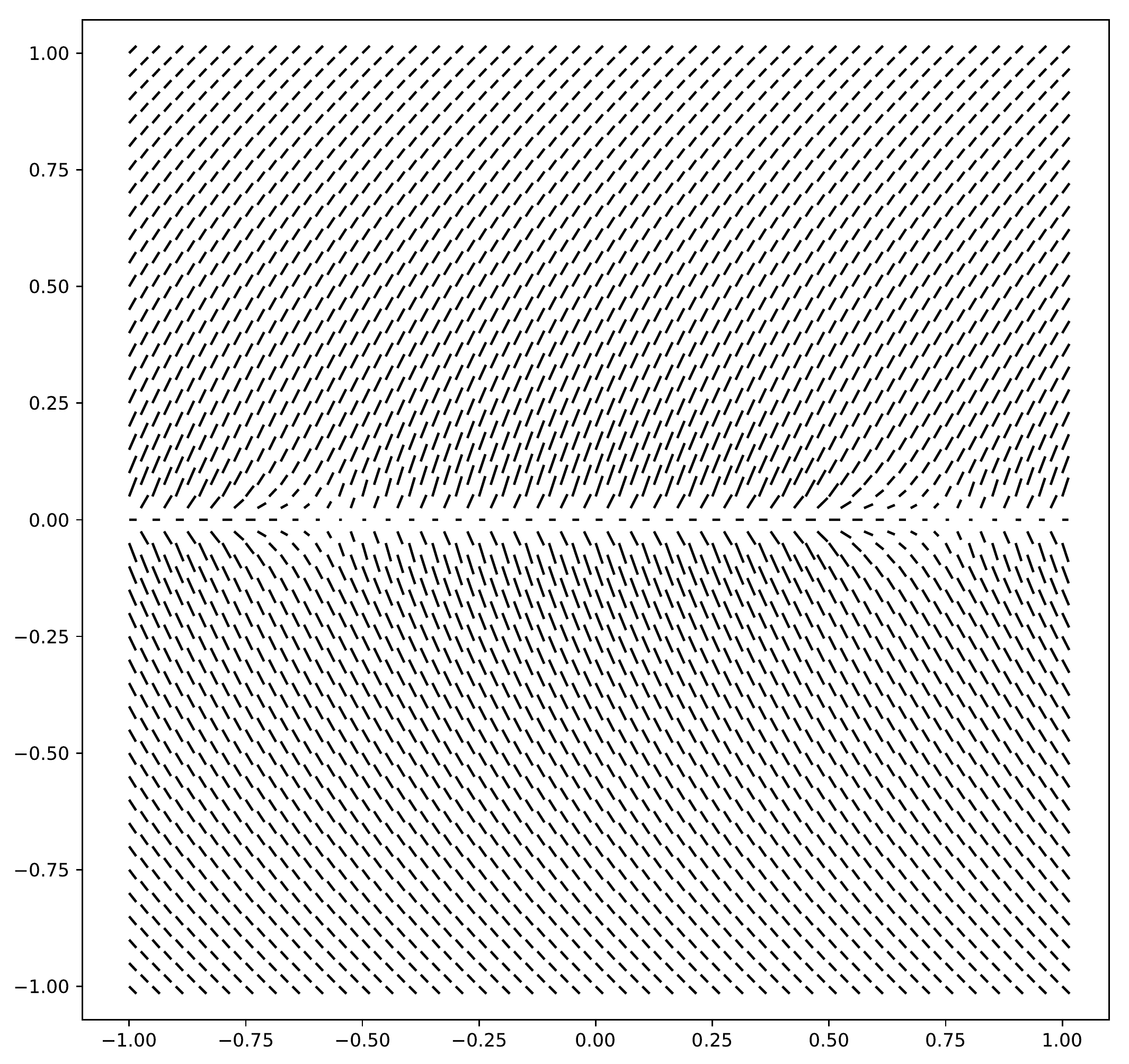}
\label{fig:smectic_k_large_res}} \qquad
\caption{Equilibria of $k$ representing smectic boundary with different defect dipole distances.}
\label{fig:smectic_k}
\end {figure}

A grain boundary in a smectic is a special `canonical' pattern of layered material systems that arises as a first bifurcation from a homogeneous state under external forcing, giving rise to piecewise homogeneously oriented domains separated by the boundary. The boundary can further reduce its energy by inducing defects within it. The existence of layers in a smectic implies that deviations of the $curl \, k$ field from $0$ are strongly energetically penalized. Thus, the parameters used in this calculation are $P_1=1$, $P_2=1$, $\alpha=50$, $K^*=5$, and $\epsilon=0.1$. Smectic boundaries can be modeled as a series of point defect pairs. To compare energies for different defect configurations, we model the smectic boundary as two defect dipoles at various distance, as illustrated in Fig \ref{fig:smectic_B_k_init}. For each configuration, $B$ is initialized following the procedure in Sec \ref{sec:half_defects}, given as (\ref{eqn:B_boundary}): 

\begin{equation}\label{eqn:B_boundary}
B = \begin{cases}
\frac{2}{a \xi}\bfe_1 \otimes \bfe_2, & \text{ if $|y|<{\frac{a\xi}{2}}$  and $x <x_{d1}$} \\
\frac{2}{a \xi}\bfe_1 \otimes \bfe_2, & \text{ if $|y|<{\frac{a\xi}{2}}$  and $x_{d2} <x < x_{d3}$} \\
\frac{2}{a \xi}\bfe_1 \otimes \bfe_2, & \text{ if $|y|<{\frac{a\xi}{2}}$  and $x > x_{d4}$} \\
0, & otherwise,
\end{cases}
\end{equation}
where $x_{d1}$, $x_{d2}$, $x_{d3}$, and $x_{d4}$ represent $x$ coordinates of four defect cores from left to right. Both $B$ and $k$ evolve following \eqref{eqn:grad_dimen}. Fig \ref{fig:smectic_k} shows equilibria of $k$ corresponding to different defect dipole configurations. 

\begin{figure}
\centering
\subfigure[Equilibrium smectic grain boundary without defect.]{
\includegraphics[width=0.4\linewidth]{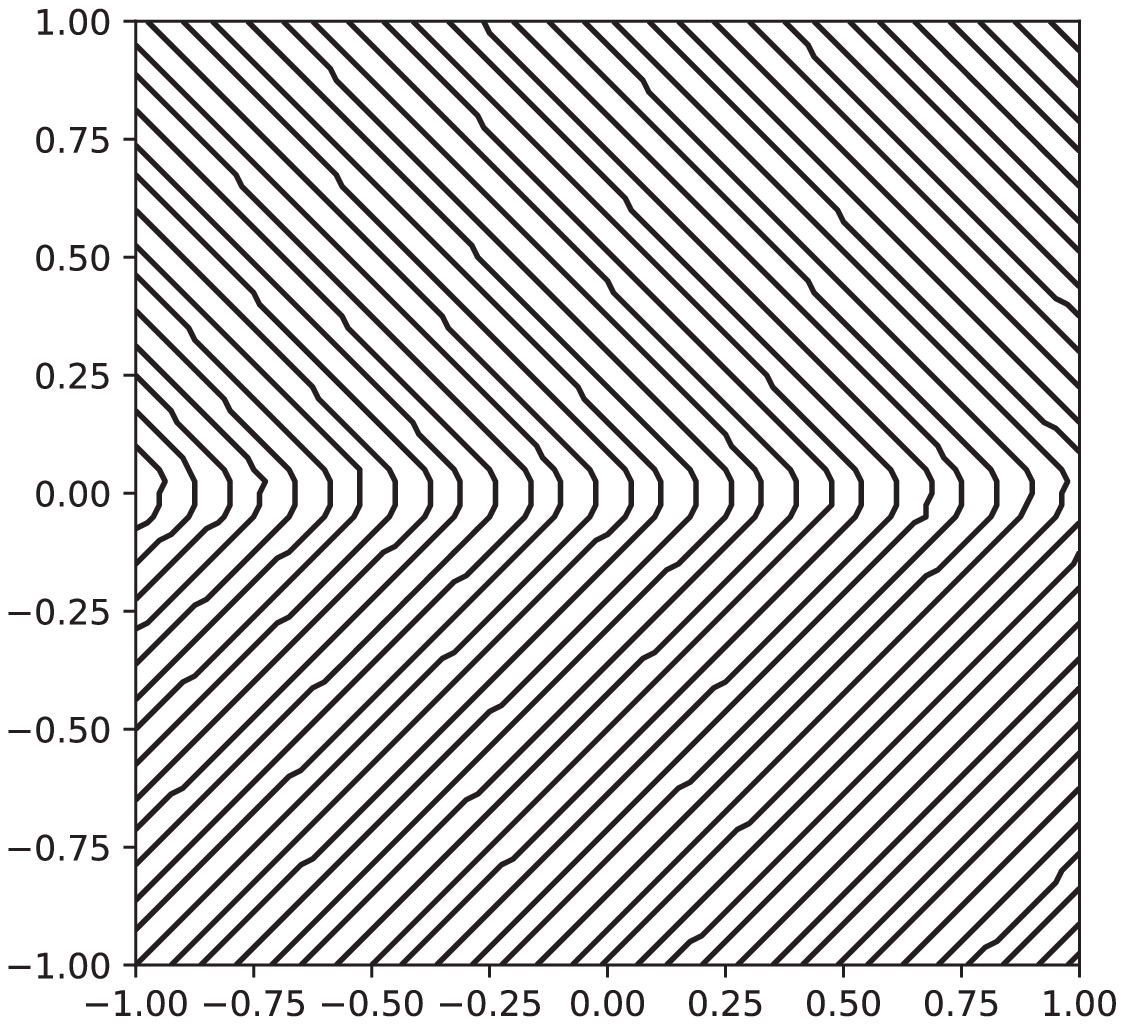}
\label{fig:smectic_layer_wo}} \qquad
\subfigure[Equilibrium smectic layers with small defect dipole distance.]{
\includegraphics[width=0.4\linewidth]{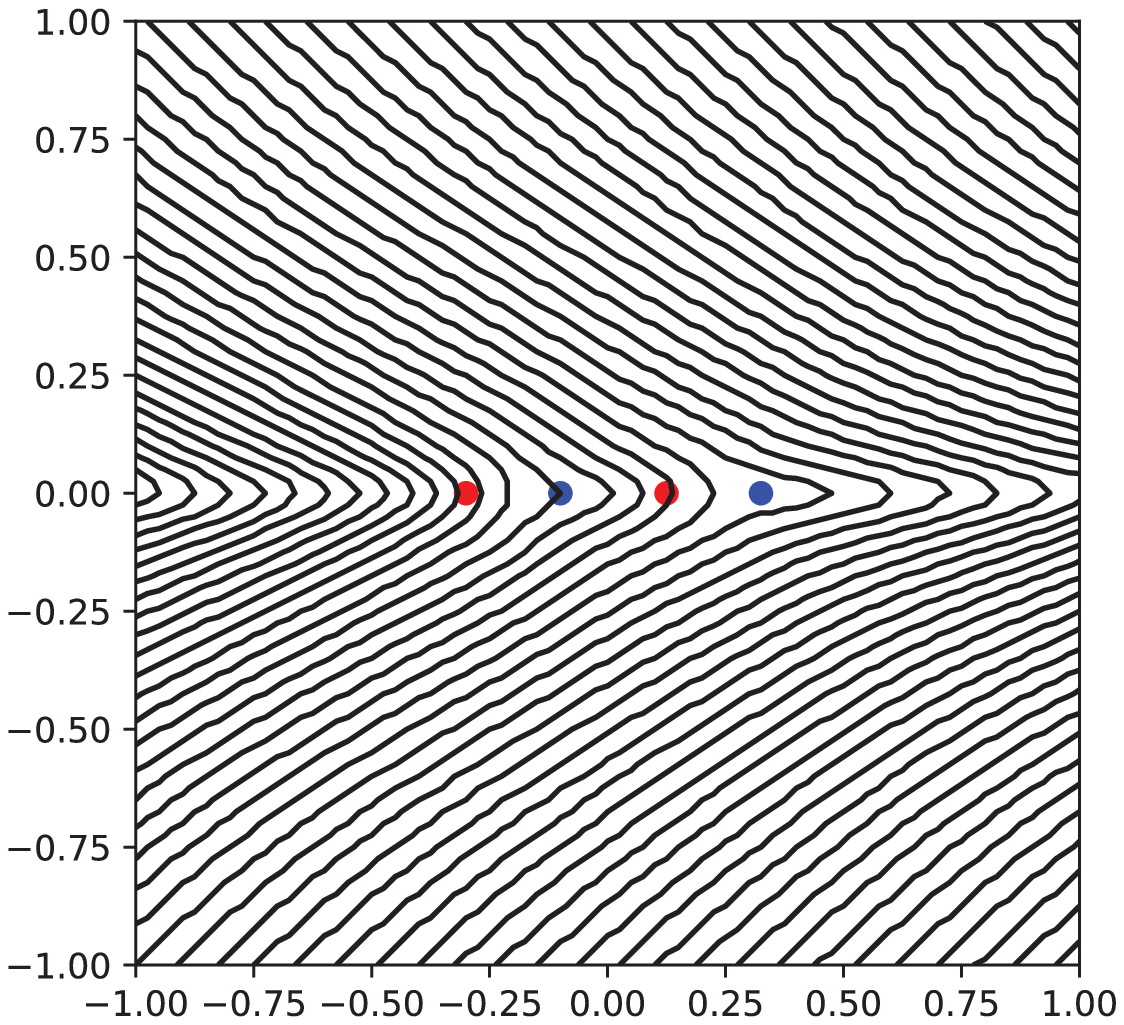}
\label{fig:smectic_layer_small}} \qquad
\subfigure[Equilibrium smectic layers with mid defect dipole distance.]{
\includegraphics[width=0.4\linewidth]{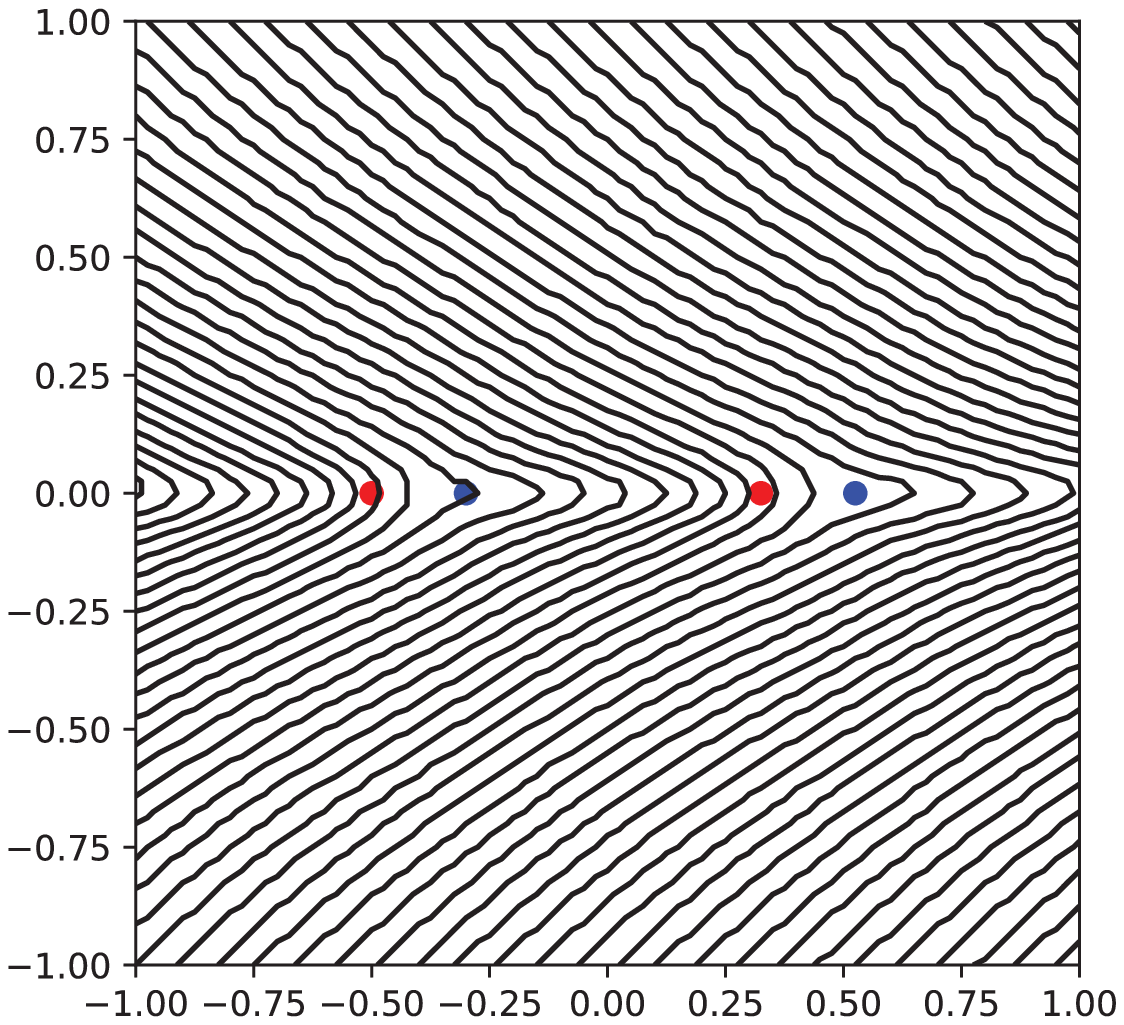}
\label{fig:smectic_layer_mid}} \qquad
\subfigure[Equilibrium smectic layers with large defect dipole distance.]{
\includegraphics[width=0.4\linewidth]{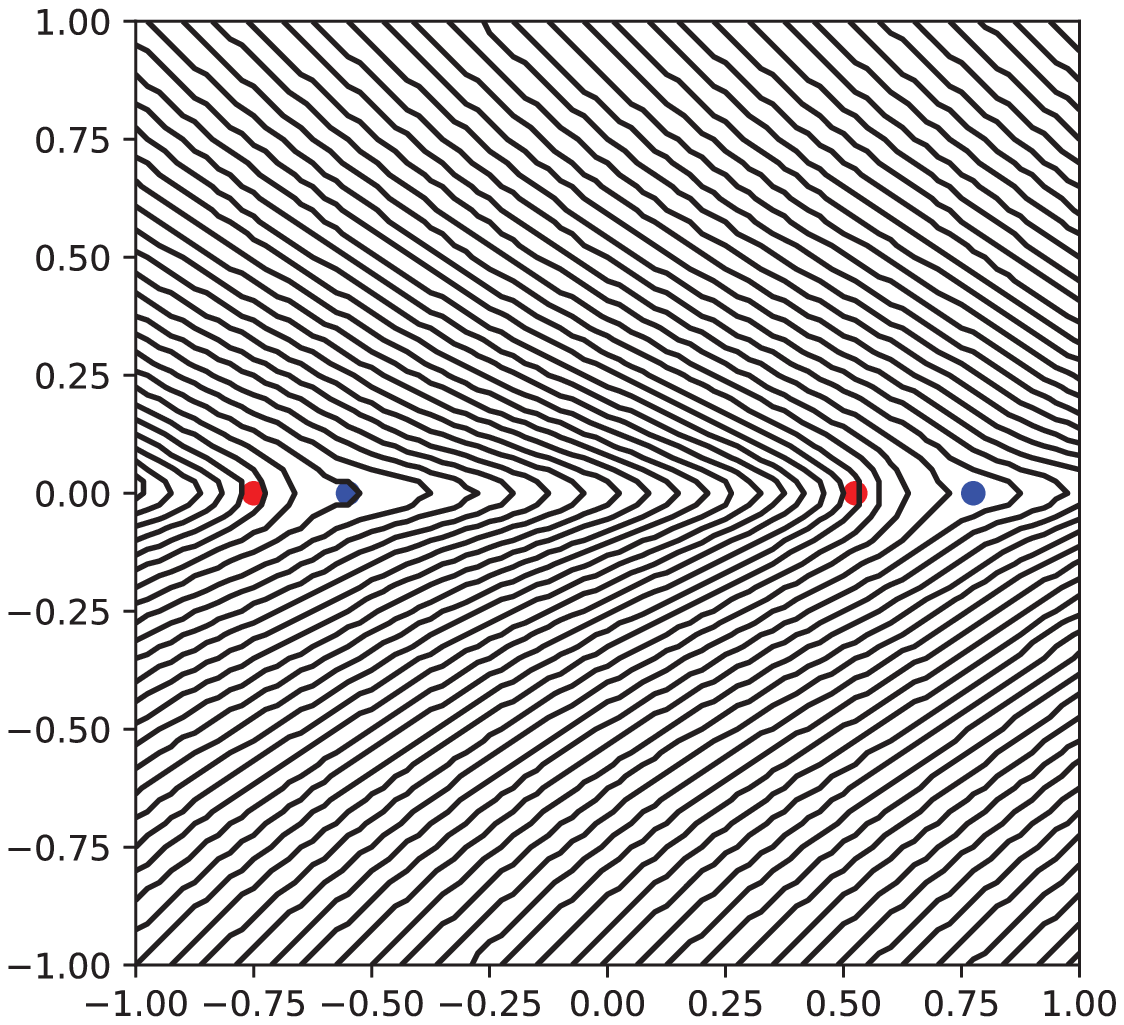}
\label{fig:smectic_layer_large}} \qquad
\caption{Equilibria of smectic layers with different defect dipole distances and without defect. Along the grain boundary, the red dots represent $+ \half$ disclination cores and the blue dots represent $- \half$ disclination cores. Each $\pm \half$ dipole may be considered a dislocation.}
\label{fig:smectic_layer}
\end {figure}

Since layers are of more interest in understanding smectic boundaries, we calculate the phase field $\theta$ from $k$ from the  following equations,
\begin{equation} \label{eqn:phase_theta}
div\, D\theta = div\, k \qquad \mbox{on} \ B
\end{equation}
with Dirichlet boundary condition $D \theta \cdot t = k \cdot t$ on $\partial B$, where $t$ is the unit tangent field on $\partial B$ (i.e., a Helmholtz decomposition of $k$). The contour plots of the phase field $\theta$ of grain boundaries without and with defects are provided in Fig. \ref{fig:smectic_layer}. In Fig. \ref{fig:smectic_layer}, black lines represent phase field layers, red dots represent $+ \half$ disclination cores, and blue dots represent $-\half$ disclination cores.

\begin{figure}
\centering
\subfigure[Prescription of $B$. $B$ is nonzero across the layer.]{
\includegraphics[width=0.4\linewidth]{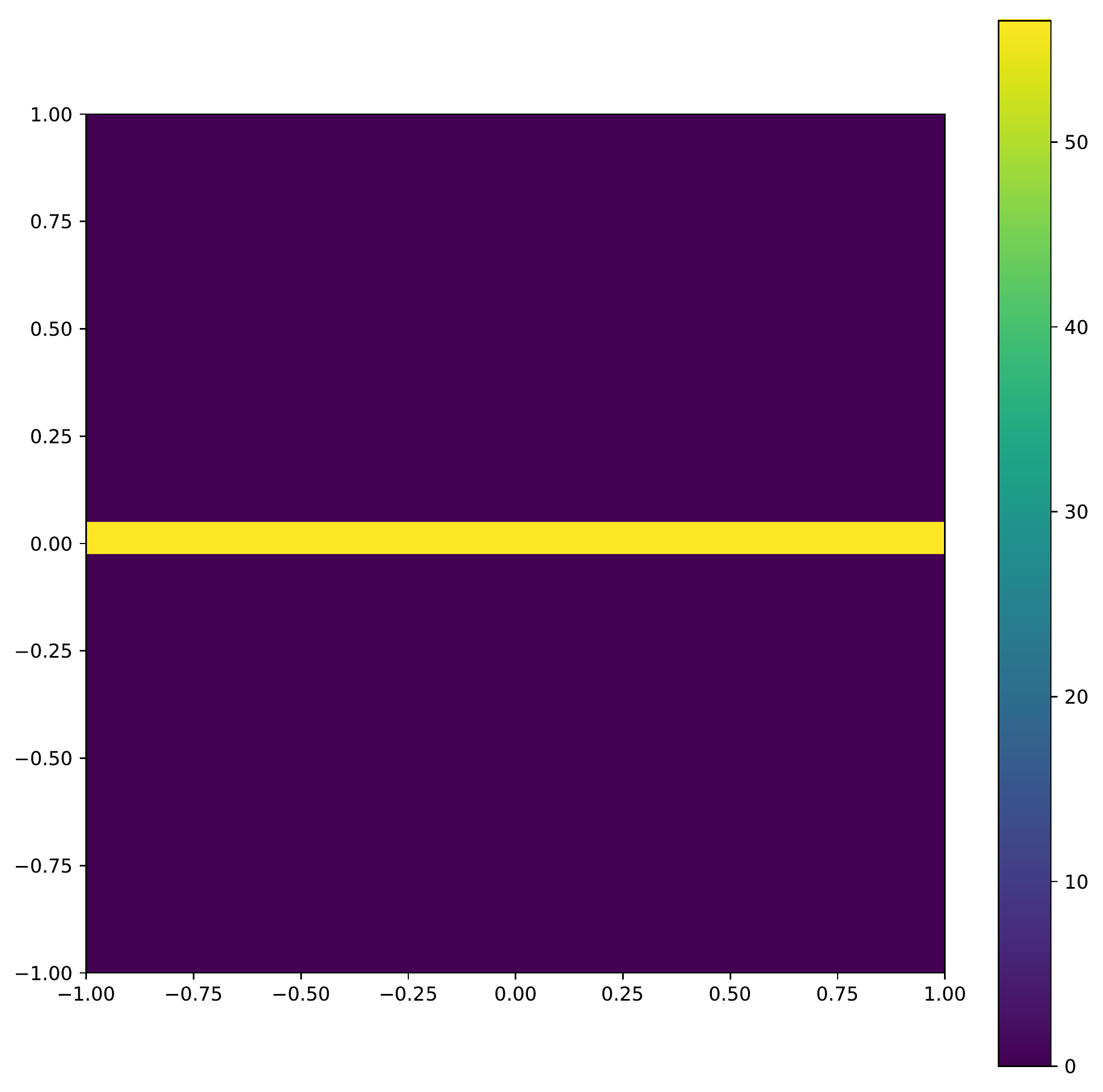}
\label{fig:smectic_layer_defect_B}} \qquad
\subfigure[Equilibrium of smectic layers with through nonzero layer field.]{
\includegraphics[width=0.4\linewidth]{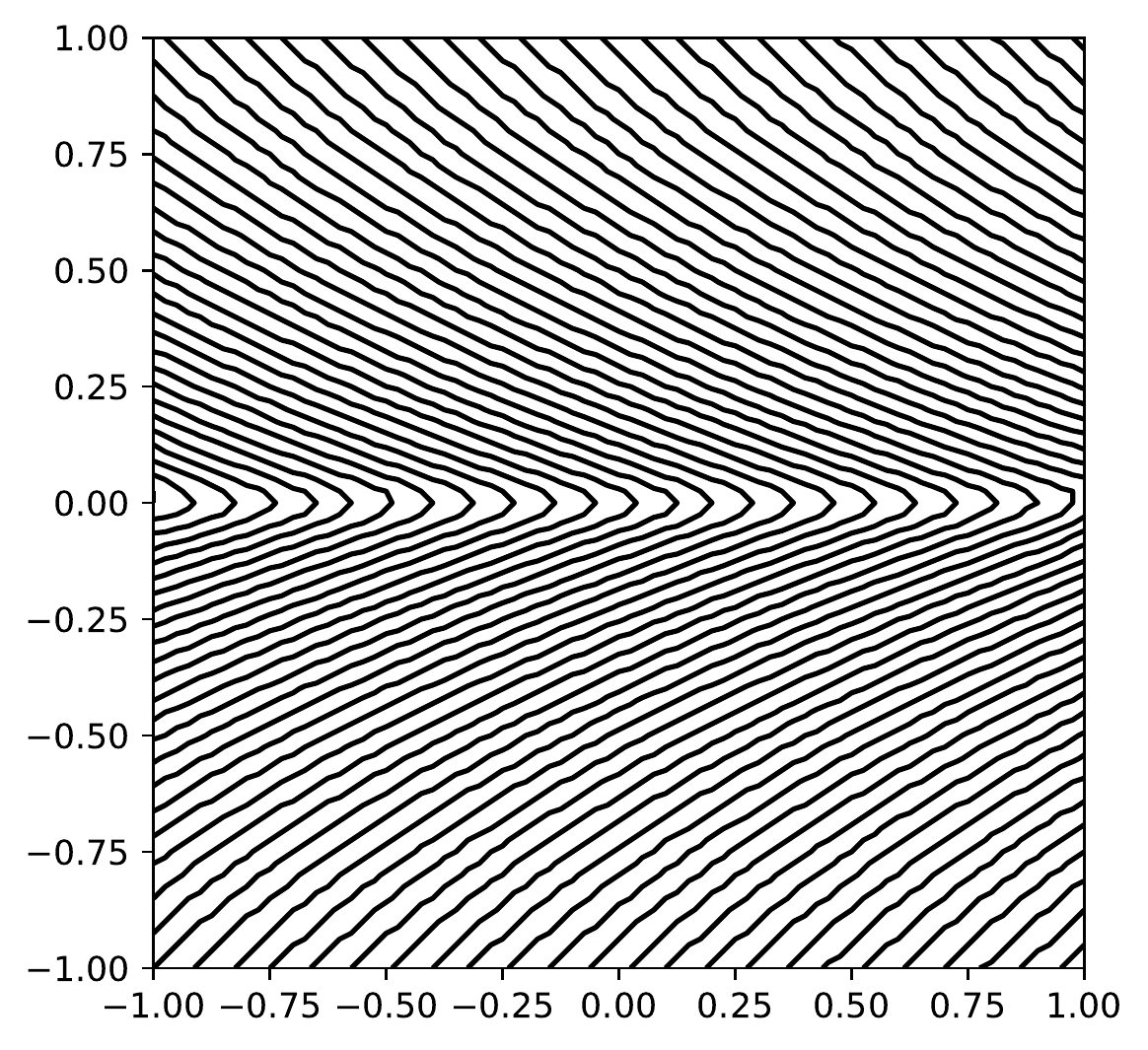}
\label{fig:smectic_layer_defect_stream}} \qquad
\caption{Prescription of $B$ and static equilibrium of smectic layers for a through defect layer field without defect dipoles.}
\label{fig:smectic_layer_defect}
\end {figure}

In this example, the total energy for the defect-free smectic boundary is $8.596\times 10^4$. The non-dimensionalized total energies for the defected boundary configurations with small, medium, and large inter-dipole separations are $0.892$, $0.891$, and $0.881$ respectively, normalized by the total energy of the defect-free boundary configuration. In addition, we consider another smectic boundary representation where defect dipoles repel and finally move out of body leaving a through layer field across the body. Figure \ref{fig:smectic_layer_defect_B} shows the corresponding prescription of $B$. Instead of defect dipoles, $B$ is nonzero within the entire layer. Figure \ref{fig:smectic_layer_defect_stream} shows its static equilibrium of smectic layers. The normalized non-dimensionalized total energy for this static equilibrium is $2.2 \times 10^{-3}$.

Thus, larger inter-dipole separations are energetically favorable and it indicates that defected configurations are more energetically stable in the considered example.

\section{Defects in a scalar field and analysis of natural stripe patterns}\label{sec:scalar}

We now turn to defects in natural stripe patterns, and use them to illustrate all the key points of our framework. To recapitulate, they are:
\begin{enumerate}
    \item Defects in ordered media are robust and can be understood as singularities, often with an associated `topological index', in the context of a coarse-grained macroscopic theory.
    \item While coarse-grained theories might predict the appropriate defects, they do not, in general, give the correct energetics that  drives the interactions and the dynamics of the defects.
    \item The energetics are properly formulated in an `enriched' theory by incorporating some of the microscopic physics through additional fields that effectively resolve defects on a fine scale.
\end{enumerate}
In addition, we also use this system as an example to argue that  
the particular details, of which microscopic aspects are `upscaled', are crucial for the resulting enriched theory to accurately capture defects and their dynamics.

A microscopic model for the evolution of convection patterns in large Prandtl number fluids is the Swift-Hohenberg equation \cite{swift1977hydrodynamic},
\begin{equation}
    u_t = R u - (\Delta + k_0^2)^2 u - u^3,
    \label{eq:sh}
\end{equation}
where $R > 0$ is the forcing parameter (analog of the Rayleigh number for convection), $u$ is a proxy for the vertically averaged temperature and $k_0$ is the preferred wavenumber of the roll patterns. 

Although convection patterns are two dimensional, the Swift-Hohenberg equation is defined in an arbitrary number of dimensions. For our purposes, we consider domains $\Omega \subset \mathbb{R}^d$, $d = 2,3$, and let $x$ denote an arbitrary point in $\Omega$. The
$L^2$ gradient flow $u_t = -\frac{\delta}{\delta u} E$ for the SH energy 
\begin{equation}
E[u(x,t)] = \int _{\Omega}\frac{|(\Delta + k_0^2) u|^2}{2} + \left(\frac{u^4}{4} - \frac{R u^2}{2}\right) \, dx.
\label{eq:sh_energy}
\end{equation}
is the Swift-Hohenberg equation~\eqref{eq:sh} with natural boundary conditions. $E[u(x,t)]$ is monotonically non-increasing along solutions. The asymptotic states are thus critical points for the non-convex energy~\eqref{eq:sh_energy}. Ignoring the boundaries,  the ground states for this energy 
are periodic stripe patterns with a range of allowed wave numbers in the vicinity of $k_0$ and all possible orientations \cite{ColletEckmannBook}. These periodic ground states are given by $u(x,t) = w_0(q\cdot x + \theta_0,|q|^2)$ where $w_0$ is $2 \pi$ periodic in its first argument, and is ``normalized" so that it's maxima (resp. minima) occur at $q \cdot x + \theta_0 = 2 n \pi$ (resp. $(2n+1) \pi$). $w_0$ describes the profiles of stable stripe patterns with a constant wave-vector $q$.

We view the Swift-Hohenberg equation as a `microscopic' model for stripe patterns since it resolves solutions on the pattern wavelength $k_0^{-1}$, which we will take as our microscale. There are, of course, smaller scales in the problem, and we can view the roll patterns as collective behavior of these even smaller units. For our purposes, this is not directly relevant and $k_0^{-1}$ is the smallest length scale of interest. A macroscopic, coarse-grained theory for stripes was obtained by Cross and Newell \cite{cross1984convection} who showed that, away from defects, natural stripe patterns are modulations of stable periodic profiles $w_0$. These modulated pattern states are given by
$u(x,t) = w_0(\theta(x,t),|k|^2)$, where $k = D \theta$ varies slowly in space and time. 

Away from defects, $\|D k\| \simeq O(\epsilon) \ll 1$ where $\epsilon$, the ratio of the pattern wavelength to the size of the system, is the appropriate small parameter that allows for the coarse-graining. The wave vector $k$ is the appropriate macroscopic (i.e. slowly varying) order parameter, and a given $k$ is consistent with distinct ``microstates" corresponding to different choices of $\theta_0$, the constant of integration needed to recover the phase from the equation $D \theta = k$. This continuous (macroscopic) translation invariance $\theta(x,y) \to \theta(x,y) + \theta_0$ implies that the linearization of the dynamics in~\eqref{eq:sh} about a modulated stripe pattern has a non-trivial kernel, and the corresponding solvability condition yield, at lowest order, the {\em (unregularized) Cross-Newell equations} \cite{cross1984convection}
\begin{equation}
    k_i = \partial_i \theta,  \qquad \langle ( w_0')^2 \rangle \theta_t = - \partial_i\left[ k_i B(k^2)\right], \qquad B(k^2) = \frac{1}{2} \frac{d}{dk^2} \langle w_0^4 \rangle,  
    \label{eq:cn}
\end{equation}
where the angle brackets $\langle \cdot \rangle$ denote averaging over one period of the stable periodic profiles $w_0(\cdot,k^2)$. The unregularized Cross-Newell equation, Eq.~\eqref{eq:cn}, is a gradient flow that describes the macroscopic dynamics for the phase $\theta(x,t)$ and this also the wave-vector field $k(x,t)$. These equations lead to the formation of shocks, so they need to be regularized by higher order effects in the small parameter $\epsilon$ \cite{newell1996defects,ercolani2000geometry}. This is, as we discussed above, an enrichment of a macroscpic theory by effects that have a microscopic origin. An alternative to employing the Fredholm alternative/solvability is to directly compute an effective energy $\mathcal{E}[k(x,t)]$  
by averaging the energy~\eqref{eq:sh_energy} over all the microstates that are consistent with a given macroscopic field $k(x,t)$ \cite{newell2017elastic}. This is equivalent to averaging over the phase shift $\theta_0 \in [0,2 \pi]$, and yields the Regularized Cross-Newell (RCN) energy 
\begin{equation}
\mathcal{E}[k(x)] = \int_{\Omega} \epsilon^2 (\nabla \cdot k)^2 + W(|k|^2) \, dx,
\label{eq:rcn}
\end{equation}
where $W$ is a nonconvex ``well potential" in $k$. 
For our purposes, $W(|k|^2) = (|k|^2-1)^2$ is an adequate approximation. In the context of patterns, we also need that $k^{\perp}$ should yield a {\em measured foliation} \cite{poenaru1981some}. 
Consequently, a necessary condition for $k$ to describe a smooth stripe pattern is that $curl \, k = 0$ \cite{fathi2012thurston}. With the substitution $k = D \theta$ (equivalent to the constraint $curl \, k = 0$), we recognize that the RCN energy is equivalent to the {\em Aviles-Giga} (AG) functional \cite{aviles1987mathematical}
\begin{equation}
\mathcal{F}[\theta] =  \int_{\Omega} \epsilon^2 (\nabla \nabla \theta)^2 + (1-|\nabla \theta|^2)^2 \, dx,
    \label{eq:AG}
\end{equation}
where $\nabla \nabla \theta$ is the $2 \times 2$ Hessian matrix of second derivatives. The AG functional was initially introduced as a model for smectic liquid crystals \cite{aviles1987mathematical}.

\begin{figure}[htb]
\centering
\subfigure[Pattern for the AG minimizer]{
\includegraphics[trim=0 100 40 80, clip, width=0.47\linewidth]{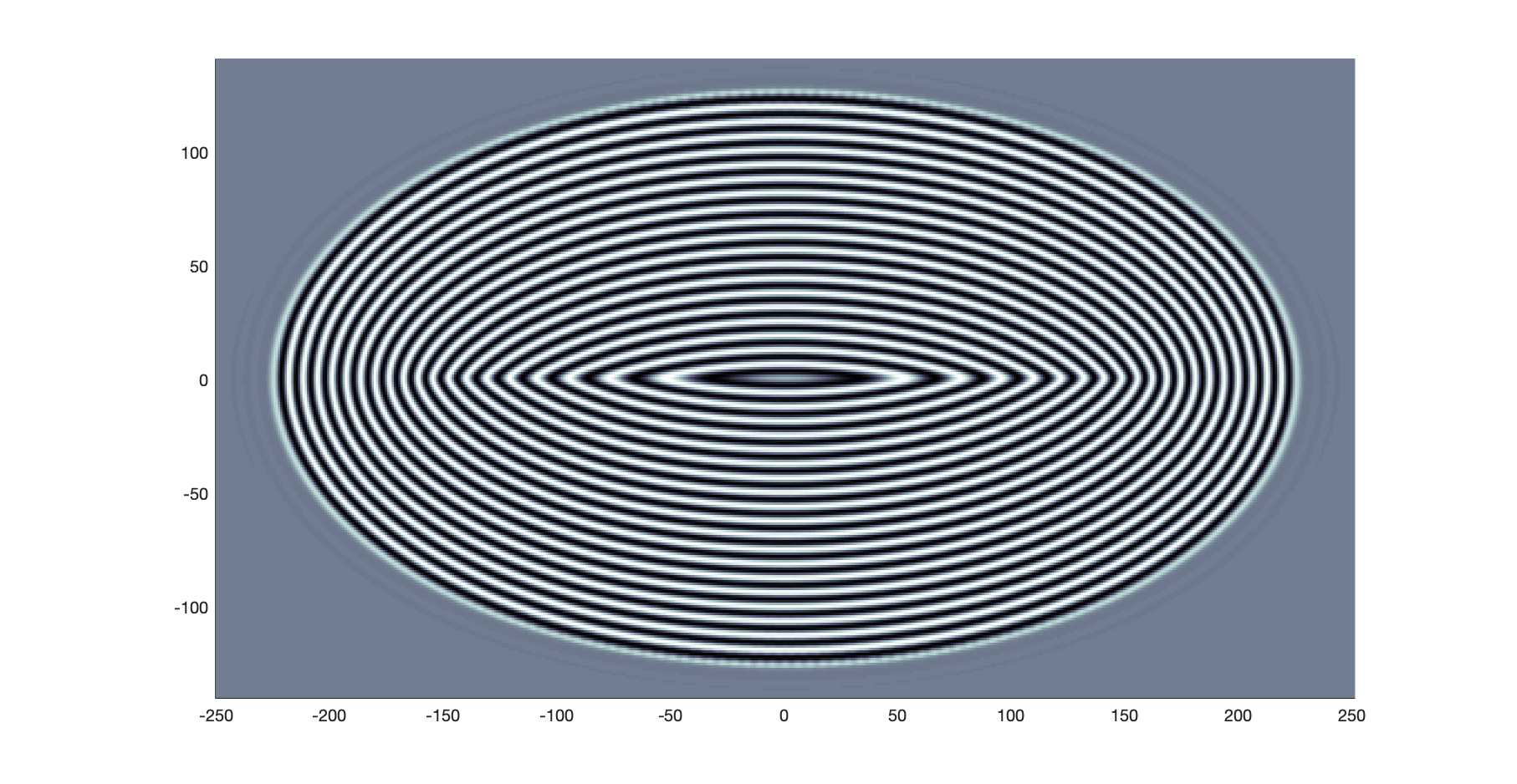}
\label{fig:eikonal}}
\subfigure[Candidate global SH minimizer on an ellipse.]{
\includegraphics[trim=0 200 0 160, clip, width=0.47\linewidth]{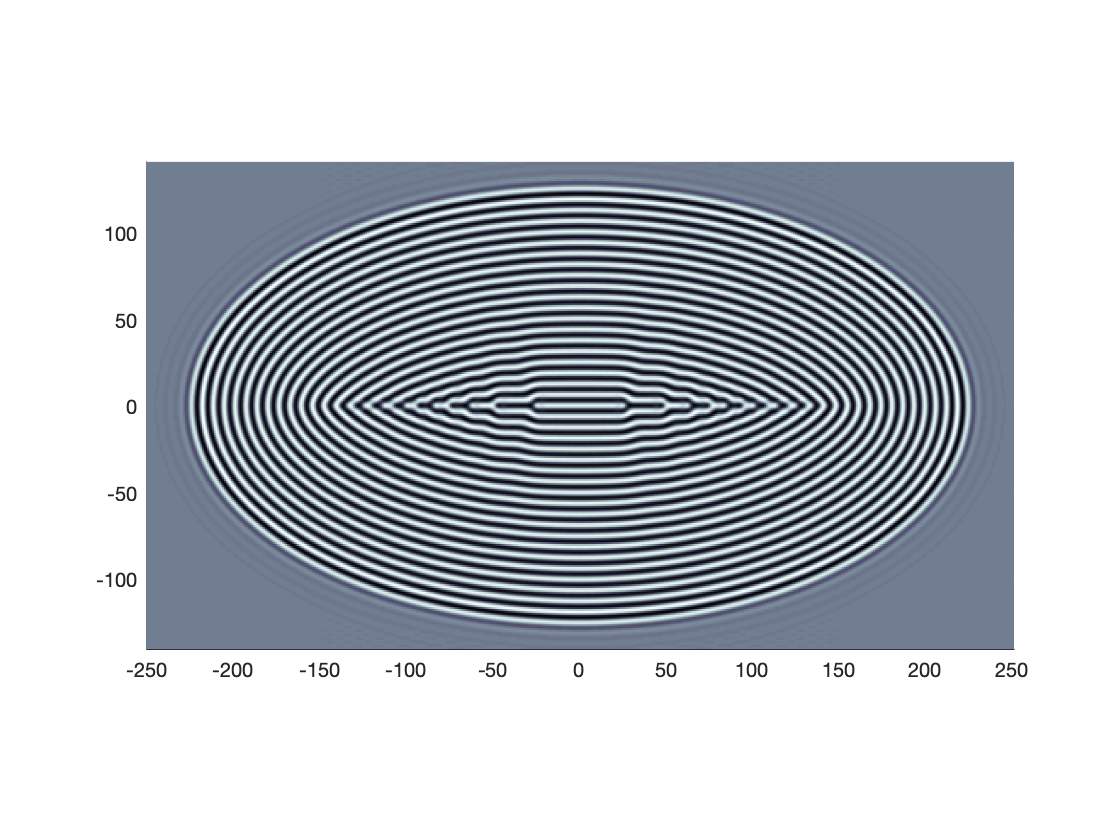}
\label{fig:pearls}}
\subfigure[Random stripe pattern.]{
\includegraphics[trim=0 200 0 160, clip, width=0.47\linewidth]{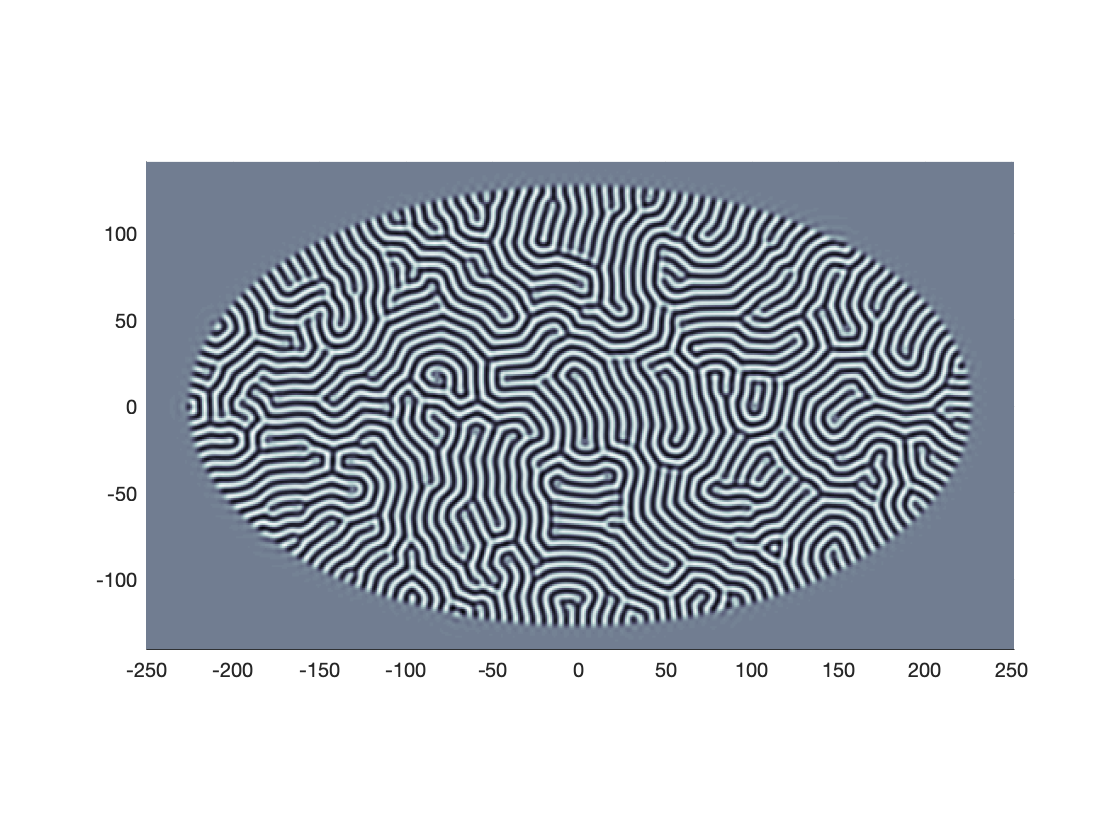}
\label{fig:random}}
\caption{(a) $\cos(\epsilon^{-1} \theta)$ where $\theta$ is a minimizer of the Aviles-Giga energy. 
(b) Long time solutions of the Swift-Hohenberg equation starting with an initial condition given by the state in (a). Along the major axis of the ellipse, we now get a string of point defects -- {\em disclination pairs}. A magnified view of a disclination pair is shown in Fig.~\protect{\ref{fig:bcs}}. (c) Long time solutions of the Swift-Hohenberg equation starting with random initial conditions. 
(a-b) $k$ is set to the unit normal on the boundary of the domain.
}
\label{fig:swift-hohenberg}
\end{figure}

\begin{figure}[tbh]
 \centering
 \begin{subfigure}[Convex disclination]{
  \includegraphics[width = 0.22 \linewidth]{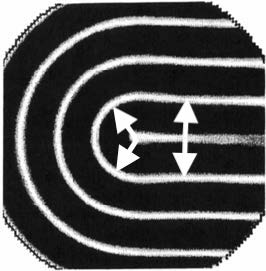}
  \label{fig:convex}}
  \end{subfigure}
  \quad
  \begin{subfigure}[Concave disclination]{
  \includegraphics[width = 0.22 \linewidth]{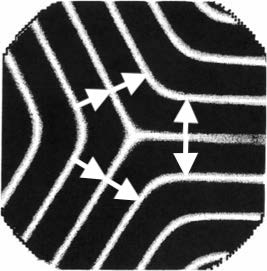}
  \label{fig:concave}}
  \end{subfigure}
  \quad
  \begin{subfigure}[A pair of disclinations.]{
  \includegraphics[height = 0.22 \linewidth]{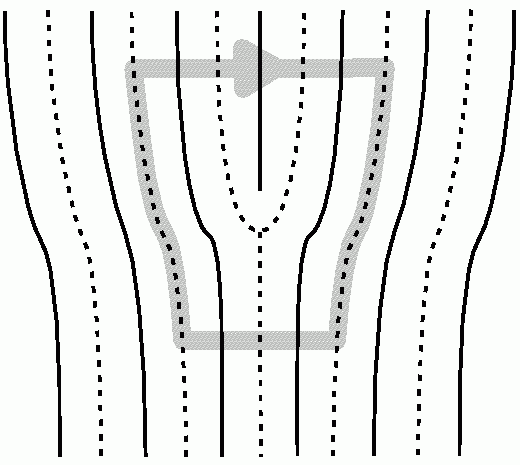}
  \label{fig:disloc}}
  \end{subfigure}
 \caption{(a-b) Disclinations are point defects with nontrivial monodromy for $k$ which as indicated by the flip in the arrow  when it is transported continuously around the defect at the center. They can only occur on the particular contours corresponding to the local maximum and minimum of the pattern field. (c) The convex disclination is on a maximum contour (solid) and the concave disclination is on a minimum contour (dashed), so they cannot annihilate each other, although their strengths add to zero. This configuration has a non-zero ``Burgers scalar" as evident by tracing the closed (gray) loop.}
 \label{fig:disclination} 
 \end{figure}

Jin and Kohn have proved that, for elliptical domains, as the small parameter $\epsilon \to 0$, the minimizers of the Aviles-Giga (AG) functional converge, $\theta^\epsilon(x) \to d(x,\partial \Omega)$, to the distance to the boundary $\partial \Omega$ \cite{Jin2000Singular}. Figure~\ref{fig:eikonal} depicts the  corresponding (putative) steady-state solution of the Swift-Hohenberg equation~\eqref{eq:sh} (as suggested by the AG model), given by the modulation ansatz $u \approx \cos(\epsilon^{-1} \theta)$. One would expect that the modulation of the AG minimizer should give a good approximation to the SH minimizer. However, if we start with the modulated AG minimizer as an initial condition, we find that the SH evolution~\eqref{eq:sh} produces a marked `topological' change to the pattern along the major axis of the elliptical domain. This is evident from comparing the modulated AG minimizer and its steady state under SH evolution, as shown in Fig.~\ref{fig:swift-hohenberg}. The key difference is that the contours (bright and dark) for the AG minimizer are all simple closed curves, while the corresponding contours for the SH solutions have distinctive point defects, {\em convex disclinations} where a contour ``ends" and {\em concave disclinations} that are ``triple points" on contours (See Fig.~\ref{fig:disclination}.) 

These differences are entirely due to fact that the local phase of a pattern is ``multi-valued" \cite{kleinert2008multivalued}. The phase $\theta$ is not directly observable, unlike the pattern field $u(x,t)$ in~\eqref{eq:sh}. Thus, we have to identify different phase functions $\theta$ that give the same field $u = w_0(\theta,|k|^2)$ where $k = D \theta$. Since $w_0$ is an even, $2 \pi$ periodic function of the first argument, we have the identifications  $\theta \to \theta + 2 n \pi, k \to k$ where $n$ is any integer (periodicity or symmetry under translations by multiples of $2 \pi$) and $\theta \to  - \theta, k \to -k$ (evenness or head-tail symmetry). 
In particular, the values $\theta = m \pi$ with integer values for $m$ are distinguished, since we can apply a combination of the two symmetries to achieve $\theta \to - \theta \to -\theta + 2 m \pi = \theta, k \to -k \to -k$, so the contours $\theta = m \pi$ are the locations which can support {\em disclinations}, i.e. flips $ \theta \to \theta, k \to - k$ 
as illustrated in Figs.~\ref{fig:convex}~and~\ref{fig:concave}. 

Disclinations arise from {\em non-orientability} of the order parameter $k$ and are thus point defects with nontrivial monodromy for the map $x \mapsto k = D \theta$. 
They are necessarily codimension 2, i.e. points in 2D and lines in 3D. These arguments were used to develop a variational theory for disclinations in stripe patterns \cite{ercolani2009variational}. These conclusions were also obtained independently in \cite{Chen2009Symmetry,Pevnyi2004Modeling,Aharoni2017Composite} for smectic liquid crystals using different arguments.  

\subsection{Beyond Aviles-Giga: Models incorporating the defect densities} 

It is an interesting question as to how one computes gradient flows for functionals that depend on such ``multi-valued" fields $\theta$ and (the director) $k$. One approach is to introduce branch cuts to obtain a single valued phase $\theta$ and a vector field $k = D \theta$ on 
a branched double cover \cite[Chap. 1]{Lando2004Graphs} of $\Omega$ where the ramification points are the locations of the convex and concave disclinations that carry the nontrivial monodromy of the map $x \mapsto k(x)$ \cite{Ercolani2003Global}. 

An alternative approach, in the spirit of the framework in~\S\ref{sec:NPSN}, is to introduce additional `defect fields' that encode the topological charges at the disclinations. Indeed, in \S\ref{sec:computations2d}, we have successfully applied this approach to the related problem of non-orientable defects in smectic liquid crystals.
In the following discussion, we will outline how one might build such a theory. We emphasize that this is still very much work in progress. We begin with describing the appropriate defect fields for pattens. Following the discussion in \S \ref{sec:model}, we set $k$ as the absolutely continuous part of $\nabla \theta$ (the rehabilitated gradient) and $A$ as the absolutely continuous part  of the distributional gradient $Dk$. 
In analogy with 
the definitions of elastic defects (cf. \cite[\S1]{zhang2018relevance}), the quantity $[k]_\gamma := \int_\gamma A \cdot dx$ is 
the net disclination density enclosed by $\gamma$. If the net disclination density within a curve $\gamma$ is zero, the corresponding {\em Burgers scalar} \cite{Kamien2016topology,Aharoni2017Composite} is well defined and given by $\displaystyle{[\theta]_\gamma:= \sum_{i=0}^{n-1} \int_{\gamma_i} k \cdot dx =  - \sum_i \llbracket \theta \rrbracket_i}$. $[\theta]_\gamma$ is the negative of the sum of the (oriented) jumps in $\theta$ at the points where $\gamma$ intersects $S$. Since the branch cut set $S$ consists of curves where $\theta$ is a multiple of $\pi$ when approached from $\Omega \setminus S$, it follows that $[\theta]_\gamma = m \pi$, where $m$ is an integer that depends on $\gamma$.

\begin{figure}[tbh]
 \centering
  \includegraphics[width = 0.55 \linewidth]{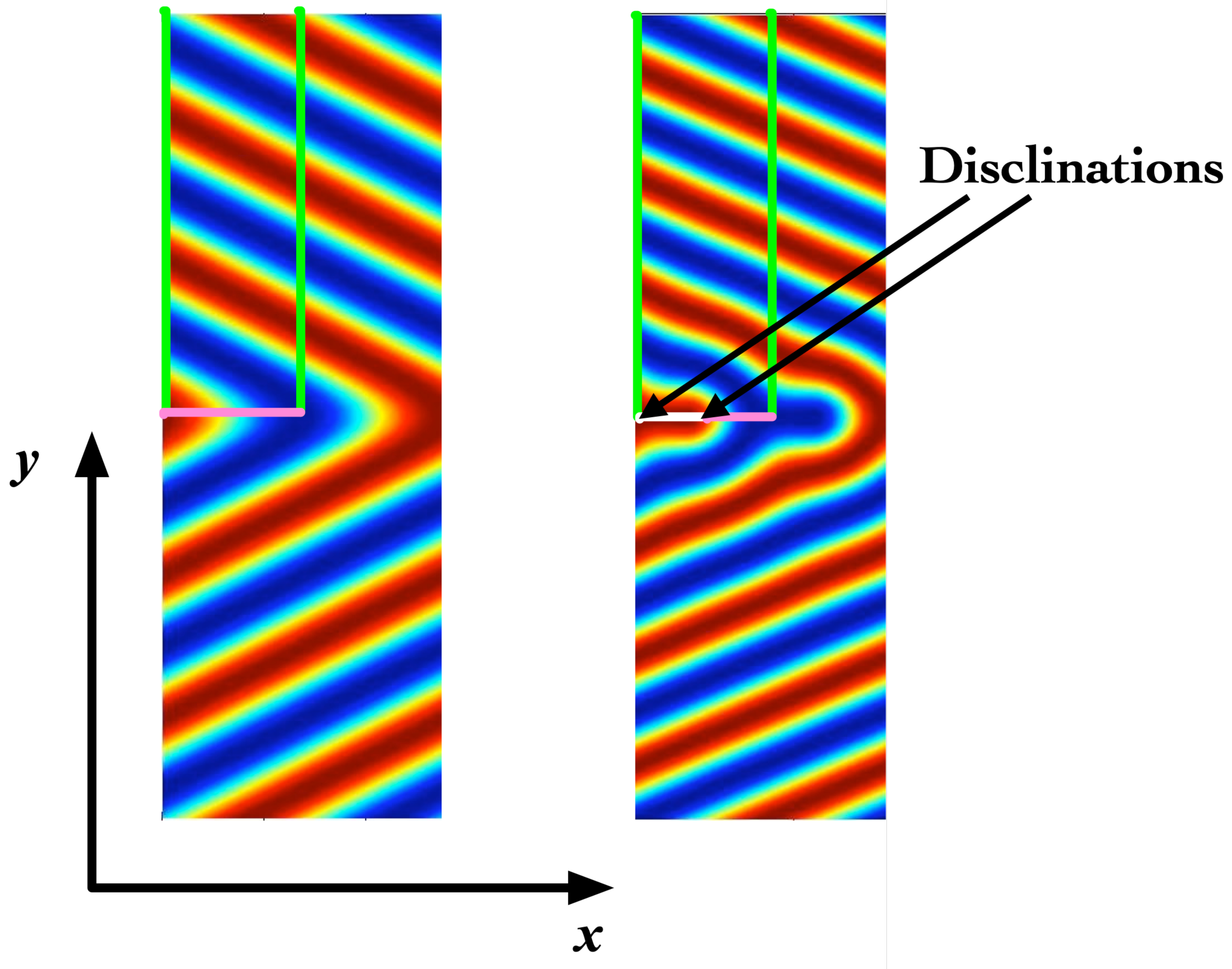}
  \caption{Modifying the boundary conditions to allow for disclination dipoles. The figures show pattern states that are symmetric under $y \to -y$ and shift-invariant under translations in $x$. On the left, we have the boundary condition $\theta_y(x,0) = 0$ implied by the symmetry $\theta(x,-y) = \theta(x,y)$. On the right, $\theta_y$ is not necessarily continuous across $y=0$ due to the possibility of disclinations.}
   \label{fig:bcs}
 \end{figure}

In what follows, we adopt the viewpoint that arbitrary (composite) defects and defect distributions  can be decomposed into sums of elementary defects, corresponding to convex and concave disclinations. This approach has proved very successful in building a topological classification of smectics in 3D \cite{Aharoni2017Composite,Machon2019Aspects}. 
The role of disclinations in stripe patterns within a {\em variational} framework was considered, by Ercolani and the last author of this work, in \cite{ercolani2009variational}. In \cite{ercolani2009variational}, we consider patterns that are 
    (i) symmetric in $y$, $\theta(x,-y) = \theta(x,y)$, and
    (ii) shift-translation invariant in $x$, $\theta(x+l,y) = \theta(x,y) + \pi$, where $l$, the fundamental period of the pattern in the horizontal direction.
The horizontal period is given by $l=\frac{\pi}{k_0 \cos \alpha}$ where $\alpha$ is the ``far-field" inclination of the stripes with the horizontal axis \cite{ercolani2009variational}. Interpreting the results of \cite{ercolani2009variational} in light of the framework that we have developed in \S\ref{sec:model}, we can conclude
\begin{enumerate}
    \item If $\theta(x,-y) = \theta(x,y)$ and $k = \nabla \theta$ is continuous, then we have $\theta_y = 0$ at $y=0$. Considering the variational problem on a ``unit cell" $\mathcal{S} = [0,l] \times [0,\infty)$ (see the figure on the left in Fig.~\ref{fig:bcs}) the Cross-Newell energy is indeed given by the Aviles-Giga expression. 
    \item In the presence of disclinations, $k$ is not necessarily continuous, and head-tail symmetry allows $k \to -k$ across $y=0$. This gives $\theta_y =0$ corresponding to a Dirichlet boundary condition at $y=0$. In general one expects both Dirichlet and Neumann regions to coexist (see the figure on the right in Fig.~\ref{fig:bcs}), so the appropriate variational formulation, allowing for disclination dipoles, is to generalize the boundary conditions on the unit cell to a mixed/free bc, $\theta(x,0) = m \pi$ or $\theta_y(x,0) = 0$ \cite{ercolani2009variational}.
\end{enumerate} 
We can, formally, write down such a theory. A jump in the far-field gradient between $\nabla\theta^{\pm} = (\cos \alpha,\pm \sin \alpha)$ costs an energy $|[\nabla \theta]|^3 \sim k_0^3 \sin^3 \alpha$ per unit length along the defect for the Aviles-Giga energy \cite{aviles1987mathematical,Jin2000Singular}. The alternative is to introduce one disclination pair per unit cell, and as proved in~\cite{ercolani2009variational}, this costs an energy $\sim c k_0^2$ where $c$ will denote an $O(1)$ constant, whose precise value we do not track, and can change from one line to the next. If we assume that the wavelength $k_0^{-1}$ is small, so that we can replace a string of discrete defects by an equivalent defect density, the energy in point defects per unit length is $k_0^2/l \sim c k_0^3 \cos \alpha \sim c k_0^2 (4 k_0^2 - [\nabla \theta]^2)^{1/2}$. This suggests an effective energy, that allows for disclination dipoles, given by a line-defect energy density 
\begin{equation}
    E = \min( |[\nabla \theta]|^3,ck_0^2(4k_0^2-|[\nabla \theta]|^2)^{1/2}).
    \label{eq:defect-RCN}
\end{equation}
Note that, if the jump $|[\nabla \theta]| =2k_0$, this energy density is zero reflecting the non-orientability/ $k\to -k$ symmetry of the field $k$ for patterns. We emphasize that Eq.~\eqref{eq:defect-RCN} is, for the moment, conjectural, although it would naturally imply the results in~\cite{ercolani2009variational} that were proven rigorously using other techniques. Abstractly, a possible approach to obtaining this effective energy might be to use the framework from \S\ref{sec:NPSN} for an upscaled/coarse-grained SH energy and then minimize over the field $B$ to obtain an effective energy functional, but there are significant challenges in carrying out the analysis needed for this partial minimization. A more concrete and plausible approach might be to determine the ``material parameters" in the general energy functional~\eqref{eqn:energy} to match with the results of a modulation analysis away from defects, as in~\eqref{eq:cn} and a numerical determination of the energy landscape for defects \cite{lloyd2017continuation}. We are currently pursuing these ideas.

\subsection{Additional defect densities for patterns in 2D}  

The dislocation and disclination densities given by $[\theta]_\gamma$ and $[k]_\gamma$ are ``linear" defect measures, in that they depend linearly on (jumps in the fields) $\theta$ and $k = D \theta$ outside $S$. 
For two dimensional stripe patterns there is  a further 
defect measure $J = \mathrm{Det}(D^2 \theta) = \theta_{xx} \theta_{yy} - \theta_{xy}^2$, the Gaussian curvature of the phase surface $(x,y) \mapsto \theta(x,y)$ \cite{newell1996defects,newell2017elastic}. If $k = 1$ a.e., $J$ is related to the field $curl\, \lambda$ in the {\em angle parametrization} (cf. Appendix \ref{app:angle_3D}) which was introduced in \cite{zhang2016non} as a representation of defects in planar nematic director fields. 
\begin{equation}
  T = \frac{1}{2\pi} \iint_D J \, dA = \frac{1}{2\pi}\oint_\gamma \theta_y d \theta_x - \theta_x d \theta_y = \frac{1}{2 \pi}\oint_\gamma \frac{ \theta_y d \theta_x - \theta_x d 
\theta_y}{|D \theta|^2}  = \frac{[\tan^{-1}(\theta_y/\theta_x)]_\gamma}{2 \pi}  
\label{eq:twist}
\end{equation}
for a domain $D$ whose boundary $\gamma = \partial D$ intersect $S$ transversally, relating the {\em total twist} of the angle parametrization along $\gamma$ to the mass of $T$ of the measure $J$. 

$J$ is one of an entire family of such measures that can encode this twist. For any (nonlinear) scalar valued function $f(k)$, that has a branch cut intersecting the unit circle $|k| = 1$ transversally, integrating the absolutely continuous (regular) part of $D f\circ k$ along a closed curve in the real domain picks up a jump of $f\circ k$ when $Df \circ k$ has a singularity in the domain inside the curve. 
The function $f(k) = \tan^{-1} ( \frac{k \cdot e_2}{k \cdot e_1} )$, where $(e_1,e_2)$ is an arbitrary orthonormal frame, gives the defect measure $J$. This specific function $f$, the angle parametrization of the planar director field, was used as the fundamental field in the model in \cite{zhang2016non}. For our framework in \S\ref{sec:NPSN}, we instead use the entire vector $k$, because the resulting theory generalizes easily to 3 dimensions, and has closer connections to defects in solids, as we discuss in \S\ref{sec:orderparameters}. For completeness, we present an extension of the angle parametrization to 3 dimensions in Appendix~\ref{app:angle_3D}.

To summarize our arguments in this section, in order to obtain disclinations, the relevant topological defects in natural patterns, it does not suffice to introduce ``some" microscopic physics (regularization) into a macroscopic model. In particular, only adding a microscopic regularization through the Laplacian/Hessian of the phase leads to the RCN/AG energy and {\em these energies do not adequately capture the effects of disclinations and disclination dipoles in the pattern}. The coarse-graining leading to RCN/AG explicitly depends on the existence of a well defined (macroscopic) phase function $\theta(x,t)$. There isn't such a phase function in the presence of disclinations, as we argue in \S\ref{sec:layers} (Also see Fig.~\ref{fig:disclination}). Consequently, without further `enrichment', RCN/AG is not equipped to model disclinations \cite{newell1996defects,Ercolani2003Global}, which are ubiquitous in the full `microscopic' theory given by the Swift-Hohenberg equation. We need a theory that accounts for {\em microscopic non-orientability} and involves the corresponding defect field, {\em viz}. the {\em disclination density} measured by $[k]_\gamma$. The defect energy Eq.~\eqref{eq:defect-RCN} is a conjecture for one possible form of this enriched theory. More generally, the defect fields corresponding to the Burgers scalar $[\theta]_\gamma$, the net disclination $[k]_\gamma$, and the twist measure $J$, might be necessary ingredients in macroscopic models that adequately capture complex behaviors seen in natural patterns ({\em cf}. Fig.~\ref{fig:random}). Building such theories, from the framework outlined in \S\ref{sec:NPSN}, is a task for the future.

\section{Discussion} \label{sec:discuss}


Patterns (ordered microstructures) and defects (breaking of an ordered pattern) are ubiquitous in extended systems. They arise from the interplay between two ``universal" mechanisms, the tendency of systems towards order as their energy/temperature is lowered, and the tendency towards disorder from entropic considerations and the likelihood of ``getting stuck" in ``local" metastable states, that precludes perfect ordering. These defects play a big role in the properties of extended systems and understanding  the birth, disappearance and dynamics of defects is critical in explaining a range of phenomena from plasticity, solid-solid phase transitions, fracture, convective transport, and complex fluids. It is thus of considerable interest to develop modeling and numerical methods to explain, analyze and predict the behaviors of defects and the extended systems they live in. 

Fortunately, defects have universal features, independent of the underlying physics, reflecting their topological origins. A primary thrust of this work is to exploit this universality to develop a modeling framework and associated numerical methods that are applicable to computing defect driven behaviors in a wide range of systems of interest in materials science and continuum mechanics. We describe a common language for defects in natural patterns, smectics, and nematics, that draws on classical ideas for defects in solids \cite{volterra,weingarten} which have been incorporated into practically computable modern continuum mechanical theory recently \cite{acharya2015continuum,zhang2018relevance,zhang2018finite,arora_acharya_ijss, acharya2020continuum, arora2020finite, arora2020unification}. In this language, we develop a modeling framework that captures the dynamics of defects in terms of {\em integrable energy densities}, an important consideration for having a good numerical formulation. Our models can handle order parameters that have a head-tail symmetry, i.e. director fields, in systems with a continuous translation symmetry ({\em e.g.} nematic liquid crystals) and in systems where this symmetry is broken and replaced by a discrete translation symmetry ({\em e.g.} smectics and convection patterns). The framework we develop gives entire classes of models, and it allows for a natural incorporation of thermodynamic principles, the balance laws of mechanics and/or other physical principles like conservation laws for topological charges of defects. 

We illustrate our methods with explicit computations for equilibrium configurations of nematic and smectic/pattern systems in 2-d (cf. \S\ref{sec:computations2d}) and in 3-d, illustrating the scope of our method. Now that we can adequately capture the equilibrium behavior of defects, an outstanding challenge is to capture the dynamics of defects. Some success has been shown in \cite{zhang2016non}, but much remains to be done in this regard, {\em e.g.} in modeling the relaxation of the random stripe pattern in Fig,~\ref{fig:random}  to the global ground state in Fig.~\ref{fig:pearls}. Another area of exploration is the coupling of our model to fluid flow to model liquid crystalline polymer flows with a high density of defects.

In a speculative vein, the philosophy guiding our work might be extendable to various scenarios, some of which are not normally thought of as related to ``defects." An example of such a situation is high Reynolds number Navier-Stokes flows in which intense vortex sheet and filament structures emerge resembling Euler flows. A heavy dose of artificial viscosity can of course not represent what is physically observed in such flow regimes as well as in direct numerical simulations (DNS). This is analogous, in a way, to not including the `correct' micro-physics in obtaining the AG functional as a limiting energy for stripe patterns.  Is it possible that accounting for such vortex sheet and filament structures by our kinematical constructs - along with the additional elasticity of short-range ordering that all fluids possess and the dissipation that is plausibly associated with the motion of such filamentary structures in disrupting such ordering through the generation of incoherent atomic motions sensed as temperature in a pde representation - can help in a better representation of high  Reynolds number flows? Systematic models along these lines can definitely be developed \cite{acharya2019some}, and their consequences remain to be explored.

\section*{Acknowledgments}
AA, SCV, and ACN were supported by the NSF Growing Convergence Research award 2021019. SCV is partially supported by the Simons Foundation through award 524875 and by the National Science Foundation through award DMR 1923922. This work was initiated during a visit by AA to the Dept.\! of Mathematics at the University of Arizona; their hospitality is much appreciated. Portions of this work were carried out when SCV was visiting the Center for Nonlinear Analysis at Carnegie Mellon University, and their hospitality is gratefully acknowledged.

\bibliographystyle{alpha}\bibliography{pattern_ref}

\appendix
\section*{Appendix}

\section{Angle parametrization of director field}\label{app:angle_3D}

Motivated by the demonstrations of representing energetic and dynamics of a planar director field with angle parametrization in \cite{zhang2016non}, we introduce a way to parametrize director field $k$ and show the relation between angle parametrization and the full 3d model proposed in this work.

\begin{figure}[hbtp]
    \centering
    \subfigure[Parametrization of $k$ in terms of $\eta$ and $\phi$.]{
    \includegraphics[width=0.4\linewidth]{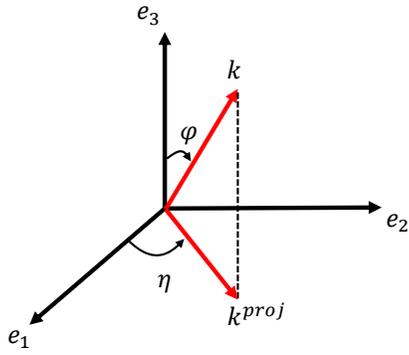}
    \label{fig:append_para1}} \qquad
    \subfigure[Constructing a 3D space by rotating a half plane $2\pi$.]{
    \includegraphics[width=0.4\linewidth]{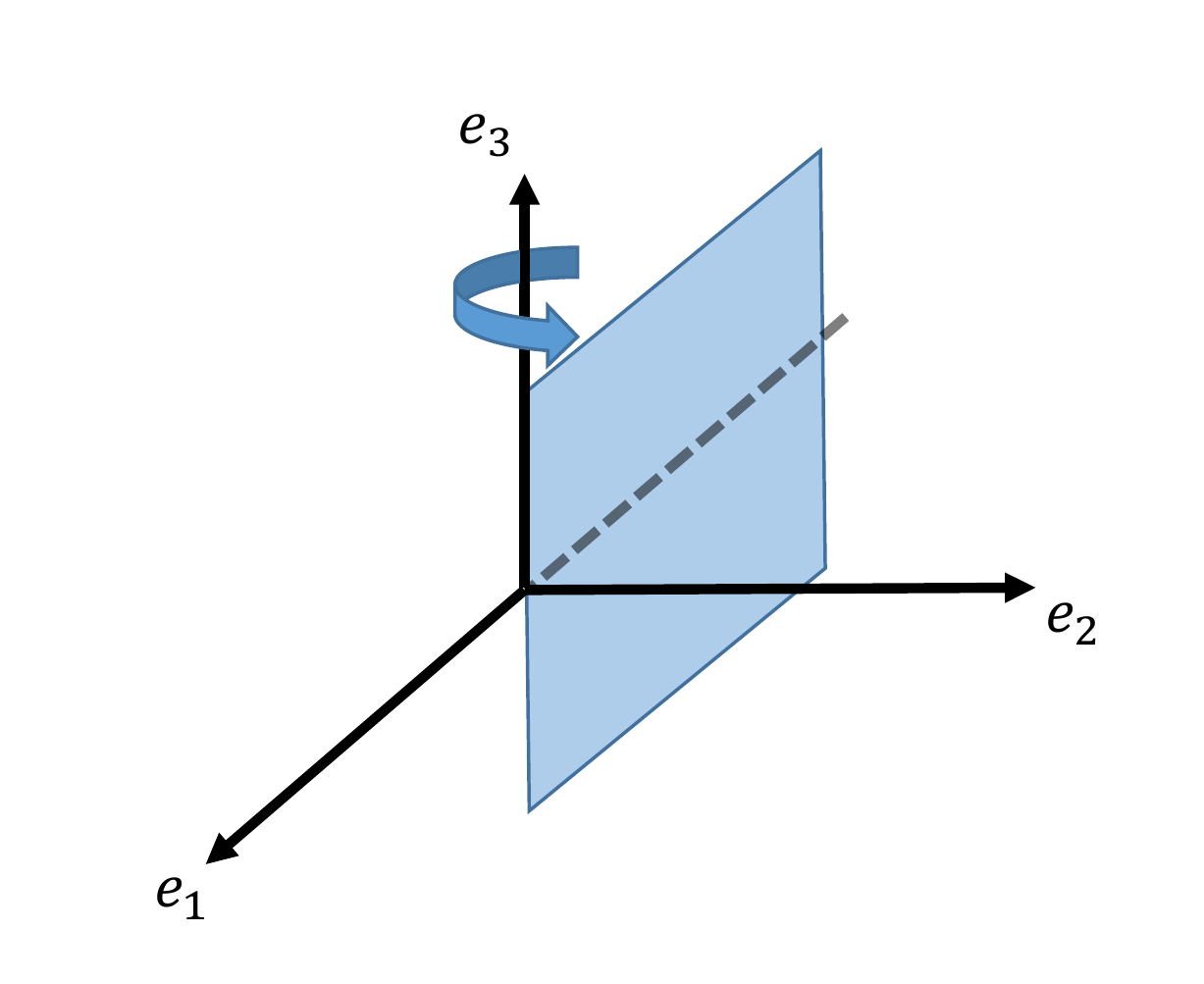}
    \label{fig:append_para2}} \qquad
    \subfigure[Two directors with opposite directions in a defect. The discontinuity can be represented by the jump between angle $\phi$.]{
    \includegraphics[width=0.4\linewidth]{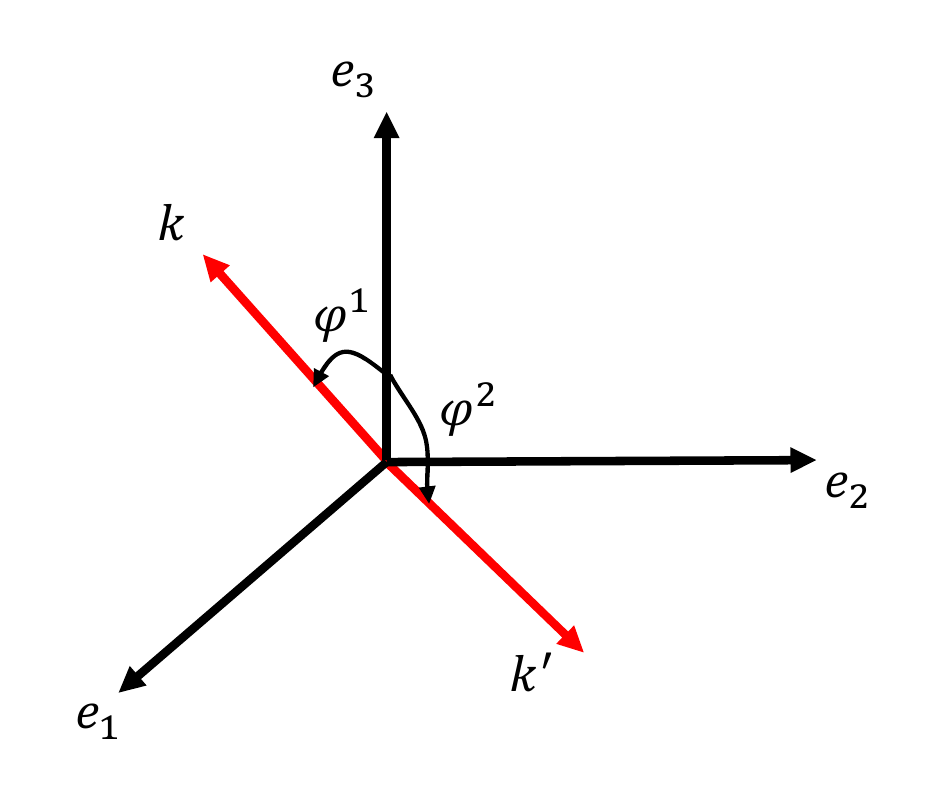}
    \label{fig:append_para3}}
    \caption{Illustrations of parametrization of $k$ in $\eta$ and $\phi$.}
\end{figure}

Consider a parametrization of $k$ as the representation of angle fields $\eta$ and $\phi$ in any Cartesian coordinates, as shown in Fig.~\ref{fig:append_para1}. The 3D space is constructed by rotating a half plane (the shaded plane shown in Fig.~\ref{fig:append_para2} along axis $\bfe_3$. Given the coordinate in Fig.~\ref{fig:append_para2}, $\eta$ is defined as the rotation angle between $\bfe_1$ and the half plane, within range between $-\pi$ and $\pi$. Namely $\eta$ is the angle between the projection of $k$ on $\bfe_1-\bfe_2$ plane and $\bfe_1$ axis. Similarly, we define $\phi$ as the angle between $k$ and $\bfe_3$, ranging from $-\pi$ to $\pi$.Thus, given a director $k$, $\eta$ and $\phi$ can be calculated as
\begin{align*}
    \eta = \arctan(k \cdot \bfe_2, k \cdot \bfe_1) \\
    \phi = \mathrm{sign}(k \cdot \bfe_2) \arccos(k \cdot \bfe_3),
\end{align*}
where $\arccos$ is the inverse cosine function whose range is from $0$ to $\pi$, $\arctan(y, x)$ is the inverse tangent function returning angle ranging between $-\pi$ and $\pi$ whose tangent value is $\frac{y}{x}$, and $sign$ is a function returning the sign of $\phi$. On the other hand, given a pair of $(\eta, \phi)$, the director $k$ can be written as 
\[
k = \cos \eta \sin |\phi| \bfe_1 + \sin \eta \sin |\phi| \bfe_2 + \cos \phi \bfe_3.
\]

Based on above parametrization, the jump of director $k$ in a defect can be interpreted in terms of $\phi$. For example, for a half strength defect, the director $k$ changes its direction shown in Fig.~\ref{fig:append_para3}, and the difference between $\phi^1$ and $\phi^2$ is $\pi$. To demonstrate the connection with planar cases discussed in \cite{zhang2016non}, we adopt same notation $\lambda$ to represent director discontinuity in $\phi$. Then the regular part of elastic distortion gradient $A$ can be written as
\[
A = \partial_\eta k \otimes D \eta + \partial_\phi k \otimes (D \phi - \lambda) = Dk - \partial_\phi k \otimes \lambda.
\]
It is easy to verify that $A = Dk$ in defect-free cases where $\lambda = 0$. In addition, $\partial_\phi k$ can be calculated as
\begin{align*}
    \partial_\phi k &= \tanh \eta \cos \eta \cos \phi \bfe_1 + \tanh \eta \sin \eta \cos \phi \bfe_2 - \sin \phi \bfe_3 \\
    &= k_1 k_3 \tanh k_2 \bfe_1 + k_2 k_3 \tanh k_2 \bfe_2 - |k - k_3 \bfe_3| \bfe_3.
\end{align*}

In terms of the fields $k$, $A$, and $\lambda$, an augmented Oseen-Frank energy density that views $\beta := curl \, \lambda$ as a disclination defect density field for nematics and smectics is as follows:
\begin{align*}
\psi = \frac{1}{2\rho}[K_1|A:I|^2 + K_2|k\cdot (X:A)|^2 + K_3 | k \times (X:A)|^2 + (K_2+K_4)(|A|^2 - |A:I|^2) \\
+ \epsilon |\beta|^2] + P_1(|k|-1)^2 + \alpha K^* g(|\lambda|),
\end{align*}
where $g$ is a nondimensional nonconvex function of $|\lambda|$ with wells at all integer multiples of $\pi$. For isolated disclinations, $\lambda$ may be specifed on a (non-planar) terminating layer around a surface with unit normal field $\nu$: $\lambda = \frac{\llbracket \phi \rrbracket}{l} \nu$ with support on the layer, and it can be shown that the $\beta$ field in that case is localized at the termination of the layer. For example, for the 3D squared loop defect in \S\ref{sec:squared_loop}, $\phi$ of directors on top layer is $\frac{\pi}{2}$ while $\phi$ of directors on bottom layer is $-\frac{\pi}{2}$. Thus, $\lambda$ can be prescribed as
\begin{equation*}\label{eqn:lambda_3d}
\lambda(x,y,z) = \begin{cases}
\frac{\pi}{2a \xi}\bfe_3, & \text{ if $|z| \le {\frac{a\xi}{2}}$, $|x|\le d$, and $|y| \le d$} \\
0, & otherwise.
\end{cases}
\end{equation*}

The model above, while confirming to conventional intuition on thinking about disclinations in nematics and smectics, however is not, at least manifestly, invariant to the choice of the arbitrary orthonormal frame used in the definition of $A$. 

Assuming $k$ to be a unit vector, we note that $B = \partial_\phi k \otimes \lambda$. Then
\[
- \pi = curl \, B = D \partial_\phi k \times \lambda + \partial_\phi k \otimes curl \, \lambda.
\]
When $\lambda$ is of the form $\lambda = a \,\nu$ with support on a terminating layer around a surface with unit normal field $\nu$ and $a$ is a constant, then it can be shown that $curl \, \lambda = 0$, except at the termination of the surface. In addition, if the $k$ field does not have any longitudinal variations along the layer, then both the defect densities $\pi$ and $\beta = curl \, \lambda$ are localized at the same location. If $k$ has longitudinal variations, then $\pi$ is distributed all along the layer.

\section{Numerical formulation for the gradient flow equations \eqref{eqn:grad_flow_gov}}\label{app:numerical_derivation}

In this section, we provide the derivation of gradient flow dynamic equations \eqref{eqn:grad_flow_gov} of the energy \eqref{eqn:energy} and its discretization for our numerical computations.

From \eqref{eqn:energy}, the total energy $E$ of a given body $\Omega$ is
\begin{equation}
    E = \int_\Omega \psi dv
    = \int_\Omega \left\{P_1 \left( |k| - 1\right)^2 + P_2 |curl\, k|^2 + \alpha K^* f( |B|) + K |D k - B|^2 + \veps |\pi|^2 \right\}dv.
\end{equation}

The first variation of the energy $E$ at a state $(k,B)$ in the direction of variations $(\delta k, \delta B)$ that vanish on $\partial \Omega$ is given by 
\begin{align*}
\delta E &= \int_\Omega \left\{  2 P_1 \left( |k| -1 \right) \delta |k| + 2 P_2 curl\, k \delta curl\, k + \alpha K^* \frac{\partial f(|B|)}{\partial B} \delta B + 2 K \left(D k - B\right) \left(\delta Dk - \delta B\right) \right. \\ 
&  \qquad \left. + 2 \,\veps \,curl\, B \delta curl\, B  \vphantom{\int_\Omega} \right\} \, dv = 0.
\end{align*}
Writing out the variations in components, one has
\begin{align}\label{eqn:weak_form}
    \int_\Omega & \left\{ 2 P_1 \left( |k| -1 \right)\frac{k_i}{|k|} \delta k_i + 2 P_2 (curl\, k)_n e_{nmi} \delta k_{i,m} + \alpha K^* \frac{\partial f(|B|)}{\partial B_{ij}} \delta B_{ij} + 2 K \left(D k - B\right)_{ij} \left(\delta k_{i,j} - \delta B_{ij}\right) \right. \notag\\ 
& \quad \left.+ \, 2 \veps (curl\, B)_{im} e_{mnj} \delta B_{ij,n} \vphantom{\int_\Omega} \right\} dv = 0,
\end{align}
and after an integration by parts, 
\begin{align} \label{eq:energy_delta}
   \delta E =  \int_\Omega \left\{ 2 P_1 \left( |k| -1 \right)\frac{k_i}{|k|} \delta k_i - 2 P_2 (curl\, k)_{n,m} e_{nmi} \delta k_{i} + \alpha K^* \frac{\partial f(|B|)}{\partial B_{ij}} \delta B_{ij} - 2 K \left(D k - B\right)_{ij,j} \delta k_{i} \right. \notag \\
\left. - 2 K \left(D k - B\right)_{ij} \delta B_{ij} - 2 \veps (curl\, B)_{im,n} e_{mnj} \delta B_{ij} \vphantom{\int_\Omega} \right\} dv = 0.
\end{align}
The coefficients of $\delta k$ and $\delta B$ in \eqref{eq:energy_delta} define the variational derivatives of $E$ w.r.t $k$ and $B$, yielding the `gradient flow' equation \eqref{eqn:grad_flow_gov} given by $\partial_s k = - \frac{\delta E}{\delta k}$ and $\partial_s B = - \frac{\delta E}{\delta B}$. 

In this work, we use a Galerkin finite element discretization of the test and trial fields in space of the form
\begin{align}\label{eqn:discretization}
 (\delta) k_i(x,y,s) = (\delta)k_i^{A,s}N^A(x,y) \notag\\
  (\delta)B_{ij}(x,y,s) = (\delta)B_{ij}^{A,s}N^A(x,y),
\end{align}
where $A$ is an index that ranges over the nodes of the finite element mesh, summation over repeated node indices is implied, and $N^A(x,y)$ represents the shape function (for trial and test function) corresponding to node $A$ with the Kronecker property that $N^A(x_C,y_C) = \delta_{AC}$. Finite-element shape functions have the `localization' property that $N^A$ has support only in the elements for which node $A$ is a vertex. We use first order, bilinear shape functions. Below, upper case Latin indices will refer to node numbers on the finite element mesh (and we will avoid using the letter $B$ as an index).

The discrete governing equations for the nodal degrees of freedom $\left(k_i^{A,s}, B_{ij}^{A,s}\right)$ at any discrete instant of time `$s$' follow from multiplying the gradient flow equations by test functions and integrating by parts, resulting in evolution equations whose rhs are given by the negative of the lhs of \eqref{eqn:weak_form}:
\begin{align}\label{eqn:disc_weak_form_k}
&\int_\Omega \delta k \cdot \partial_s k \, dv = - \int_\Omega \delta k \cdot \frac{\delta E}{\delta k} \, dv \Longrightarrow M^{AC}_{ij} \left( k_j^{C,s+\Delta s} - k_j^{C,s} \right) = \Delta s R^{A,s}_{i(k)} \notag\\
&\mbox{where, for each }i,j \notag\\
&M^{AC}_{ij} := \int_\Omega N^A N^C \, dv; \notag\\
& R^{A,s}_{i(k)} := - \int_\Omega \left. \left( 2 P_1 \left( |k| -1 \right)\frac{k_i}{|k|} N^A - 2 P_2 (curl\, k)_n e_{nmi} N^A_{,m} - 2 K \left(D k - B\right)_{ij} N^A_{,j} \right)\right|_s \,dv,
\end{align}
and,
\begin{align}\label{eqn:disc_weak_form_B}
&\int_\Omega \delta B : \partial_s B \, dv = - \int_\Omega \delta B : \frac{\delta E}{\delta B} \, dv \Longrightarrow M^{AC}_{ijkl} \left( B_{kl}^{C,s+\Delta s} - B_{kl}^{C,s} \right) = \Delta s R^{A,s}_{ij(B)} \notag\\
&\mbox{where, for each }i,j,k,l \notag\\
&M^{AC}_{ijkl} := \int_\Omega N^A N^C \, dv; \notag\\
& R^{A,s}_{ij(B)} := - \int_\Omega \left. \left( \alpha K^* \frac{\partial f(|B|)}{\partial B_{ij}} N^A  - 2 K \left(D k - B\right)_{ij} N^A - 2  \veps (curl\, B)_{im} e_{mnj} N^A_{,n}\right)\right|_s \,dv
\end{align}

Standard (4-point) Gauss integration is used to evaluate the 2-d integrals arising from bilinear interpolations, with the integrands evaluated on the discrete fields at time $s$ computed from the known nodal array $\left(k^{A,s}_i, B^{A,s}_{ij}\right)$ based on \eqref{eqn:discretization}. The nodal array of field values is evolved from the discrete time $s$ to $s+ \Delta s$ based on the difference equation \eqref{eqn:disc_weak_form_k}-\eqref{eqn:disc_weak_form_B} and we use `mass lumping'  in the form of an Identity matrix to avoid any matrix solves (this is adequate since we are interested in equilibrium states in this contribution).

\end{document}